\documentclass[11pt,a4paper]{article}
\pdfoutput=1 
\usepackage[dvipdfmx]{graphicx}
\usepackage[dvipdfmx]{color} 
\usepackage{soul}
\usepackage[makeroom]{cancel}
\usepackage{mathrsfs}
\usepackage{epstopdf}
\usepackage[utf8]{inputenc}
\usepackage{hyperref}
\usepackage{slashed}
\usepackage{graphicx}
\usepackage{pict2e}
\usepackage{tikz}
\usepackage{float}
\usepackage[percent]{overpic}
\usepackage{multirow}
\usepackage{physics}
\usepackage{booktabs}
\usepackage{mwe}
\usepackage{cite}
\usepackage{bbm}
\usepackage{todonotes}

\newcommand{\beq}{\begin{eqnarray}}
\newcommand{\eeq}{\end{eqnarray}}
\newcommand{\ba}{\begin{eqnarray}}
\newcommand{\ea}{\end{eqnarray}}
\newcommand{\be}{\begin{equation}}
\newcommand{\ee}{\end{equation}}
\newcommand{\bpmatrix}{\begin{pmatrix}}
\newcommand{\epmatrix}{\end{pmatrix}}

\newcommand{\comment}[1]{\ignorespaces}

\usepackage[T1]{fontenc} 
\usepackage{mathpazo} 
\usepackage{amsmath}
\usepackage{braket}

\usepackage{physics}
\usepackage[compat=1.1.0]{tikz-feynman}
\usepackage{amssymb}

\usepackage{tabu}

\usepackage{listings} 
\let\origthelstnumber\thelstnumber
\makeatletter
\newcommand*\Suppressnumber{%
  \lst@AddToHook{OnNewLine}{%
    \let\thelstnumber\relax%
     \advance\c@lstnumber-\@ne\relax%
    }%
}

\newcommand*\Reactivatenumber[1]{%
  \setcounter{lstnumber}{\numexpr#1-1\relax}
  \lst@AddToHook{OnNewLine}{%
   \let\thelstnumber\origthelstnumber%
   \refstepcounter{lstnumber}
  }%
}
\makeatother

\usepackage{subcaption}

\newcommand{\lambdaHPQ}{\lambda_{h\phi_q}}
\newcommand{\mpq}{m_{\phi_q}}

\setstcolor{red}

\begin{document}

\title{
	\vspace*{-3cm}
	\phantom{h} \hfill\mbox{\small KA-TP-18-2023}
	\vspace*{0.7cm}
\\[-1.1cm]
	\vspace{15mm}   
	\textbf{Dark Coloured Scalars Impact on Single and Di-Higgs Production at the LHC \\[4mm]}} 
\date{}
\author{
Pedro Gabriel$^{1,2\,}$\footnote{E-mail: \texttt{pedrogabriel347@hotmail.com}},
Margarete M\"{u}hlleitner$^{2\,}$\footnote{E-mail:
	\texttt{margarete.muehlleitner@kit.edu}},
Daniel Neacsu$^{3\,}$\footnote{E-mail:  \texttt{danielstnea@gmail.com}},
Rui Santos$^{1,4\,}$\footnote{E-mail:  \texttt{rasantos@fc.ul.pt}}
\\[9mm]
{\small\it
$^1$Centro de F\'{\i}sica Te\'{o}rica e Computacional,
    Faculdade de Ci\^{e}ncias,} \\
{\small \it    Universidade de Lisboa, Campo Grande, Edif\'{\i}cio C8
  1749-016 Lisboa, Portugal} \\[3mm]
{\small\it
$^2$Institute for Theoretical Physics, Karlsruhe Institute of Technology,} \\
{\small\it Wolfgang-Gaede-Str. 1, 76131 Karlsruhe, Germany}\\[3mm]
{\small\it
$^3$LIP -
 Laboratório de Instrumentação de Física Experimental de Partículas,} \\
{\small \it  Complexo Interdisciplinar (3is),  Av. Gama Pinto, n.2, piso 3.
  1649-003 Lisboa, Portugal} \\[3mm]
{\small\it
$^4$ISEL -
 Instituto Superior de Engenharia de Lisboa,} \\
{\small \it   Instituto Polit\'ecnico de Lisboa
 1959-007 Lisboa, Portugal} \\[3mm]
}

\maketitle

%
%

\begin{abstract}
The search for Dark Matter (DM) at colliders is primarily pursued via the detection of missing energy in particular final states. These searches 
are based on the production and decay processes where final states include DM particles and at least one Standard Model (SM) particle.  DM will then reveal itself as missing energy. 
An alternative form to get a hint of a dark sector is via loop contribution to SM processes. In this case, it is not even relevant if the new particles have their origin in the dark sector of the model. 
In this work we discuss the impact of an arbitrary number of coloured
scalars in single Higgs and double Higgs production at the Large
Hadron Collider (LHC), and we show their complementarity. 
%
%
We determine the range of variation of the corrections relative to the
SM for an arbitrary number of coloured scalars $n$, and discuss in more detail the cases $n=1$ and $n=2$.   
\end{abstract}

\newpage

\section{Introduction}

Any extension of the Standard Model (SM) aiming at
  solving the Dark Matter (DM) puzzle has to include at least one DM candidate. One of the simplest ways to address this problem is to enlarge the scalar sector of the SM by including a dark sector, usually using a discrete symmetry, and a portal coupling that connects the two sectors. Once a minimal model that provides a DM candidate is built, one
needs to make sure that it is in agreement with the current measurement of the relic density and with all results from direct and indirect detection together with the constraints
imposed by collider experiments. Models with a dark sector can then be further extended to explain other unsolved issues of the SM. Ultimately, any complete extension of the SM has to be in agreement
with all available experimental data.  

In recent years many models have been proposed to solve other discrepancies between the SM predictions and the experimental results. A particular class of models
manages to solve two of these problems simultaneously: the B-physics anomalies, related essentially to the $b\to s\mu^+\mu^-$ transition~\cite{LHCb:2021trn,LHCb:2019hip} and the muon $g-2$ 
anomaly~\cite{Muong-2:2023cdq, ParticleDataGroup:2018ovx,Gorringe:2015cma, Aoyama:2020ynm, Muong-2:2006rrc} while providing a sound DM candidate. 
However, a very recent reinterpretation of the LHCb collaborations completely washed out the discrepancy with the SM prediction in the $b\to s\mu^+\mu^-$ transition~\cite{LHCb:2022zom, LHCb:2022qnv}. 
Still, this type of models can be made compatible with these new results for  $b\to s\mu^+\mu^-$  (compatible with the SM predictions) while still solving the DM and $g-2$ problems. 

The existence of this type of models prompted us to study the contribution of the new coloured scalars, that live in the dark sector, to single and di-Higgs production.
The models were discussed in great detail in~\cite{Huang:2020ris, Capucha:2022kwo} and are based on a previous model
proposed in~\cite{Cerdeno:2019vpd}. They introduce  massive coloured scalar fields which, depending on the charge assignments and $SU(2)$ quantum numbers, can lead to one or several coloured scalars.
A discrete $Z_2$ symmetry is imposed such that the new fields from the dark sector are odd under $Z_2$ while the SM fields are even under this symmetry. 
In Ref.~\cite{Huang:2020ris}, three new fields were added to the SM,  one $SU(3)_c$ coloured scalar, $\Phi_3$, one colourless scalar, $\Phi_2$, and one vectorlike fermion, $\chi$, 
with an integer electric charge of 0 or $\pm 1$. The scalars are
$SU(2)_L$ singlets and the fermions form an $SU(2)_L$ doublet. This model was dubbed Model 5. In Ref.~\cite{Capucha:2022kwo}
a different scenario was studied with the scalars as $SU(2)_L$ doublets and the fermion as an $SU(2)_L$ singlet, and called Model 3.

As the dark sector communicates with the SM via the Higgs potential,
the new scalars couple to the Higgs boson. In fact, only two types of interactions are relevant to our discussion: the Higgs couplings to the new coloured scalars and the strong couplings
of the coloured scalars with the gluons with origin in the covariant
derivative. Therefore the one-loop single Higgs and di-Higgs
production only depend on very specific terms in the Higgs potential, the ones that connect the coloured scalars
with the SM Higgs doublet. 
Besides that, the SM Higgs coupling to the fermions (and also the Higgs self-couplings) remain exactly the SM ones - there is no mixing of the Higgs with the other scalars as they have different quantum numbers.
The coloured scalars contribute to the gluon fusion single Higgs and di-Higgs production with only one coloured scalar of electric charge $2/3$, $\phi_q^{+2/3}$, in Model 5, while for Model 3 
there are two coloured scalars contributing with electric charges
$2/3$ and $5/3$, $\phi_q^{+2/3}$ and $\phi_q^{+5/3}$, respectively. We
 also generalise our results to the case of an arbitrary
number of coloured scalars. 
Note that single Higgs production is a clean probe of the Higgs portal coupling in a scenario where the extension of the SM only includes an arbitrary
number of coloured scalars. The di-Higgs cross section can then be used to further confirm the structure suggested by single Higgs production.
From now on we will drop the old nomenclature and just refer to the model by the number of coloured scalars.

The LHC has performed numerous searches for DM. The only truly model-independent bound in the case of coloured scalar production and decay (depending only on the mass of the coloured scalar) 
would be a monojet event, that relies only on the strong gauge
coupling. These bounds would be valid in a scenario where the
couplings of the coloured scalars 
to the quarks and vector-like fermions are negligible or where
branching ratios that lead to visible final states are too small
to be detected. However, according to \cite{Cerdeno:2019vpd}  the best bounds are obtained in the searches for 
 DM associated with top and bottom quarks~\cite{ATLAS:2017hoo}. These are more restrictive than a re-interpretation of the searches for squarks at the LHC. They conclude in \cite{Cerdeno:2019vpd}  that the mass of coloured
scalars have a rough lower bound of 1 TeV. We will use this bound in our analysis.

We finalise this section by noting that the only new coupling present in the processes to be analysed is the portal coupling. Hence, in the case $n=1$ all results will depend on only two variables, the portal coupling and the coloured scalar mass. 
For an arbitrary $n$ we will have $n$ portal couplings and $n$ coloured scalar masses.

The paper is organised as follows. In Sec.~\ref{sec:2} we present the single Higgs production mode, and in Sec.~\ref{sec:3} the di-Higgs production mode is discussed. In Sec.~\ref{sec:4} we compare the contributions
of the new physics models to single Higgs and double Higgs production.
Our conclusions are given in Sec.~\ref{sec:5}.

\section{Single Higgs Production}
\label{sec:2}
We consider $n$ independent coloured complex scalars $\phi_q^{i=1,...,n}$
transforming in the fundamental representation of $SU(3)_c$. After electroweak symmetry breaking, the potential relevant to this work is given by
\begin{equation}
V = \sum_{i=1}^{n} \bigg[\underbrace{ (\mu_{\phi_q^i}^2+\frac{v^2}{2}\lambda_{h \phi_q^i}) }_{m_{\phi_q^i}^2} |\phi_q^i|^2+ \frac{1}{2} \lambda_{h \phi_q^i} h^2 |\phi_q^i|^2 + v \lambda_{h \phi_q^i} h |\phi_q^i|^2 + \lambda_{\phi_q^i} |\phi_q^i|^4 + ...\bigg] + ... \ \ ,
\label{eq:potential}
\end{equation}
where the couplings $\lambda_{h \phi_q^i}$ and $\lambda_{\phi_q^i}$ are real and we have defined the masses of the fields by
\begin{equation}
	m_{\phi_q^i}^2 = \mu_{\phi_q^i}^2 + \frac{v^2}{2} \lambda_{h\phi_q^i} \ .
	\label{eq:potential:masses}
\end{equation}	
	Note that there are in total $3n$ independent parameters. If
        we also consider that these $n$ fields form an $SU(2)_L$
        multiplet, $\Phi_q = \mqty*( \phi_q^1 & \phi_q^2 & ... &
        \phi_q^n )^\text{\textbf{T}}$, this would impose the following
        constraints: $\mu_{\phi_q^i}^2=\mu_{\phi_q^k}^2 \equiv
        \mu_{\Phi_q}^2$ and $\lambda_{\phi_q^i}=\lambda_{\phi_q^k}
        \equiv \lambda_{\Phi_q}$.\footnote{We use
            uppercase $\Phi$ and lowercase $\phi$ to distinguish
            between the parameters defined for the multiplet $\Phi$
            and the scalars $\phi$.} We are now left with only $n+2$
        degrees of freedom. This implies
          that for equal portal couplings $\lambda_{h\phi_q^i}$ the masses $m_{\phi_q^i}^2$ given by Eq.~(\ref{eq:potential:masses}) are also equal and vice-versa.
For this work we will consider the more general case of $n$ independent fields but still assuming that they never have the exact same quantum numbers.

\begin{figure}[h!]
    \centering 
    \begin{subfigure}[c]{.3\linewidth}
        \begin{tikzpicture}
        \begin{feynman}
            \vertex                (l1);
            \vertex [ below right =1cm and 1cm of l1]  (l2);
            \vertex [ below  left =1cm and 1cm of l2]  (l3);
            \vertex [       right =1cm of l2]            (vf) {\(h\)};
            \vertex [        left =of l1]                (g1) {\(g\)};
            \vertex [        left =of l3]                (g2) {\(g\)};
            
            \diagram* {
            (l1) -- [ fermion, edge label = \(Q\)] (l2)
                 -- [ fermion, edge label = \(Q\)] (l3) 
                 -- [ fermion, edge label = \(Q\)] (l1),
            (l2) -- [ scalar ] (vf),
            (g1) -- [gluon] (l1), (g2) -- [gluon] (l3) };
        \end{feynman}
        \end{tikzpicture}
        \caption{ }%
        \label{subfig:singleDiagrams:SM}
    \end{subfigure}%
    \begin{subfigure}[c]{.3\linewidth}
        \begin{tikzpicture}
        \begin{feynman}
            \vertex                (l1);
            \vertex [ below right =1cm and 1cm of l1]  (l2);
            \vertex [ below  left =1cm and 1cm of l2]  (l3);
            \vertex [       right =1cm of l2]            (vf) {\(h\)};
            \vertex [        left =of l1]                (g1) {\(g\)};
            \vertex [        left =of l3]                (g2) {\(g\)};
            
            \diagram* {
            (l1) -- [ scalar, edge label = \(\phi_q^i\)] (l2)
                 -- [ scalar, edge label = \(\phi_q^i\)] (l3) 
                 -- [ scalar, edge label = \(\phi_q^i\)] (l1),
            (l2) -- [ scalar ] (vf),
            (g1) -- [gluon] (l1), (g2) -- [gluon] (l3) };
        \end{feynman}
        \end{tikzpicture}
        \caption{}%
        \label{subfig:singleDiagrams:BDM1}
    \end{subfigure}%
    \begin{subfigure}[c]{.3\linewidth}
        \begin{tikzpicture}
        \begin{feynman}
            \vertex                (l1);
            \vertex [       right =1cm of l1]  (l2);
            \vertex [       right =1cm of l2]               (vf) {\(h\)};
            \vertex [       above left =1cm and 1cm of l1]  (g1) {\(g\)};
            \vertex [       below left =1cm and 1cm of l1]  (g2) {\(g\)};
            
            \diagram* {
            (l1) -- [ scalar, out=90, in=90, looseness=1.6, edge label = \(\phi_q^i\)] (l2)
                 -- [ scalar, out=-90, in=-90, looseness=1.6, edge label = \(\phi_q^i\)] (l1),
            (l2) -- [ scalar ] (vf),
            (g1) -- [gluon] (l1), (g2) -- [gluon] (l1) };
        \end{feynman}
        \end{tikzpicture}
        \caption{}%
        \label{subfig:singleDiagrams:BDM2}
    \end{subfigure}
    \caption{ Generic single Higgs production diagrams. (a) - SM quark loops; (b)/(c) - BDM coloured scalars loops. } \label{fig:singleDiagrams}  
    \end{figure}
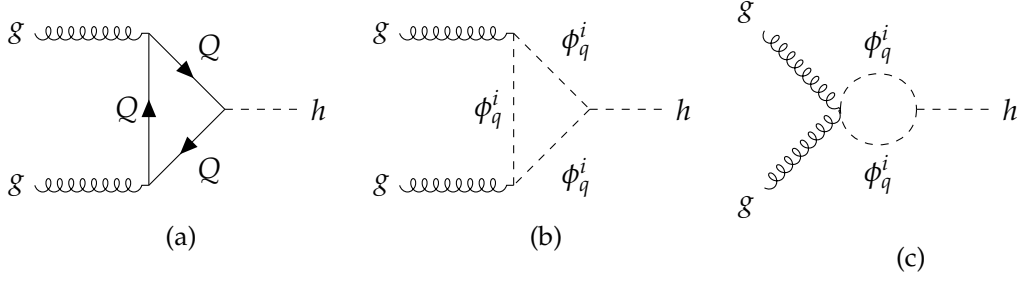

Single Higgs production via gluon fusion, which is the main production
process at the LHC, 
proceeds at leading order (LO) in the SM via quark loops
\cite{Georgi:1977gs} as shown in Fig.~\ref{subfig:singleDiagrams:SM}, with the heavier quarks giving the major contribution.
In the new models, which we will refer to as BDM models, two new
diagrams emerge as shown in Figs.~\ref{subfig:singleDiagrams:BDM1}
and \ref{subfig:singleDiagrams:BDM2}  . 

\begin{figure}[h!]
            \centering
            \includegraphics[width=7cm]{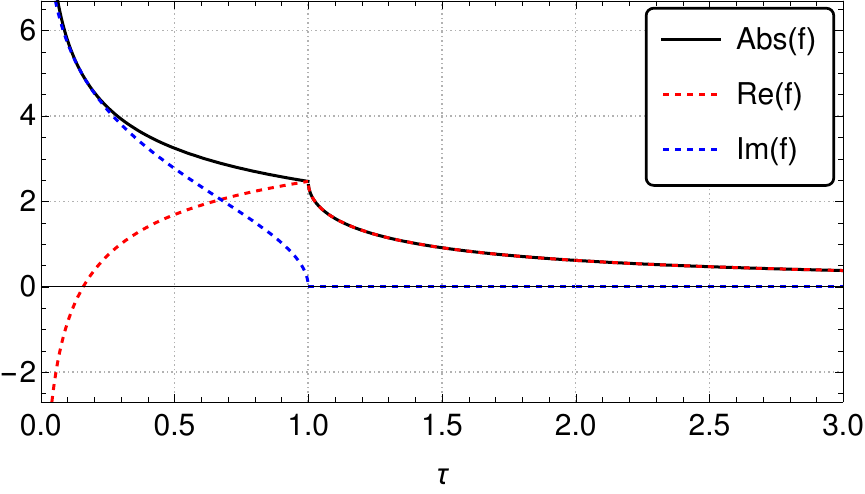}
            \hspace{.5cm}
            \includegraphics[width=7cm]{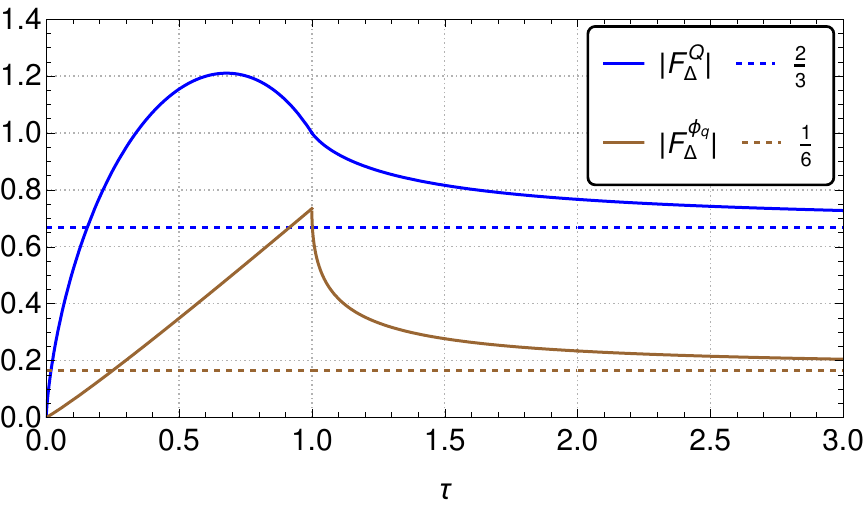}
            \caption{ Left: $f(\tau)$ as a function of $\tau$; right: quark and scalar form factors as a function of $\tau$.  }
            \label{subfig:form}
        \end{figure}

The amplitude for this process can be cast into the form
    \begin{equation}
        \mathcal{M}_\triangle^{ gg \rightarrow h } 
                        = \frac{ g_s^2 m_h^2 }{ 16 \pi^2 } \bigg(
                                  \sum_Q           g_{Q}^h         F_\triangle^Q
                                  +\sum_{\phi_q^i}  g_{\phi_q^i}^h  F_\triangle^{\phi_q^i}
                                  \bigg) 
                                \ A_{1\mu\nu} \ \epsilon_a^\mu \epsilon_b^\nu \ \delta_{ab} ,
        \label{eq:singleProd:Amplitude}
    \end{equation}
    where the indices $a$ and $b$ are associated with the incoming gluons, $A_1^{\mu\nu}= g^{\mu\nu} - p_b^\mu p_a^\nu/p_a \cdot p_b$
    and the quark and scalar form factors are given
    by~\cite{Spira:2016ztx}
    \begin{align}
        F_\triangle^{Q} &=  \tau_Q \left( 1 + (1 - \tau_Q) f(\tau_Q) \right), \qquad         & g_Q^h &= \frac{1}{v}, \label{eq:SingleProd:FormFactorSM}\\
        F_\triangle^{\phi_q^i} &= -\frac{1}{2} \tau_{\phi_q^i} \left( 1- \tau_{\phi_q^i} f(\tau_{\phi_q^i}) \right), \qquad 
                & g_{\phi_q^i}^h &=\frac{\lambda_{h \phi_q^i} v}{ 2 m_{\phi_q^i}^2},
          \label{eq:SingleProd:FormFactorNP}
    \end{align}
with  $\tau_X=4 m_{X}^2 / m_h^2$ ($X=Q,\phi_q^i$)  and $f(\tau)$ defined as
    \begin{equation}
        f(\tau)= \left\{ \begin{array}{ll}
                                    \arcsin( \frac{1}{\sqrt{\tau}} )^2 & \tau \geq 1  \\
                                     -\frac{1}{4} \left[ \log( \frac{1+\sqrt{1-\tau}}{1-\sqrt{1-\tau}}) -i \pi \right]^2 & \tau < 1
                                 \end{array} \right.
		\ \ .
        \label{eq:ftau}
    \end{equation}
In the limit of large masses the form factors approach a constant value,
    \begin{eqnarray}
        \lim_{m_Q^2\rightarrow \infty} F_\triangle^{Q} &=& \frac{2}{3}
        \ , \label{eq:SingleProd:FormFactor:limits1} \\
        \lim_{m_{\phi_q^i}^2 \rightarrow \infty} F_\triangle^{\phi_q^i} &=& \frac{1}{6} \ ,
        \label{eq:SingleProd:FormFactor:limits2}
    \end{eqnarray}
  and therefore the large mass behaviour is determined solely by the
 coupling pre-factors  $g_X^h$. Consequently, for large masses, the scalar loop contribution to the amplitude is suppressed by a factor of $1/m_{\phi_q^i}^2$. 
  Because the quark Yukawa couplings are proportional to their masses, the quark loop contribution approaches a constant value for large masses. 
  Thus, although this process can be used to determine how many heavy quarks are present in the model the same is not true for the coloured scalars.
In Fig.~\ref{subfig:form} we present $f(\tau)$ as a function of $\tau$
in the left plot and the quark and scalar form factors as a function
of $\tau$ in the right plot, which nicely shows
  that the two form factors approach constant values in the large mass limit.

\subsection{The LHC Production Cross Section}
    The calculation of the gluon fusion production cross section is performed at LO by implementing the new form factors for the coloured scalars (Eq.~(\ref{eq:SingleProd:FormFactorNP})) in the program \texttt{HIGLU}~\cite{Spira:1995mt}
which can be used to calculate the single Higgs production cross section at the LHC in the SM and in the Minimal Supersymmetric extension of the SM (MSSM). In the SM the $gg$ initiated production is much larger than its quark counterpart making the latter negligible in SM-like
models, such as the ones discussed in this work. We can therefore write the hadronic cross section as
    \begin{equation}
        \sigma( pp \rightarrow h ) = \sigma_0^h \tau_h \dv{\mathcal{L}^{gg}}{\tau_h} \label{eq:SingleProd:sig}, \qquad \qquad
        \sigma_0^h = \frac{\pi}{16 m_h^4} \left| \mathcal{M}_\triangle^{gg \rightarrow h} \right|^2 ,
    \end{equation}
where $\dv{\mathcal{L}^{gg}}{\tau_h}$ is the gluon luminosity
and $\tau_h=m_h^2/s$, with $s$ denoting the total hadronic c.m.~energy squared.
In order to reduce the impact of the important
  higher-order (HO) effects we calculate the relative deviation of the new physics
(NP) cross section in our model from the SM cross section, defined
as
\begin{equation}
\delta_h = \frac{ \sigma_{NP}-\sigma_{SM} }{ \sigma_{SM} }.
\end{equation} 
We hence assume that the relative HO corrections to the new physics cross section in our model do not deviate significantly from those of the SM case which can safely be assumed for the QCD corrections\footnote{The gluon fusion cross section is known at next-to-leading order (NLO) QCD including the full mass dependences \cite{DJOUADI1991440,DAWSON1991283,Graudenz:1992pv,Spira:1995rr,Harlander:2005rq,Anastasiou:2006hc,Aglietti:2006tp,Anastasiou:2009kn}. Within the heavy top-quark limit the next-to-next-to-leading order (NNLO) \cite{Catani:2001ic,Harlander:2001is,Harlander:2002wh,Anastasiou:2002yz,RAVINDRAN2003325,MARZANI2008127} and next-to-next-to-next-to leading order (N$^3$LO) \cite{Gehrmann:2011aa,Anastasiou:2013srw,Anastasiou:2013mca,Kilgore:2013gba,Li:2014bfa,Anastasiou:2014lda,Anastasiou:2015vya,Anastasiou:2016cez,Anastasiou:2016cez,Mistlberger:2018etf,Duhr:2022cob,Baglio:2022wzu} QCD corrections have been calculated. An explicit large top-mass expansion has estimated the missing quark-mass effects beyond NLO to be less than 1\% \cite{Harlander:2009bw,Harlander:2009mq,Harlander:2009my,Pak:2009dg}.} while for the EW corrections\footnote{The NLO EW corrections have been calculated in \cite{Djouadi:1994ge,Chetyrkin:1996ke,Chetyrkin:1996wr,Degrassi:2004mx,Aglietti:2006yd,Actis:2008ug,Actis:2008ts} and the mixed QCD-EW corrections in \cite{Anastasiou:2008tj}.} this is not necessarily the case. The latter are, however, small compared to the QCD corrections. Using
Eqs.~(\ref{eq:singleProd:Amplitude}--\ref{eq:SingleProd:FormFactorNP}) and Eq.~(\ref{eq:SingleProd:sig}), we can write $\delta_h$ as
    \begin{equation}
        \delta_h
        = \frac{ 
                  \left| 
                       \sum_Q F_\triangle^{Q}
                    +v \sum_{\phi_q^i} g_{\phi_q^i}^h F_\triangle^{\phi_q^i} 
                  \right|^2 
                - \left| \sum_Q F_\triangle^Q \right|^2  
               }{ 
                  \left| \sum_Q F_\triangle^Q \right|^2 
               }
        =
            2 v \sum_{\phi_q^i} g_{\phi_q^i}^h 
                  \operatorname{Re}\left[ \frac{ F_\triangle^{\phi_q^i} }{ \sum_Q F_\triangle^Q  } \right]
        + v^2 \frac{ \left|\sum_{\phi_q^i} g_{\phi_q^i}^h F_\triangle^{\phi_q^i} \right|^2 }{ \left| \sum_Q F_\triangle^Q \right|^2 } .
        \label{eq:singleProd:deltah:Full}
    \end{equation}
    
For the following numerical analysis we include the bottom, charm and
top quark loops in single Higgs production, while in double Higgs production
only top and bottom quark loops are taken into account. We use the following 
input values for the Higgs, top, bottom and charm quark masses, respectively:
\beq
m_h = 125 \mbox{ GeV}, \quad m_t = 172.5 \mbox{ GeV}, \quad m_b = 4.75 \mbox{ GeV}, \quad m_c = 1.43 \mbox{ GeV}
.
\eeq
We use the LO pdfs NNPDF40\_lo\_as\_01180 \cite{Butterworth:2015oua,NNPDF:2021njg} and
the LO strong coupling constant 
\beq
\alpha_s =  0.118 \;.
\eeq
The cross sections are calculated for a c.m. energy of
$\sqrt{s}=14$~TeV. Note that the dependence on $\sqrt{s}$ 
cancels out in $\delta_h$. 

\subsection{Model with One Scalar versus a Model with Two Scalars}
    Let us start by considering the scenarios $n=1$ (just one coloured scalar) and  $n=2$ (two coloured scalars). 
    As already discussed, all scalar masses will be taken to be above 1 TeV. 
    In the case $n=1$ and considering here, for the sake of the discussion, 
    only the top quark contribution 
    (the bottom contribution only ranges at the percent level), the
    following simplified form for $\delta_h$ is obtained
    \begin{equation}
        \delta_h
            = \lambda_{h \phi_q^1} \frac{v^2}{m_{\phi_q^1}^2} 
                \left( 
                    \frac{ F_\triangle^{\phi_q^1} }{ F_\triangle^{Q} } 
                \right) 
            + \lambda_{h \phi_q^1}^2 \frac{v^4}{4 m_{\phi_q^1}^4 } 
                \left( 
                    \frac{ F_\triangle^{\phi_q^1} }{ F_\triangle^{Q} } 
                \right)^2 .
        \label{eq:singleProd:deltah:Mod5}
    \end{equation}
Any extension with more than one coloured scalar will have one more effective Higgs-scalar coupling $\lambda_{h \phi_q^i}$ and one more scalar mass $m_{\phi_q^i}$ for each new scalar added to the model. 
Thus, in order to simplify the presentation of the results, we impose the constraint of equal coloured scalar masses for any extension with more than one coloured scalar. As we will show later, for masses
above 5 TeV the cross sections will be very small unless the number of
scalars becomes very large. So the interesting range for the mass is
indeed very small. Note that in the plots presented later we will
always include the bottom, charm and top contributions.

In the BDM models, the quartic coupling $\lambda_{h\phi_q^i}$
that enters the calculation of the cross section is an effective
coupling in the following sense: in the case $n=1$ it is just the portal 
coupling between the Higgs and the singlet coloured scalar; for $n=2$, the two effective couplings are the sum of combinations of three portal couplings (in the case of an $SU(2)$ representation). 
In more detail, for $n=1$ the coloured scalar is an $SU(2)$ singlet and the portal coupling with the Higgs doublet can be written as
\begin{equation}
	V_{\text{portal}}^{n=1} =\lambda_{H \Phi_q} |H|^2 |\Phi_q|^2 \;,
\end{equation}
and the effective coupling takes the form
\begin{equation}
	\lambda_{h\phi_q^1} = \lambda_{H \Phi_q} \, .
\end{equation}
In the scenario $n=2$ the coloured scalar is an $SU(2)$ doublet  and the portal couplings are now
\begin{equation}
	V_{\text{portal}}^{n=2} =\lambda_{H \Phi_q} |H|^2 |\Phi_q|^2 +  \lambda'_{H \Phi_q} |H^\dagger \Phi_q|^2 + y_{H \Phi_q} |H^\dagger i \sigma_2 \Phi_q|^2 ,
\end{equation}
which results in two effective couplings, 
\begin{equation}
	\lambda_{h\phi_q^1} = \lambda_{H \Phi_q} + \lambda'_{H \Phi_q}
	\ , \qquad 
	\lambda_{h\phi_q^2} = \lambda_{H \Phi_q} + y_{H \Phi_q} .
\end{equation}
We have also checked that the same applies to the triplet representation of $SU(2)$ \cite{Crivellin:2021ejk}. However, one should stress that what is relevant here
is that we will discuss any type of model with an arbitrary number of scalars, each with an effective portal coupling and a given mass. The results can then be translated to any specific model of this kind.

Since all form factors are positive and strictly decreasing for $m_{\phi_q^i} > 1$ TeV, the highest contributions to the cross sections will be achieved when these form factors are at their highest value corresponding to the lowest mass for all the scalars. Under 
the equal masses constraint ($m_{\phi_q^i}=m_{\phi_q^j}\equiv m_{\phi_q} \Rightarrow F_\triangle^{\phi_q^i}=F_\triangle^{\phi_q^j}\equiv F_\triangle^{\phi_q}$) we can write
%
    \begin{equation}
        \delta_h
            = \left( \sum_{i}
              \lambda_{h\phi_q^i} \right)  
                \frac{v^2}{\mpq^2} 
                \left( 
                    \frac{ F_\triangle^{\phi_q} }{ F_\triangle^{Q} } 
                \right) 
            + \left(\sum_{i}
            \lambda_{h\phi_q^i} \right)^2 
                \frac{v^4}{4 \mpq^4 } 
                \left( 
                    \frac{ F_\triangle^{\phi_q} }{ F_\triangle^{Q} } 
                \right)^2 .
        \label{eq:singleProd:deltah:Mod3}
    \end{equation}
With all masses equal, $\delta_h$ is not sensitive to individual couplings but only to their total sum, $\sum_i \lambda_{h\phi_q^i}$. Further, taking all couplings equal, $\lambda_{h\phi_q^i} = \lambda_{h\phi_q^j} \equiv \lambdaHPQ$, we still cover the full range of possible values for $\delta_h$ because for any particular choice of couplings $\{ \lambda_{h\phi_q^1}, ..., \lambda_{h\phi_q^n}  \} $ there is always a single coupling $\lambdaHPQ$ such that $\sum_{i} \lambda_{h\phi_q^i} = n \lambdaHPQ$ which will give equivalent results for $\delta_h$. 
With this approximation the coupling  $\lambdaHPQ$ will just be rescaled by a factor $n$ when going from the case $n=1$ to arbitrary $n$. 

Before presenting the results we will discuss the allowed values for the couplings. 
As the upper bound we will consider the perturbativity bound of $4\pi$. For the lower bound, one of the conditions for the potential (Eq.~(\ref{eq:potential})), to be bounded from 
below, following the same procedure as in \cite{Ferreira:2004yd}, gives rise to the following constraint
    \begin{equation}
        \lambda_{h\phi_q^i} \geq - \frac{m_h}{v} \sqrt{2\lambda_{\phi_q^i} } \,,
        \label{eq:bfb}
    \end{equation}
where $m_h$ is the SM Higgs boson mass and $\lambda_{\phi_q^i}$ is the $\phi_q^i$ quartic self coupling parameter that must be positive, $\lambda_{\phi_q^i} \geq 0$, and obey the perturbativity bound of $\lambda_{\phi_q^i} \leq 4\pi$. Therefore we will vary the
relevant parameters $\lambda_{h\phi_q^i}$ between the lower value given by Eq.~(\ref{eq:bfb}) and the upper value $4\pi$. 

    \begin{figure}[h!]
        \centering
            \includegraphics[width=10cm]{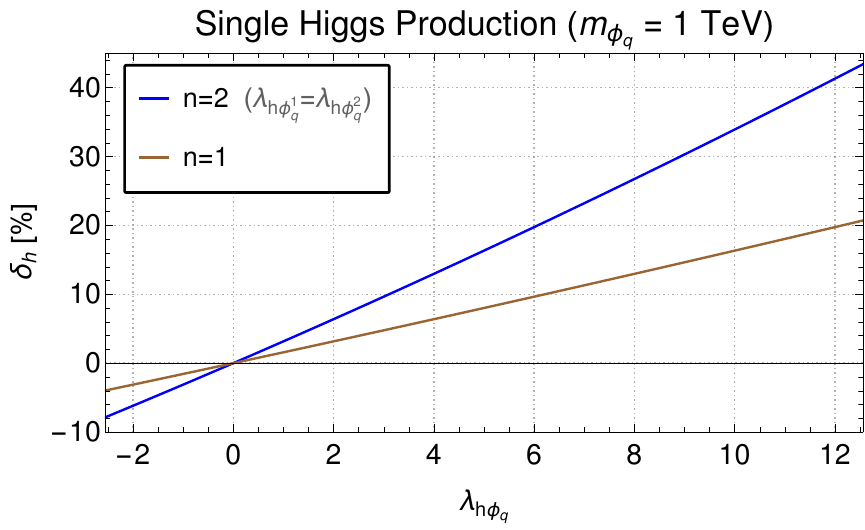}
            \caption{ $\delta_h$ as a function of the effective portal
              coupling $\lambdaHPQ$ for a mass of $\mpq = 1$ TeV and
              for $n=1$ and $n=2$.}
                      \label{subfig:singleProd:lambda}
        \end{figure}

In Fig.~\ref{subfig:singleProd:lambda} we present the results for
$\delta_h$ as a function of the effective portal coupling $\lambdaHPQ$
for a mass of $\mpq = 1$ TeV and for $n=1$ and $n=2$. The single Higgs
cross section was calculated with \texttt{HIGLU} for a c.m. energy of
14 TeV resulting in a SM LO cross section of $\sigma^{h}_{SM}= 15.76$
pb for the above given input values.
It is evident that  $\delta_h$ varies linearly with the effective coupling $\lambdaHPQ$, which means that, in this range, the interference term between the SM and NP form factors is dominant. 
The large scalar masses we are working with and the
  fact that the interference term is proportional to only $1/\mpq^2$
  while the purely NP contributions are suppressed by a factor of
  $1/\mpq^4$ (cf.~Eq.~(\ref{eq:singleProd:deltah:Mod5})), are the reason behind this behaviour.

        \begin{figure}[h!]
            \centering
            \includegraphics[width=6.5cm]{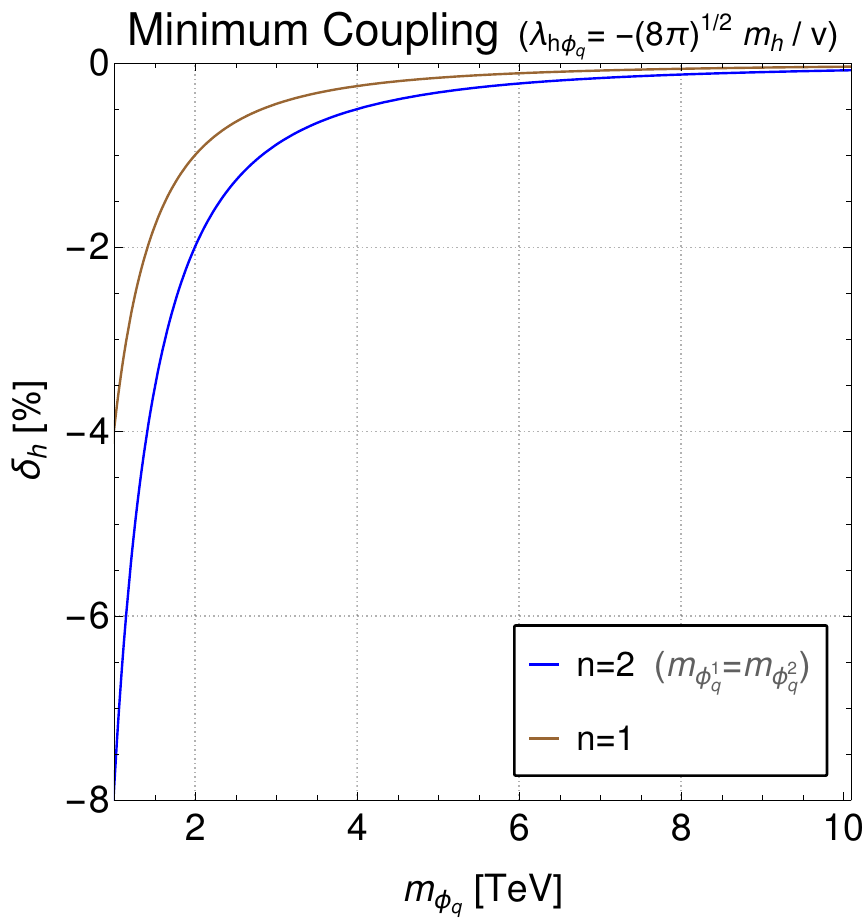}
               \hspace{1cm} 
            \includegraphics[width=6.5cm]{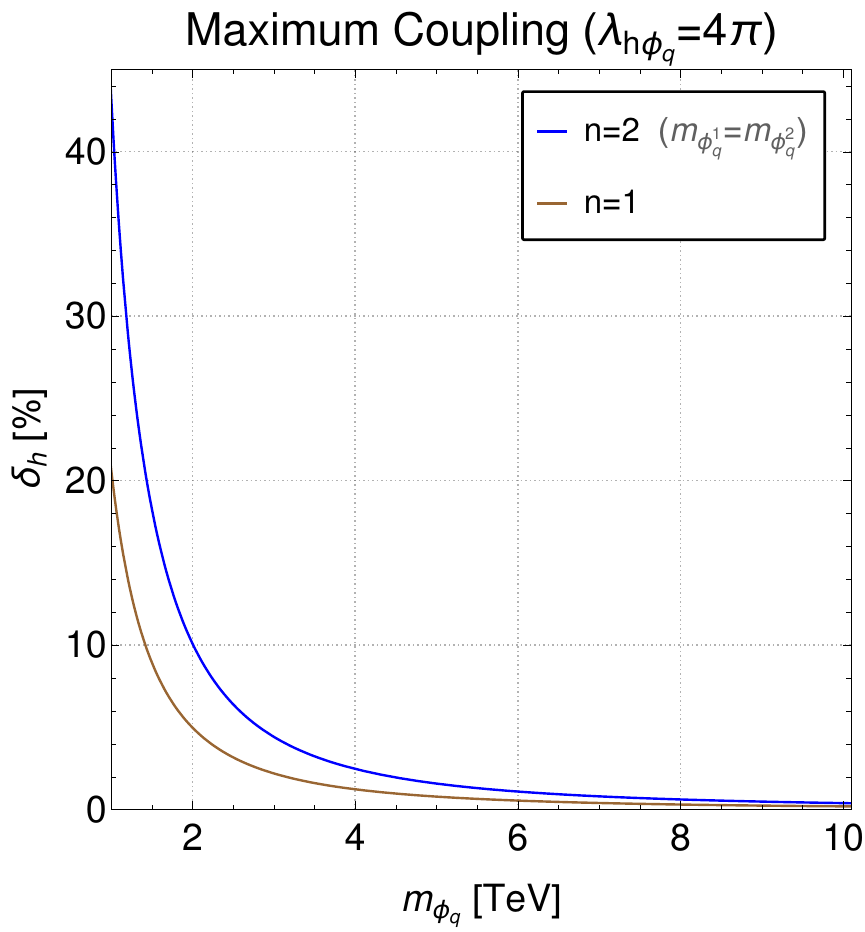}
            \caption{ $\delta_h$ as a function of the coloured scalar
              mass for the minimum value of the coupling (left) and
              the maximum value of the coupling (right) and for $n=1$ and
              $n=2$. }
            \label{subfig:singleProd:mass}
        \end{figure}        

In Fig.~\ref{subfig:singleProd:mass} we show the results for $\delta_h$ as a function of the coloured scalar mass for the minimum value of the coupling (left) and the maximum value of the coupling (right) and for $n=1$ and $n=2$.
Since, as argued above, the interference term
is dominant, $\delta_h$ behaves approximately as $ 1/\mpq^2$ for fixed
    $\lambdaHPQ$. For the allowed range of variation the maximum
  value of variation relative to the SM is
  between about -10\% and +40\%.

The NP term only becomes comparable to the interference term in the
limit   
    \begin{equation}
        \lambdaHPQ=\frac{ 4 \mpq^2 }{ v^2 }
        \frac{F_\triangle^Q}{F_\triangle^{\phi_q}} \xrightarrow[
        \mpq\to\infty ]{ m_{Q}\to\infty } \frac{ 16 \mpq^2 }{ v^2 } ,
        \label{eq:singleProd:deltah:Mod3:middlePoint}
    \end{equation}
which means that for a mass of $\mpq=1$ TeV $\lambdaHPQ\approx 260$ for $n=1$. As more scalars are added the picture can change. As the interference term scales with $n$ and the NP term scales as $n^2$, for a number of scalars above 20 and
all masses equal to 1 TeV the NP term starts to
dominate.

\subsection{ Models with $n$ Coloured Scalars }
    In the previous section we have set all masses to be
    equal. Relaxing this condition forces us to return to the more
    general expression given in
    Eq.~(\ref{eq:singleProd:deltah:Full}). However, we can follow a
    different approach in order to simplify the final expression by taking
    advantage of the large scalar masses and using the limit for
    $F_\triangle^{ \phi_q^i }$ given in Eq.~(\ref{eq:SingleProd:FormFactor:limits2}). For scalar masses above 1 TeV the error in $F_\triangle^{ \phi_q^i }$ by using this limit is only about~0.2\%. With this approximation  $\delta_h$ can be written as
    \begin{equation}
        \delta_h
        = \frac{1}{ \left|\sum_Q F_\triangle^Q \right| }   \frac{v^2}{6} 
        \sum_i \frac{ \lambda_{h\phi_q^i} }{ m_{\phi_q^i}^2 }
        + \frac{1}{ \left|\sum_Q F_\triangle^Q \right|^2 } \frac{v^4}{144} 
        \left( \sum_i \frac{ \lambda_{h\phi_q^i} }{ m_{\phi_q^i}^2 } \right)^2 ,
        \label{eq:singleProd:deltah:formAprox}
    \end{equation}
    where we have $\left|\sum_Q F_\triangle^Q \right| \approx 0.641$ when including the top, bottom and charm quarks. Including only the top quark and the limit in Eq.~(\ref{eq:SingleProd:FormFactor:limits1}) would imply an error in $\left|\sum_Q F_\triangle^Q \right|$ of around ~4\%. 
    This approximation has the advantage of allowing us to write the results as a function of the ratio $x_i = \lambda_{h\phi_q^i}/m_{\phi_q^i}^2$ where the index $i$ represents each 
    scalar\footnote{ This approximation is not strictly necessary. In the general case the ratio would be $x_i = \lambda_{h\phi_q^i} \frac{F_\triangle^{\phi_q^i}}{m_{\phi_q^i}^2}$. All conclusions in this section are only dependent on the fact
     that $x_i$ decreases with mass, a behaviour present whether we
     use the approximation or not since $F_\triangle^{\phi_q^i}$ approaches a constant value for large masses and $\lambda_{h\phi_q^i}$ takes a constant value between its boundaries.}.  
 It is now clear that we can show $\delta_h$ as a function of the sum $(\sum_i x_i)$. As previously discussed, as long as we span all possible values for this sum we will also have fully explored all values that $\delta_h$ can take. 
 In order to do this let us first note that the minimum and maximum of $(\sum_i x_i)$ are achieved when all $x_i$ are at their minimum and maximum values, respectively. Hence, to generate all values for the sum and consequently for $\delta_h$, we can make the simple choice of $x_i=x_j$ with the limits of $\min(x_i)=\min\left( \lambda_{h\phi_q^i}/m_{\phi_q^i}^2 \right)= - (8\pi)^{1/2} m_h/v \,  \text{TeV}^{-2}$ and $\max(x_i)=\max\left( \lambda_{h\phi_q^i}/m_{\phi_q^i}^2 \right) = 4\pi \ \text{TeV}^{-2}$ where we have considered $\min(m_{\phi_q^i})= 1 \ \text{TeV}$. 
 

        \begin{figure}[h!]
            \centering
            \includegraphics[width=7cm]{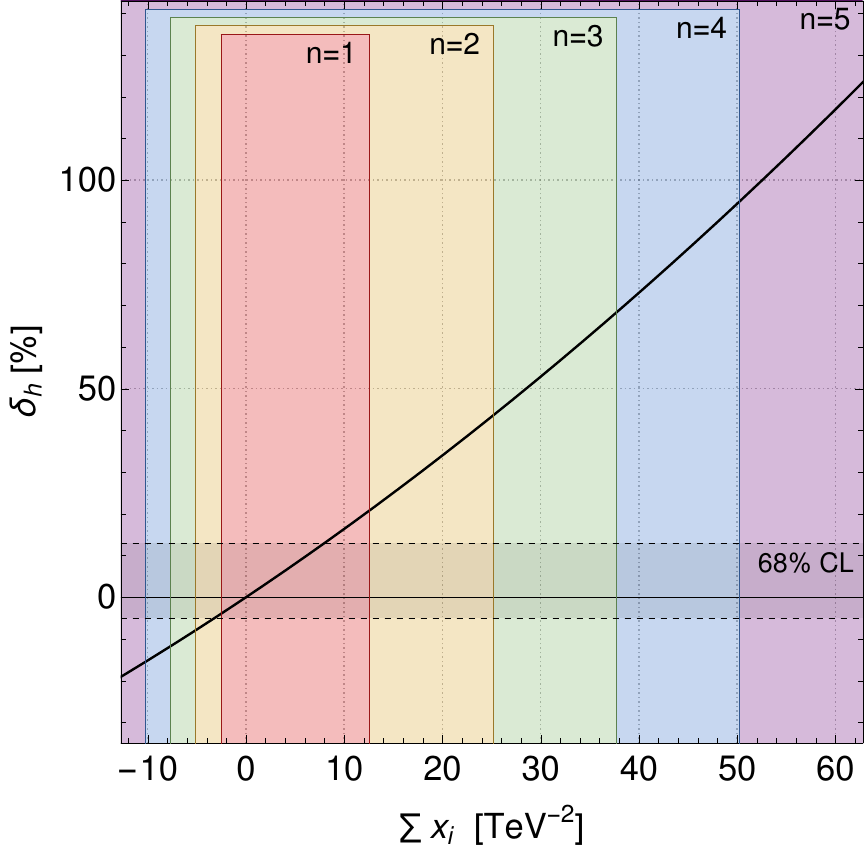}
             \hspace{1cm} 
            \includegraphics[width=7cm]{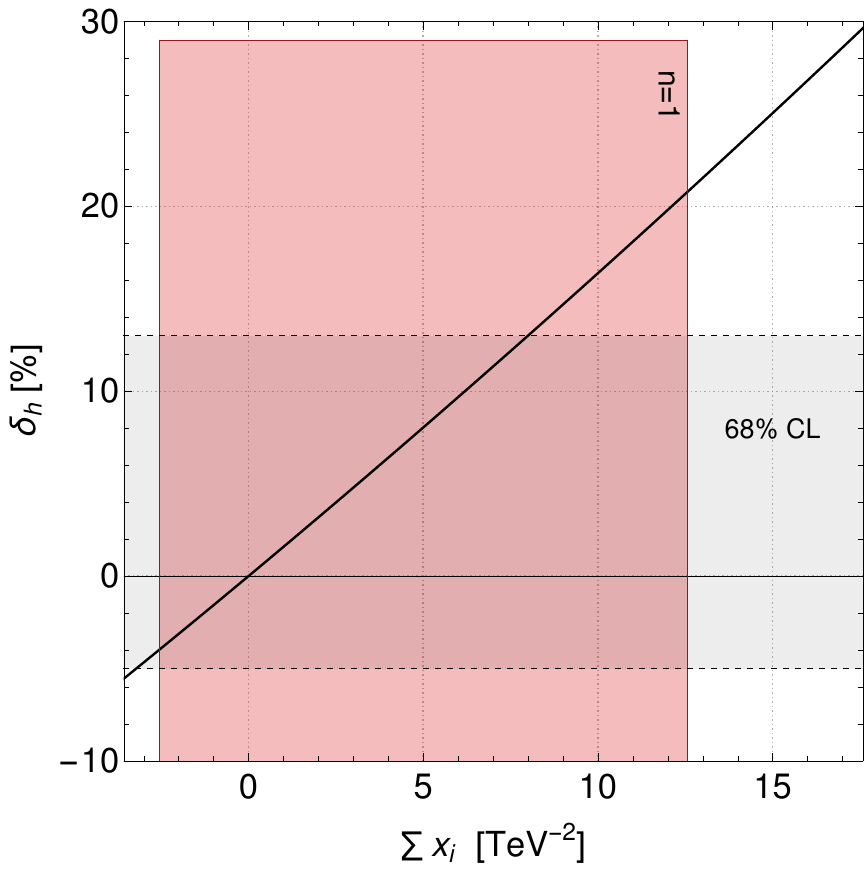}
            \caption{ $\delta_h$ as a function of $\sum_i x_i$. The minimum and maximum limits for a model with $n$ scalars are indicated by the coloured zones, where a minimum mass of $m_{\phi_q^i}=1$ TeV is considered 
            and the couplings are varied between the  lower bound of $ - (8\pi)^{1/2} m_h/v$ and the perturbativity upper bound of $4 \pi$. In the left plot the horizontal lines are taken from  the ATLAS combination~\cite{ATLAS:2019nkf} and show
            the $1\sigma$ results for Higgs production via gluon fusion. In the right plot we present just the case $n=1$ for a better understanding of the bounds on $\sum_i x_i$ for $n=1$.}
            \label{subfig:plotn1}
        \end{figure}        

In Fig.~\ref{subfig:plotn1} we show  $\delta_h$ as a function of
$\sum_i x_i$. The minimum and maximum limits for a model with $n$
scalars are indicated by the coloured zones, where a minimum mass of 1
TeV is considered and  the couplings are varied between their minimum
and maximum allowed values.
The horizontal lines represent the
  relative experimental uncertainty of the experimental results for Higgs production via gluon fusion at $1\sigma$. In the left plot the lines are taken from the ATLAS combination~\cite{ATLAS:2019nkf} at 13 TeV and 80 fb$^{-1}$, leading to  $\delta_h \in [-5,13]$.
In the right plot we present just the case $n=1$ for a better understanding of the bounds on $\sum_i x_i$ for $n=1$. Considering $n=1$ we can see that $-3.2 < \sum_i x_i  < 8$ approximately. This in turn means that for a mass of 1 TeV the coupling is also constrained
to be $-3.2 < \lambda_{h\phi_q^1} < 8$. Therefore the bounds are not very strong at the moment but are already better than the perturbative limit for the upper bound. Still, as the mass grows the bound on the coupling gets weaker. For $n>1$ if the couplings are all of the same order,
the constraints will be stronger if again the masses are all of the order 1 TeV. But there is always the possibility of having all couplings very small except one, recovering the $n=1$ constraints for the larger coupling. Furthermore, if the couplings have different signs we end up with a larger freedom than for the case  $n=1$. These scenarios will have to be studied for the specific model in question using every other
information on the model.

        \begin{figure}[h!]
            \centering
            \includegraphics[width=7cm]{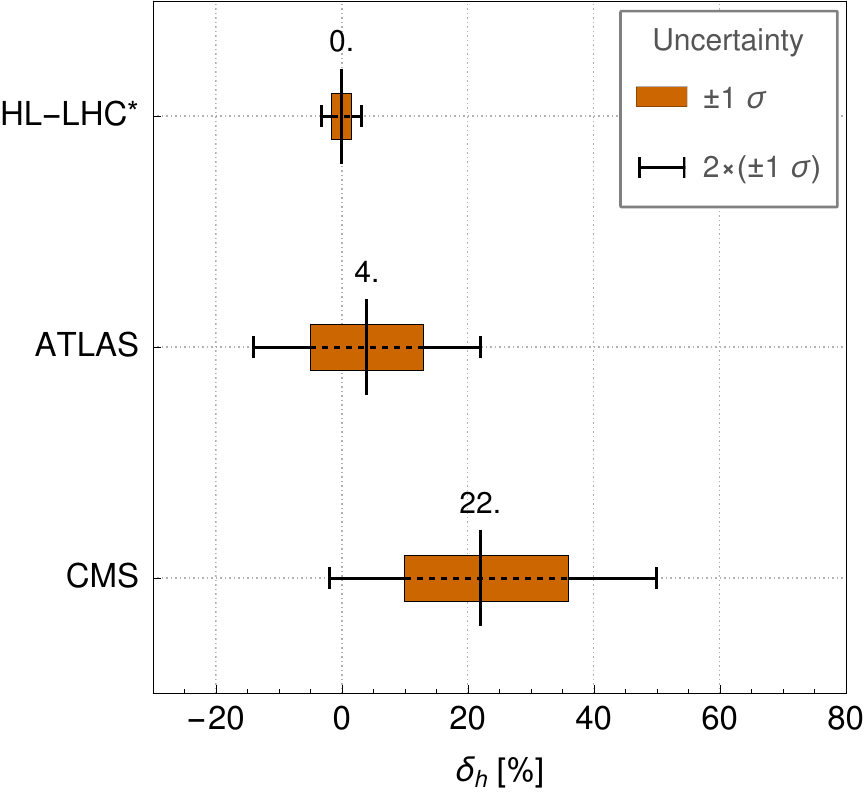}
             \hspace{1cm} 
            \includegraphics[width=7cm]{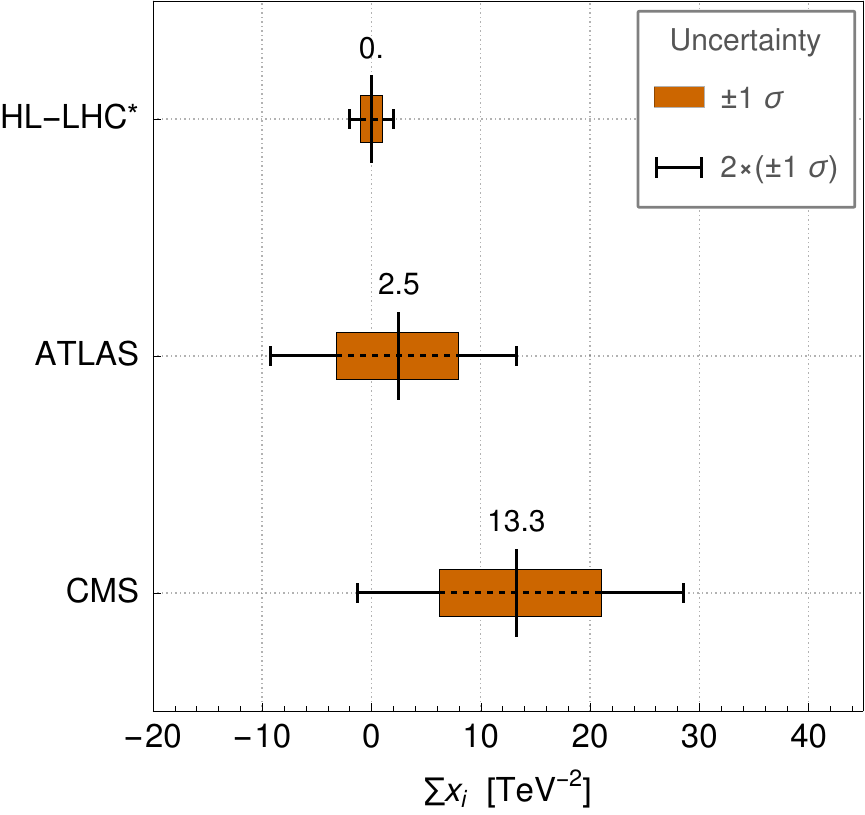}
            \caption{ Allowed values of $\delta_h$ (left)   and
              $\sum_i x_i$ (right)  at $1 \sigma$ and $2 \times 1\sigma$
              using the present experimental limits from ATLAS~\cite{ATLAS:2019nkf},  CMS~\cite{CMS:2018uag}, 
            and the predictions for the future HL-LHC \cite{cepeda2019higgs}.}
            \label{subfig:plotn2}
        \end{figure}        
 
 In Fig.~\ref{subfig:plotn2} we show the allowed values of $\sum_i
 x_i$ (left) and $\delta_h$ (right) at $1 \sigma$ and $2 \times 1\sigma$ using the present experimental limits from ATLAS~\cite{ATLAS:2019nkf},  CMS~\cite{CMS:2018uag}, 
 and the predictions for the future HL-LHC \cite{cepeda2019higgs}. The predictions for the HL-LHC show that we will attain a result of the order  $\delta_h \in [-1.6,1.6]$.

\section{Double Higgs Production}
\label{sec:3}
Similar to the single Higgs case, the production of a pair of Higgs bosons is dominated by the gluon fusion process, which at LO is given by a triangle and a box diagram with heavy quarks running in the loop \cite{Glover:1987nx}. The new coloured scalars 
    will contribute to di-Higgs production by similar loop
    diagrams. Due to the new
      2 gluon-2 coloured scalars and 2 Higgs-2 coloured scalars
      couplings, however, there are now additional topologies that
      contribute to the process.

\subsection{The Leading-Order Amplitude}
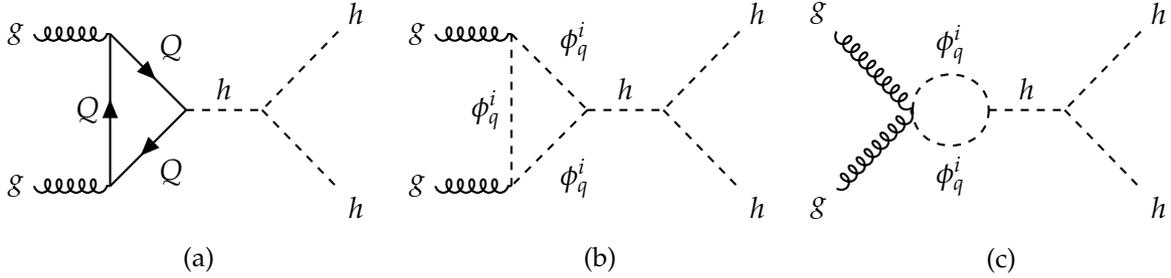
\begin{figure}[h!]
    \centering 
    \begin{subfigure}[c]{.33\linewidth}
        \begin{tikzpicture}
        \begin{feynman}[large]
            \vertex                (l1);
            \vertex [ below right =1cm and 1cm of l1]  (l2);
            \vertex [ below  left =1cm and 1cm of l2]  (l3);
            \vertex [       right =1cm of l2]          (vf);
            \vertex [        left =1cm of l1]                (g1) {\(g\)};
            \vertex [        left =1cm of l3]                (g2) {\(g\)};
            \vertex [ above right =1cm and 1cm of vf]  (h1) {\(h\)};
            \vertex [ below right =1cm and 1cm of vf]  (h2) {\(h\)};
            
            \diagram* {
            (l1) -- [ fermion, edge label = \(Q\)] (l2)
                 -- [ fermion, edge label = \(Q\)] (l3) 
                 -- [ fermion, edge label = \(Q\)] (l1),
            (l2) -- [ scalar, edge label = \(h\)](vf),
            (g1) -- [gluon] (l1), (g2) -- [gluon] (l3),
            (h1) -- [scalar] (vf) -- [scalar] (h2) };
        \end{feynman}
        \end{tikzpicture}
        \caption{ }%
        \label{subfig:pairDiagrams:triangle:SM}
    \end{subfigure}%
    \begin{subfigure}[c]{.33\linewidth}
        \begin{tikzpicture}
        \begin{feynman}[large]
            \vertex                (l1);
            \vertex [ below right =1cm and 1cm of l1]  (l2);
            \vertex [ below  left =1cm and 1cm of l2]  (l3);
            \vertex [       right =1cm of l2]          (vf);
            \vertex [        left =1cm of l1]                (g1) {\(g\)};
            \vertex [        left =1cm of l3]                (g2) {\(g\)};
            \vertex [ above right =1cm and 1cm of vf]  (h1) {\(h\)};
            \vertex [ below right =1cm and 1cm of vf]  (h2) {\(h\)};
            
            \diagram* {
            (l1) -- [ scalar, edge label = \(\phi_q^i\)] (l2)
                 -- [ scalar, edge label = \(\phi_q^i\)] (l3) 
                 -- [ scalar, edge label = \(\phi_q^i\)] (l1),
            (l2) -- [ scalar, edge label = \(h\)](vf),
            (g1) -- [gluon] (l1), (g2) -- [gluon] (l3),
            (h1) -- [scalar] (vf) -- [scalar] (h2) };
        \end{feynman}
        \end{tikzpicture}
        \caption{}%
        \label{subfig:pairDiagrams:triangle:BDM1}
    \end{subfigure}%
    \begin{subfigure}[c]{.33\linewidth}
        \begin{tikzpicture}
        \begin{feynman}[large]
            \vertex                (l1);
            \vertex [       right =1cm of l1]  (l2);
            \vertex [       right =1cm of l2]  (vf);
            \vertex [       above left =1cm and 1cm of l1]                (g1) {\(g\)};
            \vertex [       below left =1cm and 1cm of l1]                (g2) {\(g\)};
            \vertex [ above right =1cm and 1cm of vf]  (h1) {\(h\)};
            \vertex [ below right =1cm and 1cm of vf]  (h2) {\(h\)};
            
            \diagram* {
            (l1) -- [ scalar, out=90, in=90, looseness=1.6, edge label = \(\phi_q^i\)] (l2)
                 -- [ scalar, out=-90, in=-90, looseness=1.6, edge label = \(\phi_q^i\)] (l1),
            (l2) -- [ scalar, edge label = \(h\)](vf),
            (g1) -- [gluon] (l1), (g2) -- [gluon] (l1),
            (h1) -- [scalar] (vf) -- [scalar] (h2) };
        \end{feynman}
        \end{tikzpicture}
        \caption{}%
        \label{subfig:pairDiagrams:triangle:BDM2}
    \end{subfigure}
    \caption{Generic diagrams contributing to double Higgs production
      involving the trilinear Higgs self-coupling: (a) - SM quark loop; (b)/(c) - coloured scalars loop.} \label{fig:pairDiagrams:triangle}  
    \end{figure}

\begin{figure}[h!]
    \centering 
    \begin{subfigure}[c]{.45\linewidth}
        \begin{tikzpicture}
        \begin{feynman}[large]
            \vertex                (l1);
            \vertex [ right=of l1] (l2);
            \vertex [ below=of l2] (l3);
            \vertex [  left=of l3] (l4);
            \vertex [  left=of l1]  (g1) {\(g\)};
            \vertex [  left=of l4]  (g2) {\(g\)};
            \vertex [ right=of l2]  (h1) {\(h\)};
            \vertex [ right=of l3]  (h2) {\(h\)};
            
            \diagram* {
            (l1) -- [ fermion, edge label = \(Q\)] (l2) 
                 -- [ fermion, edge label = \(Q\)] (l3) 
                 -- [ fermion, edge label = \(Q\)] (l4) 
                 -- [ fermion, edge label = \(Q\)] (l1),
            (g1) -- [gluon] (l1), (g2) -- [gluon] (l4),
            (l2) -- [scalar] (h1), (l3) -- [scalar] (h2) };
        \end{feynman}
        \end{tikzpicture}
        \caption{ }%
        \label{subfig:pairDiagrams:box:SM}
    \end{subfigure}%
    \begin{subfigure}[c]{.45\linewidth}
        \begin{tikzpicture}
        \begin{feynman}[large]
            \vertex                (l1);
            \vertex [ right=of l1] (l2);
            \vertex [ below=of l2] (l3);
            \vertex [  left=of l3] (l4);
            \vertex [  left=of l1]  (g1) {\(g\)};
            \vertex [  left=of l4]  (g2) {\(g\)};
            \vertex [ right=of l2]  (h1) {\(h\)};
            \vertex [ right=of l3]  (h2) {\(h\)};
            
            \diagram* {
            (l1) -- [ scalar, edge label = \(\phi_q^i\)] (l2) 
                 -- [ scalar, edge label = \(\phi_q^i\)] (l3) 
                 -- [ scalar, edge label = \(\phi_q^i\)] (l4) 
                 -- [ scalar, edge label = \(\phi_q^i\)] (l1),
            (g1) -- [gluon] (l1), (g2) -- [gluon] (l4),
            (l2) -- [scalar] (h1), (l3) -- [scalar] (h2) };
        \end{feynman}
        \end{tikzpicture}
        \caption{}%
        \label{subfig:pairDiagrams:box:BDM1}
    \end{subfigure}

    \begin{subfigure}[c]{.33\linewidth}
        \begin{tikzpicture}
        \begin{feynman}[large]
           \vertex  (l1);
            \vertex [ above left =1cm and 1.5cm of l1]   (g1) {\(g\)};
            \vertex [ below left =1cm and 1.5cm of l1]   (g2) {\(g\)};
            \vertex [ above right =1cm and 1cm of l1]  (l2);
            \vertex [ below right =1cm and 1cm of l1]  (l3);
            \vertex [ right =1.5cm of l2]  (h1) {\(h\)};
            \vertex [ right =1.5cm of l3]  (h2) {\(h\)};
            
            \diagram* {
            (g1) -- [gluon] (l1) -- [gluon] (g2),
            (l1) -- [ scalar, edge label = \(\phi_q^i\)] (l2)
                 -- [ scalar, edge label = \(\phi_q^i\)] (l3) 
                 -- [ scalar, edge label = \(\phi_q^i\)] (l1),
            (h1) -- [scalar] (l2), (h2) -- [scalar] (l3) };
        \end{feynman}
        \end{tikzpicture}
        \caption{ }%
        \label{subfig:pairDiagrams:box:BDM2}
    \end{subfigure}%
    \begin{subfigure}[c]{.36\linewidth}
        \begin{tikzpicture}
        \begin{feynman}[large]
            \vertex                (l1);
            \vertex [ below right =1cm and 1cm of l1]  (l2);
            \vertex [ below  left =1cm and 1cm of l2]  (l3);
            \vertex [        left =of l1]                (g1) {\(g\)};
            \vertex [        left =of l3]                (g2) {\(g\)};
            \vertex [ above right =1cm and 1.5cm of l2]  (h1) {\(h\)};
            \vertex [ below right =1cm and 1.5cm of l2]  (h2) {\(h\)};
            
            \diagram* {
            (l1) -- [ scalar, edge label = \(\phi_q^i\)] (l2)
                 -- [ scalar, edge label = \(\phi_q^i\)] (l3) 
                 -- [ scalar, edge label = \(\phi_q^i\)] (l1),
            (g1) -- [gluon] (l1), (g2) -- [gluon] (l3),
            (h1) -- [scalar] (l2) -- [scalar] (h2) };
        \end{feynman}
        \end{tikzpicture}
        \caption{}%
        \label{subfig:pairDiagrams:box:BDM3}
    \end{subfigure}%
    \begin{subfigure}[c]{.36\linewidth}
        \begin{tikzpicture}
        \begin{feynman}[large]
           \vertex  (l1);
            \vertex [ above left =1cm and 1.5cm of l1]   (g1) {\(g\)};
            \vertex [ below left =1cm and 1.5cm of l1]   (g2) {\(g\)};
            \vertex [ right =1.5cm of l1]  (l2);
            \vertex [ above right =1cm and 1.5cm of l2]  (h1) {\(h\)};
            \vertex [ below right =1cm and 1.5cm of l2]  (h2) {\(h\)};
            
            \diagram* {
            (g1) -- [gluon] (l1) -- [gluon] (g2),
            (l1) -- [ scalar, out=90, in=90, looseness=1.6, edge label = \(\phi_q^i\)] (l2)
                 -- [ scalar, out=-90, in=-90, looseness=1.6, edge label = \(\phi_q^i\)] (l1),
            (h1) -- [scalar] (l2), (h2) -- [scalar] (l2) };
        \end{feynman}
        \end{tikzpicture}
        \caption{}%
        \label{subfig:pairDiagrams:box:BDM4}
    \end{subfigure}
    \caption{Generic diagrams contributing to double Higgs production
      independent of the trilinear Higgs self-coupling. (a) - SM quark loop; (b-e) - coloured scalars loop.} \label{fig:pairDiagrams:box}  
    \end{figure}
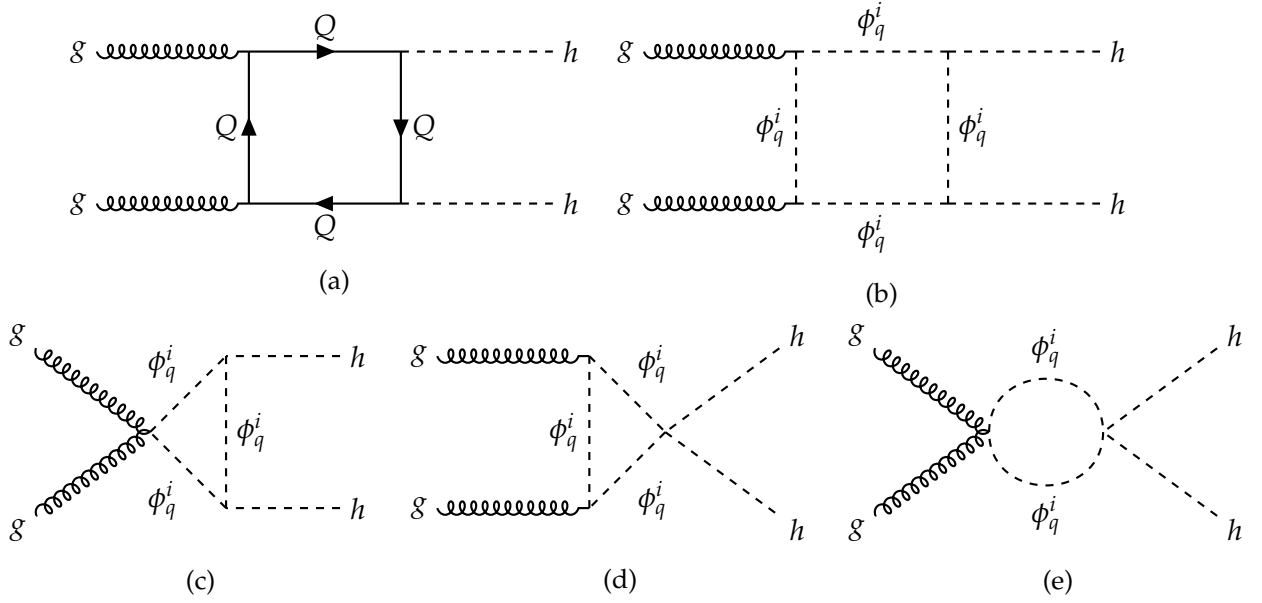

The complete set of diagrams is given by the ones involving the
trilinear Higgs self-coupling, shown in
Fig.~\ref{fig:pairDiagrams:triangle}, and diagrams that do not depend
on it, depicted in Fig.~\ref{fig:pairDiagrams:box}. The new topologies arising in
our model are given in Fig.~\ref{fig:pairDiagrams:triangle} (b) and
(c) and in Fig.~\ref{fig:pairDiagrams:box} (b)-(e). As in the SM, we have triangle and box topologies and now additionally also a
  self-energy-like topology. 

%

The LO amplitude can be decomposed into two different
tensor structures, which correspond to total gluon spin 0 and 2,
respectively, along the collision axis. They are given by \cite{Plehn:1996wb}
\begin{align}
        A_1^{\mu\nu}&= g^{\mu\nu} - \frac{ p_b^\mu p_a^\nu }{ \left( p_a p_b \right) } \label{eq:projOperators:A1}\\
        A_2^{\mu\nu}&= g^{\mu\nu} 
                            + \frac{1}{ p_T^2 ( p_a p_b ) } 
                              \left[
                                    \left(  p_c^2  \right) p_b^\mu p_a^\nu 
                                 -2 \left( p_b p_c \right) p_c^\mu p_a^\nu  
                                 -2 \left( p_a p_c \right) p_b^\mu p_c^\nu
                                 +2 \left( p_a p_b \right) p_c^\mu p_c^\nu
                              \right] 
                              \label{eq:projOperators:A2}
    \end{align}
with
\begin{equation}
A_1 \cdot A_1 = A_2 \cdot A_2 = 2\,, \quad A_1 \cdot A_2 = 0 
\end{equation}
and
\begin{equation}
        p_T^2 = 2 \frac{ \left( p_a p_c \right) \left( p_b p_c \right)
        }{ \left( p_a p_b \right) } - p_c^2 ,
\end{equation}
where $p_{a,b}$ denote the four-momenta of the two incoming gluons,
and $p_{c,d}$ those of the outgoing Higgs bosons. The LO
amplitude given by the diagrams in
Fig.~\ref{fig:pairDiagrams:triangle}, which contain the trilinear
Higgs self-coupling, can be cast in the form
   \begin{equation}
        \mathcal{M}_{hhh}^{ gg \rightarrow hh } 
                        = \frac{ g_s^2 s }{ 16 \pi^2 } \ C_\triangle
                                \left(
                                    \sum_Q         g_{Q}^h       F_\triangle^{Q} 
                                  + \sum_{\phi_q^i}  g_{\phi_q^i}^h  F_{\triangle}^{\phi_q^i }
                                \right) A_{1\mu\nu}
                                \epsilon_a^\mu \epsilon_b^\nu \ \delta_{ab}
		\ ,
            \label{eq:pairProd:Amplitude:Triangle}
    \end{equation}
where 
\begin{equation}
	C_\triangle = \frac{ 3 m_h^2/v }{ s-m_h^2 } 
	\ ,
\end{equation} 
$\epsilon_{a,b}^{\mu/\nu}$ represent the gluon polarisation vectors
and $g_s$ denotes the strong coupling constant. The first term in
Eq.~(\ref{eq:pairProd:Amplitude:Triangle}) 
corresponds to the first diagram and the second one to the last two
diagrams.\footnote{In accordance 
with the \texttt{FeynArts} \cite{HAHN2001418,feynarts_manual_link} notation, we call triangle
diagrams loops with three legs attached and box diagrams loops with
four legs attached.  }  
The form factors $F_\triangle^{Q/\phi_q^i}$ and the couplings $g_{Q/\phi_q^i}^h$
are given in Eqs~(\ref{eq:SingleProd:FormFactorSM}) and
(\ref{eq:SingleProd:FormFactorNP}).
The amplitude independent of the Higgs self-coupling can be written
as 
    \begin{align}
        \mathcal{M}_{\text{no }hhh}^{ gg \rightarrow hh }
                        = \frac{ g_s^2 s }{ 16 \pi^2 } \ C_\square 
                                \Bigg[
                                    &\sum_Q
                                    \left( 
                                        (g_{Q}^h)^2 F_\square^{Q} A_{1\mu\nu} 
                                      + (g_{Q}^h)^2 G_\square^{Q} A_{2\mu\nu}
                                    \right) \nonumber \\ 
                                  + &\sum_{\phi_q^i }
                                  \left(
                                    \left(
                                         (g_{\phi_q^i }^h)^2  F_{\square_1}^{\phi_q^i}
                                       + g_{\phi_q^i}^{hh} F_{\square_2}^{\phi_q^i}
                                    \right) A_{1\mu\nu}
+ (g_{\phi_q^i}^h)^2 G_{\square_1}^{\phi_q^i} A_{2\mu\nu}
                                  \right)
                                \Bigg] \epsilon_a^\mu \epsilon_b^\nu \ \delta_{ab}
		\ ,
            \label{eq:pairProd:Amplitude:Square}
    \end{align}
 where $C_\square=1$, the prefactors $g_{Q/\phi_q^i}^{h}$ are given in
 Eqs.~(\ref{eq:SingleProd:FormFactorSM}) and (\ref{eq:SingleProd:FormFactorNP})
 and 
\begin{equation}
g_{\phi_q^i}^{hh} =\frac{\lambda_{h\phi_q^i} }{ 2 m_{\phi_q^i}^2} \;.
\end{equation} 
The quark form factors $F_\square^Q$ and $G_\square^Q$ corresponding
to Fig.~\ref{fig:pairDiagrams:box} (a), which have been
calculated in the literature before (cf.~e.g.~\cite{Plehn:1996wb}), are deferred to the
Appendix~\ref{annex:pairProd:SMforms}, while the new form factors are
given here. The form factor 
$F_{\square_1}^{\phi_q^i}$ sums the contributions of the diagrams
Figs.~\ref{fig:pairDiagrams:box} (b) and 
(c) proportional to $A_1^{\mu\nu}$, $F_{\square_2}^{\phi_q^i}$ stems
from the sum of the contributions of Figs.~\ref{fig:pairDiagrams:box}
(d) and (e), and $G_{\square_1}^{\phi_q^i}$ is the sum of the
contributions of Figs.~\ref{fig:pairDiagrams:box}
(b) and (c) proportional to $A_2^{\mu\nu}$. They read
explicitly\footnote{See also e.g.~\cite{Kribs:2012kz} and
  \cite{Enkhbat:2013oba}. In the former paper, the authors focused on
  the impact of light coloured scalars on di-Higgs production while in the latter the effect of light coloured scalar
leptoquarks was analysed.} 
\begin{multline}
        G_{\square_1}^{\phi_q} = \frac{ 4 \mpq^4 } {s}
                        \left( \frac{1}{t u - m_h^4 } \right)
                    \Bigg( 
                         s ( t+u ) C_{ab}^{\mpq^2}
                        + ( 2 t ) ( t - m_h^2) C_{ac}^{\mpq^2}
                        + ( 2 u ) ( u - m_h^2) C_{bc}^{\mpq^2}\\
                        - ( t^2+u^2 - 2 m_h^4 ) C_{cd}^{\mpq^2} 
                        -( s t^2 +2 \mpq^2 (t u - m_h^4 )) D_{bac}^{\mpq^2}\\
                        -( s u^2 +2 \mpq^2 (t u - m_h^4 )) D_{abc}^{\mpq^2}
                        -(2 \mpq^2 (t u - m_h^4 ) ) D_{acb}^{\mpq^2}
                    \Bigg)
                    \label{eq:pairProd:Amplitude:Square:G:NP}
    \end{multline}
    \begin{multline}
        F_{\square_1}^{\phi_q} =
                    \frac{ 4 \mpq^4 } {s}
                    \Bigg( 
                          \frac{2}{s} ( t - m_h^2 ) C_{ac}^{\mpq^2}
                        + \frac{2}{s} ( u - m_h^2 ) C_{bc}^{\mpq^2}\\
                        - ( 2  \mpq^2 ) ( D_{abc}^{\mpq^2} + D_{bac}^{\mpq^2} )
                        - ( 2 \mpq^2 + \frac{1}{s}  ( t u - m_h^4 ) ) D_{acb}^{\mpq^2}
                    \Bigg)
                    \label{eq:pairProd:Amplitude:Square:F1:NP}
    \end{multline}
where we have suppressed the i index only for convenience and  
\begin{equation} 
F_{\square_2}^{\phi_q^i}=F_{\triangle}^{\phi_q^i} \;, \label{eq:fsquare2}
\end{equation}
with the latter given in Eq.~(\ref{eq:SingleProd:FormFactorNP}). The
Mandelstam variables $s,t,u$ and the scalar integrals $C_{ij}$ and $D_{ijk}$
are defined in the appendix.

\subsection{The Leading-Order Cross Section}
The amplitude squared for the computation of the cross section can be
separated into two different parts, one for each spin
projection,\footnote{The interference term 
      vanishes as for the tensor structures $A_1$ and $A_2$ we have
      $A_1 \cdot A_2 =0$.} so that the differential partonic cross
    section can be cast into the form
\begin{equation}
\frac{d\hat{\sigma}^{hh}}{d\hat{t}} = \frac{G_F^2 \alpha_s^2}{256
  (2\pi)^3} \left[\left| \mathcal{M}_F \right|^2 + \left|
    \mathcal{M}_G \right|^2 \right]  \;, \label{eq:partonic}
\end{equation}
where $G_F$ denotes the Fermi constant, $\alpha_s$ the strong coupling
constant, and $\hat{t}$ the momentum transfer squared from one of the
initial state gluons to one of the final state Higgs bosons. Each of
the partial amplitudes $\mathcal{M}_{F/G}$ contains only the terms
constructed with the $F/G$ form factors, respectively. Hence
\begin{eqnarray}
	\mathcal{M}_F \!\!&=&\!\! \sum_Q \left( C_\triangle\, g_Q^h F_\triangle^Q +
  C_\square\, (g_Q^h)^2 F_\square^Q \right) + \sum_{\phi_q^i} \left(
                 C_\triangle \,  g_{\phi_q^i}^h F_\triangle^{\phi_q^i} +
                  C_\square \, \left( (g_{\phi_q^i}^h)^2
                  F_{\square_1}^{\phi_q^i} + g_{\phi_q^i}^{hh}
                      F_{\square_2}^{\phi_q^i} \right) \right)  \\
	\mathcal{M}_G \!\!&=&\!\!  C_\square \left( \sum_Q g_Q^h G_\square^Q + 
\sum_{\phi_Q^i} (g_{\phi_q^i}^h)^2 G_{\square_1}^{\phi_q^i} \right)
	\ .
\end{eqnarray}
The total cross section for $hh$ production through gluon
fusion at the LHC is obtained by integrating Eq.~(\ref{eq:partonic})
over the scattering angle and the gluon luminosity, 
\begin{equation}
        \sigma( pp \rightarrow hh ) = \int_{4 m_h^2 /s}^{1} \dd
        \tau_h \dv{ \mathcal{L}^{gg} }{ \tau_h } \hat{\sigma}^{hh}(\hat{s} =
        \tau_h s) \,
        \label{eq:pairProd:sig}
    \end{equation}
    where $s$ is the c.m.~energy at the LHC. The numerical evaluation
    of the total production cross section is performed at LO with
    the program \texttt{HPAIR}~\cite{Plehn:1996wb, Dawson:1998py} where we have
    implemented the new form factors. The Fortran code \texttt{HPAIR}
    was originally written for the SM and the MSSM and calculates the
double Higgs production through gluon fusion at LO and NLO in the heavy quark limit.

Also for double Higgs production we present our results as a ratio
with respect to the SM value in order to minimise the contribution of
HO effects, that is $\delta_{hh}$ is defined as  
	\begin{equation}
		\delta_{hh} = \frac{ \sigma_{NP}-\sigma_{SM} }{ \sigma_{SM} } \, .
		\label{eq:singleProd:deltahh:definition}
	\end{equation}
This assumes that the HO corrections in our model do not differ significantly from those of the SM, which is a rather good approximation for the QCD corrections\footnote{After first results in the heavy-top limit \cite{Dawson:1998py}, the NLO QCD corrections including the full top quark mass dependence have been provided in 
\cite{Borowka:2016ehy,Borowka:2016ypz,Baglio:2018lrj,Baglio:2020ini,Baglio:2020wgt}. 
The NNLO corrections have been obtained in the large $m_t$ limit \cite{deFlorian:2013jea,Grigo:2014jma}, the results at
next-to-next-to-leading logarithmic accuracy (NNLL) became available in
\cite{Shao:2013bz,deFlorian:2015moa}, and the corrections up to
N$^3$LO were presented in
\cite{Banerjee:2018lfq,Chen:2019lzz,Chen:2019fhs,Ajjath:2022kpv} for the heavy
top-mass limit. For a review of higher-order corrections to SM di-Higgs production,
see e.g.~\cite{DiMicco:2019ngk}.}, whereas not necessarily for the EW corrections, for which at present only first partial results exist, however,\footnote{First results on the electroweak corrections have been provided in \cite{Muhlleitner:2022ijf,Davies:2022ram,Davies:2023obx,Davies:2023npk}.} and which are expected to be less important.
In contrast to single Higgs production we cannot find a simple
analytic formula for this quantity due to the more involved form of
the amplitudes and consequently also of the cross sections and the dependence of the form
factors on the c.m.~energy. 

\subsection{Phenomenological Analysis of the Cases $n=1$ and $n=2$}
 
    Let us start with the simpler scenarios with one or two coloured
    scalars. The dependence of the form factors on the mass is not
    trivial. Since the NP contributions should decouple from the SM
    for very large masses this means that $\delta_{hh}$ would
    eventually behave as a strictly decreasing function of
    the coloured scalar mass. 
 
    \begin{figure}[h!]
        \centering
            \includegraphics[width=9cm]{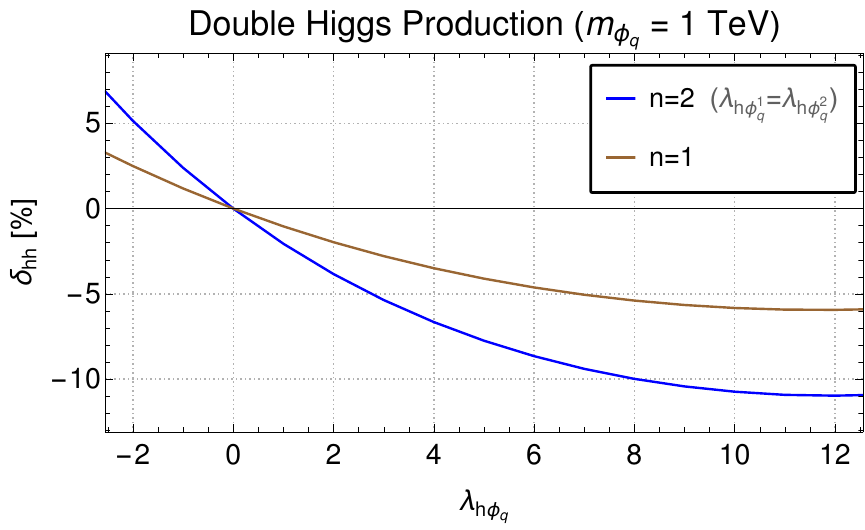}
            \caption{ $\delta_{hh} =
              (\sigma_{NP}-\sigma_{SM})/\sigma_{SM}$   as a function
              of the effective portal coupling $\lambdaHPQ$ for a mass
              of $\mpq = 1$ TeV and for $n=1$ and $n=2$. }
                      \label{subfig:doubleProd:lambda}
        \end{figure}

We will follow the same approach as for single Higgs production and choose all coloured scalar masses equal to be 1 TeV. 
    As for the couplings, while in single Higgs production with equal
    masses only the total sum of the couplings was relevant, in
    di-Higgs production the amplitude now depends on
    $\lambda_{h\phi_q^k}$ and $\lambda_{h\phi_q^k}^2$ terms. 
For now we will impose the constraint of equal couplings for $n=2$. In Fig.~\ref{subfig:doubleProd:lambda} we present  $\delta_{hh}$   as a function of the effective portal coupling $\lambdaHPQ$ for a mass of $\mpq = 1$ TeV and for $n=1$ and $n=2$.
            The double Higgs cross section was calculated with \texttt{HPAIR}
            for a c.m. energy of 14 TeV. As expected, for
            $\lambdaHPQ=0$ the NP and SM LO cross sections coincide,
            where the SM LO cross section calculated with \texttt{HPAIR}
            amounts to 16.37~fb.

        \begin{figure}[h!]
            \centering
            \includegraphics[width=6.3cm]{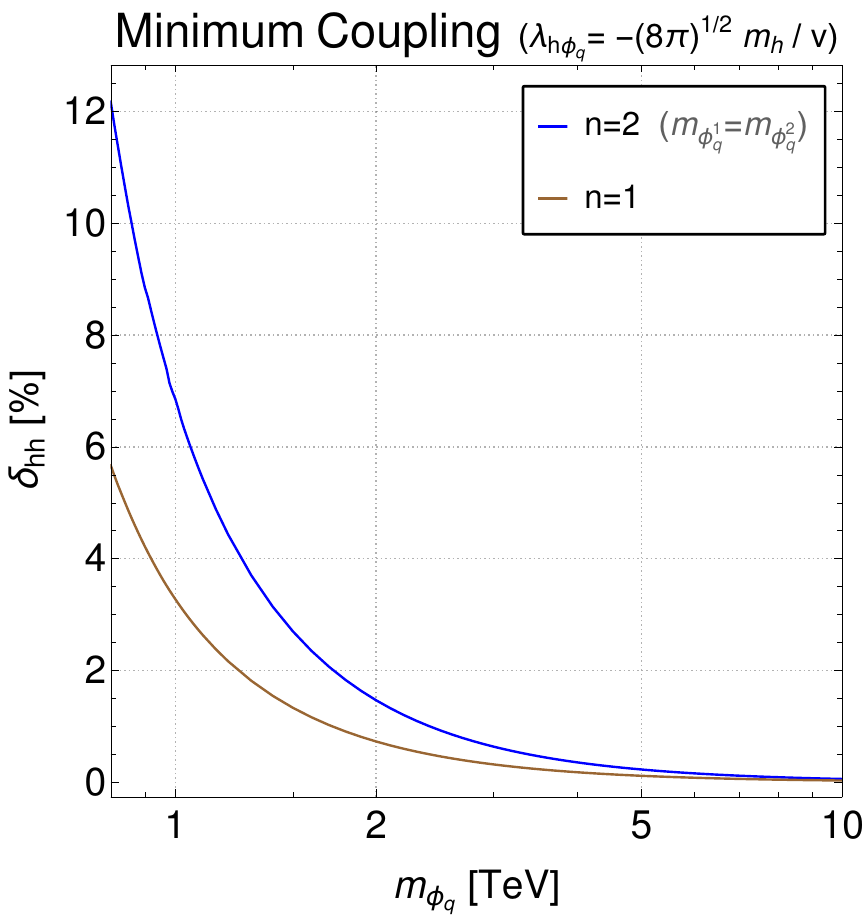}
               \hspace{1cm} 
            \includegraphics[width=6.5cm]{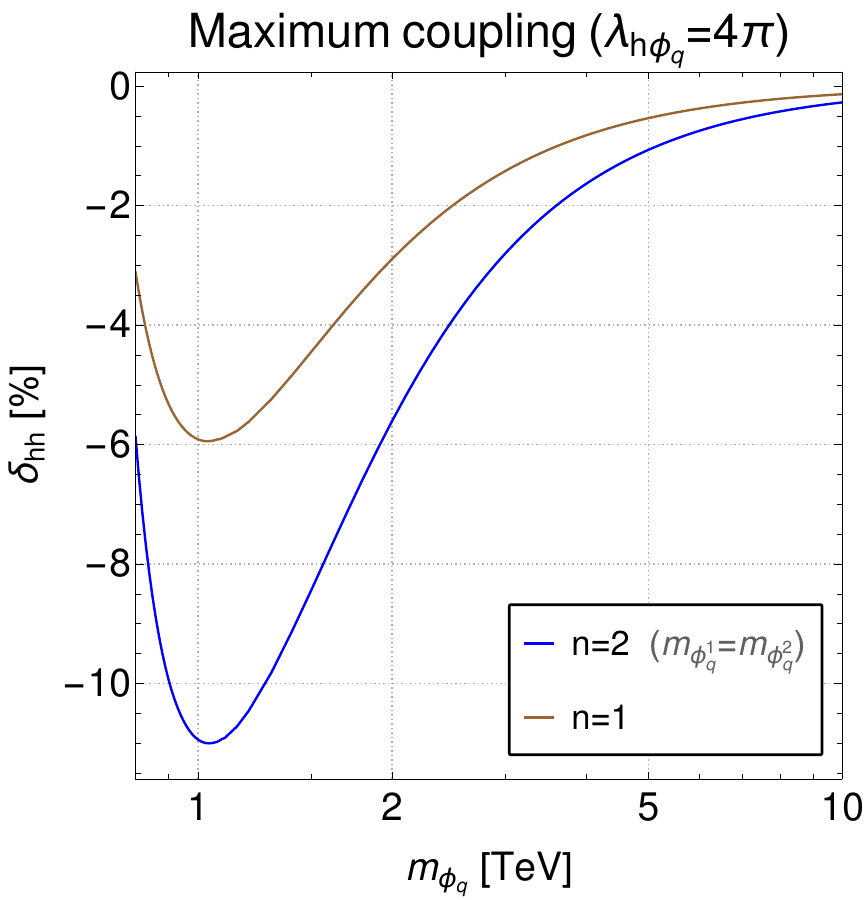}
            \caption{ $\delta_{hh} =
              (\sigma_{NP}-\sigma_{SM})/\sigma_{SM}$ as a function of
              the coloured scalar mass for the minimum value of the
              coupling (left) and maximum value of the coupling
              (right) and for $n=1$ (brown) and $n=2$ (blue). 
            The double Higgs cross section was calculated with \texttt{HPAIR}
            for a c.m.~energy of 14 TeV.}
            \label{subfig:doubleProd:mass}
        \end{figure}        
  
 In Fig.~\ref{subfig:doubleProd:mass} we now present  $\delta_{hh}$
 as a function of the coloured scalar mass for the minimum (left) and
 maximum (value) of the effective portal coupling $\lambdaHPQ$ and for
 $n=1$ and $n=2$.\footnote{Note that $\delta_{hh}$ has a
     different sign behaviour than $\delta_h$ as a function of
     $\lambdaHPQ$. This is a consequence of the destructive interference
     between trilinear and box diagrams. For details, see the
     discussion in Sec.~\ref{sec:4}. }  
As expected, the models share similar behaviours when reducing the $n=2$ case to a single coupling and mass under the equal parameters constraints which approximately double the cross section for two coloured scalars relative to the $n=1$ scenario.
Since $\delta_{hh}$ depends generally on powers of $(\lambdaHPQ)^p$
with $1 \leq p \leq 4$, this is an indication that the linear terms
seem to be the most significant ones for these results - doubling the couplings approximately doubles the cross section. 
Linear terms can only originate from the diagrams proportional to $\lambdaHPQ$ and their interference with the SM ones.
This is further supported by the observation that, when $\lambdaHPQ >
0$, the contributions to the Higgs pair production cross section are
negative and, hence, odd powers of the coupling are involved. 
    On the other hand, the shape of $\delta_{hh}$ is clearly described
    by a non-linear function in $\lambdaHPQ$. Contrary to what
    happened in single Higgs production, this is no longer necessarily
    a sign that the interference terms are insufficient to describe
    the results. This is due to the fact that a dependence on
    $\lambdaHPQ^2$ can originate from either the square of the purely
    NP diagrams proportional to $\lambdaHPQ$ (see diagrams
    \ref{subfig:pairDiagrams:triangle:BDM1},
    \ref{subfig:pairDiagrams:triangle:BDM2},
    \ref{subfig:pairDiagrams:box:BDM3},
    \ref{subfig:pairDiagrams:box:BDM4}) or from the SM interference
    with the NP diagrams proportional to $\lambdaHPQ^2$ (see diagrams
    \ref{subfig:pairDiagrams:box:BDM1},
    \ref{subfig:pairDiagrams:box:BDM2}). 
The interference term depends on $\sum_k (\lambda_{h\phi_q^k})^2$, while
the term originating from squaring the NP diagrams depends on $(\sum_k
\lambda_{h\phi_q^k})^2$.
	For $n=1$ the two dependencies are identical while for $n\geq2$ the former represents an extra degree of freedom for $\delta_{hh}$ for a fixed $\sum_k \lambda_{h\phi_q^k}$.
This is an important observation if we want to present the results as a function of the sum of the couplings, $\sum_k \lambda_{h\phi_q^k}$, as we did in the single Higgs case. 
 
 \texttt{HPAIR} has further been altered with the option to turn on or off particular sets of diagrams. Naturally, we will separate the ones proportional to $\lambdaHPQ$ and $\lambdaHPQ^2$. We further separate the two pairs of diagrams \ref{subfig:pairDiagrams:triangle:BDM1}-\ref{subfig:pairDiagrams:triangle:BDM2} and \ref{subfig:pairDiagrams:box:BDM3}-\ref{subfig:pairDiagrams:box:BDM4}, since their form factors are the same as in single Higgs production.
  The sets of diagrams chosen serve the purpose of separating the
  contributions of the form factors $F_\triangle^{\phi_q}$,
  $F_{\square_2}^{\phi_q}$, which are linear in $g_{\phi_q}^h$ and
  $g_{\phi_q}^{hh}$, respectively, and
  $F_{\square_1}^{\phi_q}$ and $G_{\square_1}^{\phi_q}$, which are
  proportional to the squared coupling $(g_{\phi_q}^h)^2$.

\begin{figure}[!htp]
        \centering
        \includegraphics[width=\linewidth]{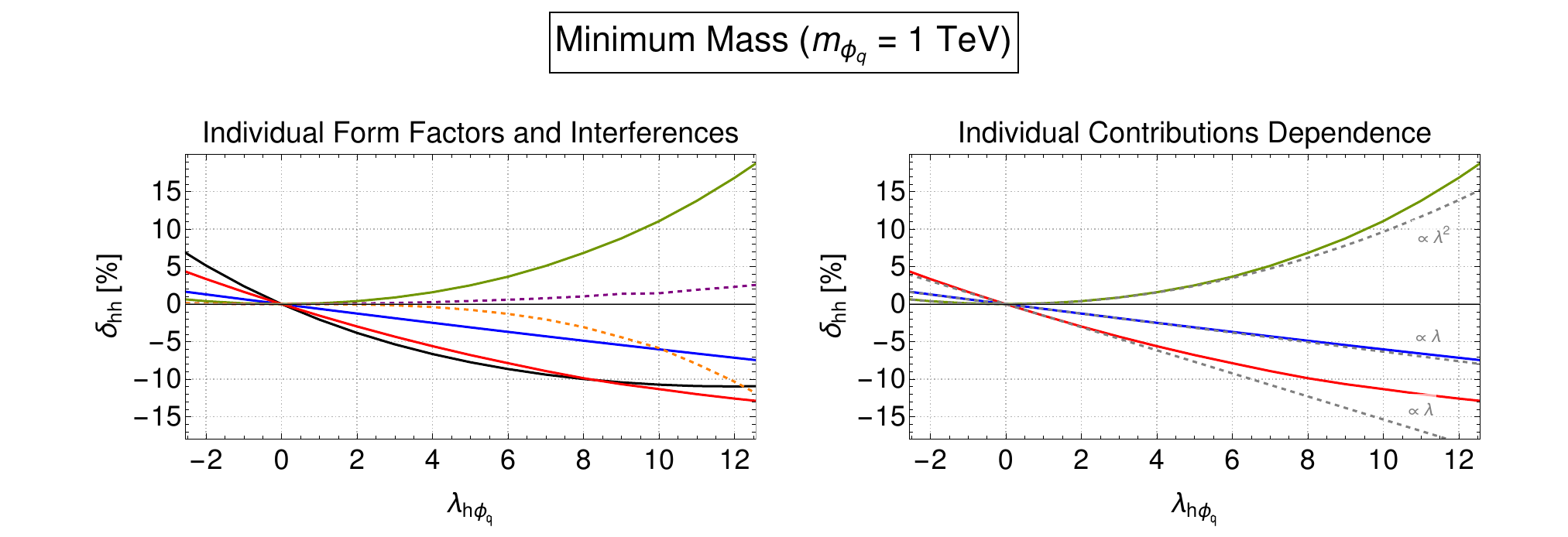}
        
        \includegraphics[width=\linewidth]{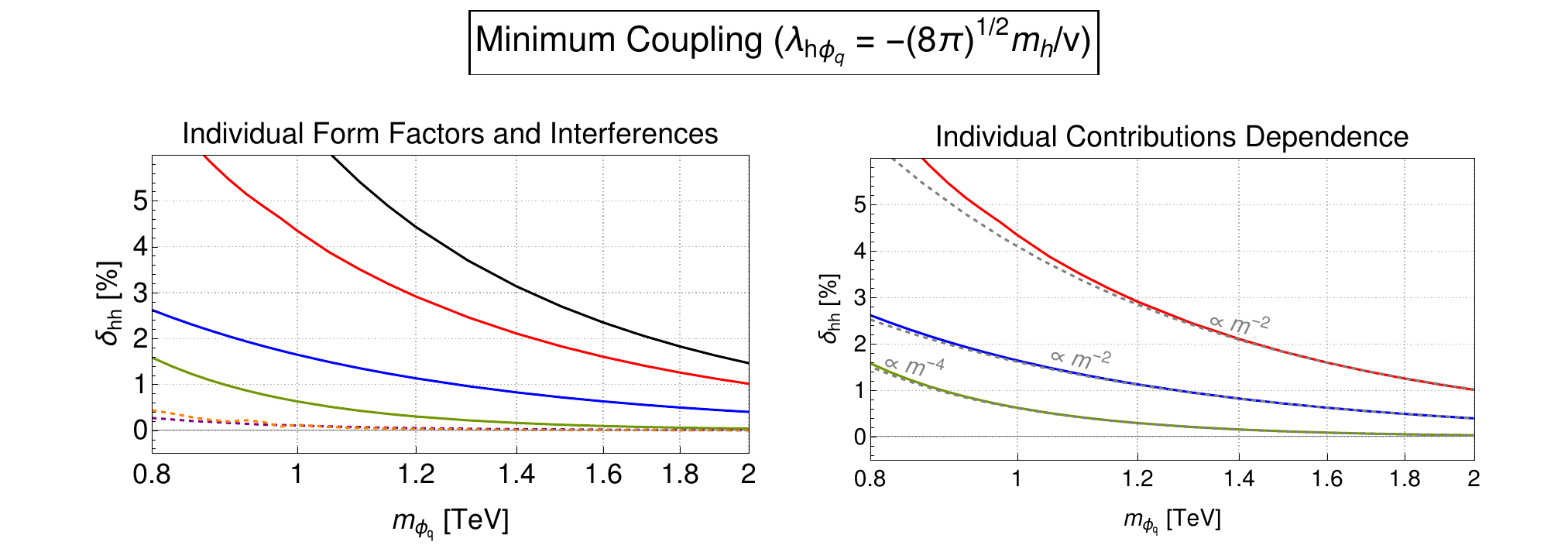}
        
        \includegraphics[width=\linewidth]{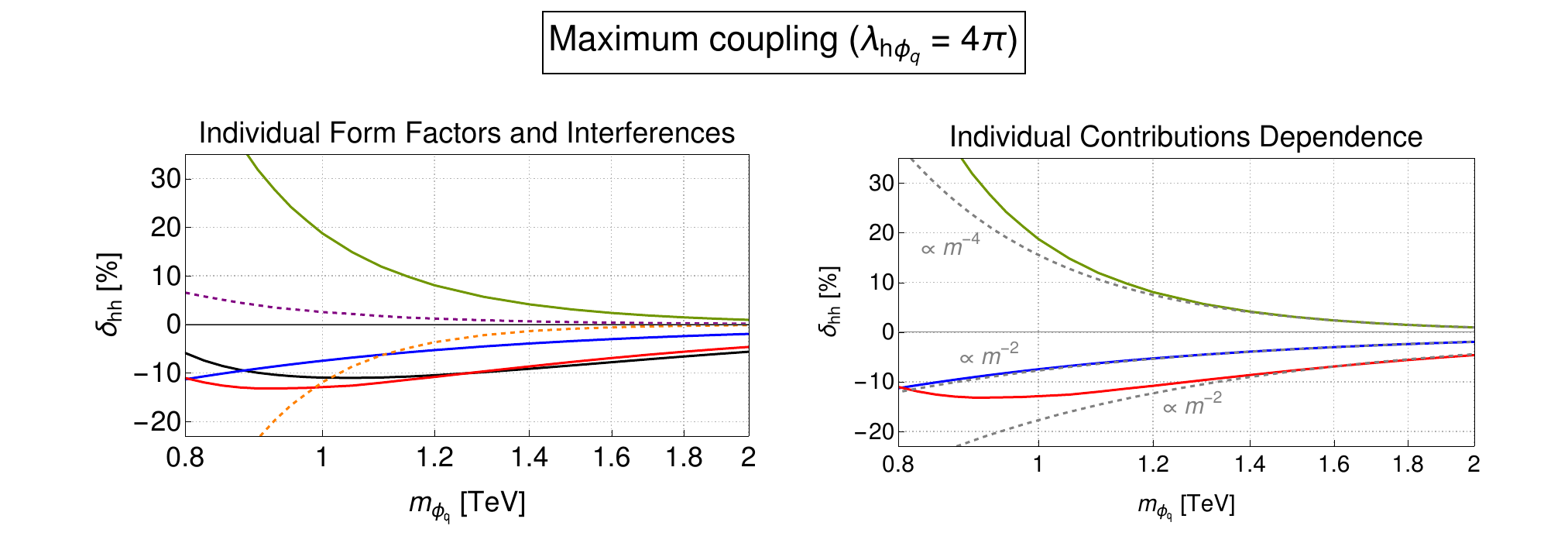}
        
        \includegraphics[width=0.85\linewidth]{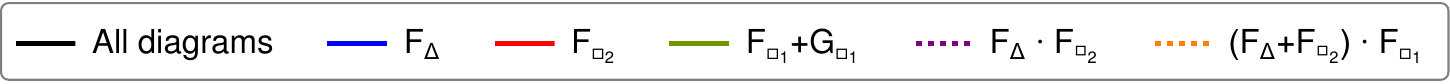}
        \caption[Double Higgs individual form factor contributions and
        interferences.]{$\delta_{hh}$ for $n=2$ and a fixed mass of 1 TeV as a function of the coupling (top), for the minimum coupling as a function of the mass (middle) and for the maximum coupling
        as a function of the mass (bottom). Left: 
        individual coloured form factor contributions and interference
        terms. Right: dependence of the individual contributions on the couplings (top) and the masses (middle and bottom).
       Black line: sum of all contributions;
       coloured lines: individual coloured scalar form factor
       contributions, separated as indicated by the legend and
       described in Eq.~(\ref{eq:formfaccontr}). 
      The dashed lines are for the interference terms.
        Grey dashed lines: asymptotic behaviour in the scenario where
        the interference terms with the SM are the dominant
        ones. The grey full line at 0 in all
        plots is there to guide the eyes.}
        \label{fig:pairProd:separatedDiag}
    \end{figure}

The results for $n=2$ are presented in Fig.~\ref{fig:pairProd:separatedDiag} for a fixed mass of 1 TeV as a function of the coupling  (top), for the minimum coupling as a function of the mass (middle) and for the maximum coupling
        as a function of the mass (bottom). The left plots show the individual contributions and the interference terms while the right plots present how the individual contributions behave with the couplings (top) and with the mass (middle and bottom).
        The black line represents the sum of all contributions, while
        the coloured lines represent the individual coloured scalar
        form factor contributions, separated as indicated by the
        legend. Note that the SM contributions drop out in $\delta_{hh}$.
More specifically, the contributions denoted by the different colours
are proportional to the following coloured form factors and couplings,
\beq
\begin{array}{lll}
\mbox{blue}/F_\triangle: & \sim \{F_\triangle^{\phi_q}, |F_\triangle^{\phi_q}|^2\}
  & \sim \{G_{\phi_q}^h, (G_{\phi_q}^h)^2 \} \\
\mbox{red}/F_{\square_2}: & \sim \{F_{\square_2}^{\phi_q}, |F_{\square_2}^{\phi_q}|^2\}
  & \sim \{G_{\phi_q}^{hh}, (G_{\phi_q}^{hh})^2
    \} \\
\mbox{green}/F_{\square_1}+G_{\square_1}: & \sim \{F_{\square_1}^{\phi_q}, G_{\square_1}^{\phi_q},
                |F_{\square_1}^{\phi_q}|^2, |G_{\square_1}^{\phi_q}|^2\}
  & \sim \{G_{\phi_q}^{h,\, 2}, G_{\phi_q}^{h,\, 2}, (G_{\phi_q}^{h,\,
    2})^2, (G_{\phi_q}^{h,\, 2})^2 \} \\
\mbox{violet}/F_\triangle \cdot F_{\square_2}: & \sim 2\mbox{Re}
  (F_\triangle^{\phi_q}  F_{\square_2}^{\phi_q*}) 
  & \sim G_{\phi_q}^h \cdot G_{\phi_q}^{hh}
  \\
\mbox{orange} &
\\
\quad /(F_{\triangle}+F_{\square_2})\cdot F_{\square_1}: &
\sim \{ 2\mbox{Re} (F_\triangle^{\phi_q} 
                 F_{\square_1}^{\phi_q*}) , 2\mbox{Re} (F_{\square_1}^{\phi_q}
                 F_{\square_2}^{\phi_q*})\} & \sim \{ G_{\phi_q}^h
                                              \cdot G_{\phi_q}^{h,\, 2},
G_{\phi_q}^{h,\, 2} \cdot G_{\phi_q}^{hh} \}
\end{array}  \label{eq:formfaccontr}
\eeq
where we introduced the abbreviations
\begin{eqnarray}
G_{\phi_q}^h \equiv \sum_{\phi_q^i} g_{\phi_q^i}^h \,, \quad
G_{\phi_q}^{hh} \equiv \sum_{\phi_q^i} g_{\phi_q^i}^{hh} \, \quad
G_{\phi_q}^{h,\, 2} \equiv \sum_{\phi_q^i}
    (g_{\phi_q^i}^h)^2 \;.
\end{eqnarray}
Note that $G_{\phi_q}^{hh}$ and $G_{\phi_q}^h$ only
  differ by a factor $1/v$.
The terms linear in the form factors of the blue, red and green
contribution stem from the interference with the SM form factors. 
The violet and orange contributions (dashed lines) hence denote the interference terms
between the coloured contributions. 
        The grey dashed lines in the right upper plot show the
        asymptotic behaviour in the coupling in the scenario where the interference
        terms with the SM are the dominant ones (where we generically
        denote by $\lambda$ the couplings $G_{\phi_q}^h$ (blue line) and
        $G_{\phi_q}^{hh}$ (red line) and by $\lambda^2$ the coupling
        $G_{\phi_q}^{h,\, 2}$ (green line)). 
We can infer from the plot that for masses of 1 TeV and higher, both the $F_\triangle$ (blue line) and the
$F_{\square_1}+G_{\square_1}$ (green line) contributions are rather well
described by only considering their interference with the SM form
factors. 
The $F_{\square_2}$ contribution
  (red line),
  however, is not well approximated by the interference with the SM
  contribution only. This observation is also confirmed by the middle and lower right plots which show the asymptotic behaviours in coloured masses for fixed coupling for the case that the interference term dominates.

\begin{figure}[!htp]
        \centering
        \includegraphics[width=0.6\linewidth]{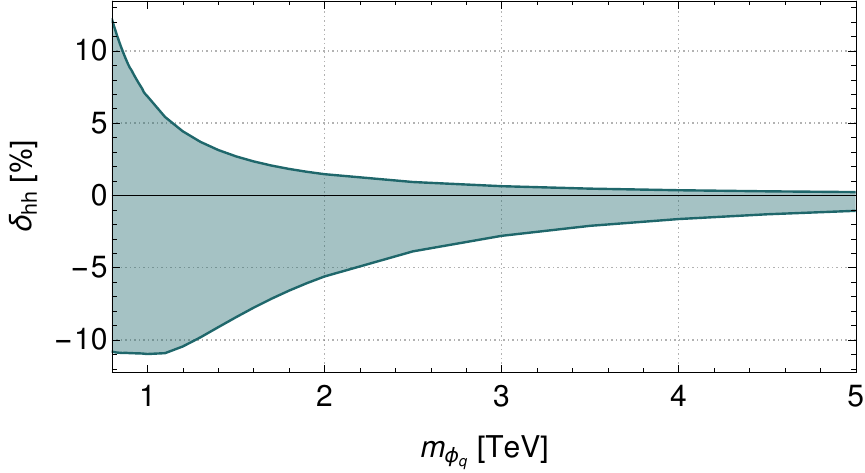}
        \caption[Double Higgs production with two scalars for equal masses but unconstrained couplings, full range as a function of mass.]{$\delta_{hh}$ as a function of the scalar mass $\mpq$ for $n=2$. The couplings are varied between the two extreme values as discussed previously. 
        \label{fig:pairProd:separatedDiag:method:fullDeltaMasses} }
    \end{figure}

We end this section by presenting in Fig.~\ref{fig:pairProd:separatedDiag:method:fullDeltaMasses} the double Higgs corrections $\delta_{hh}$ as a function of the scalar mass $\mpq$ for $n=2$. 
The couplings are varied between the two extreme values as
discussed previously. The new physics
  impact due to the additional coloured loops are
below 10 \% already for a mass of 1 TeV and fall steeply with rising mass. Therefore
the effect of two extra coloured scalar only will be extremely hard to probe even at the HL-LHC.
  
\subsection{Model $n=2$ for Different Masses}
\begin{figure}[!htp]
        	\centering
            \includegraphics[width=0.5\linewidth]{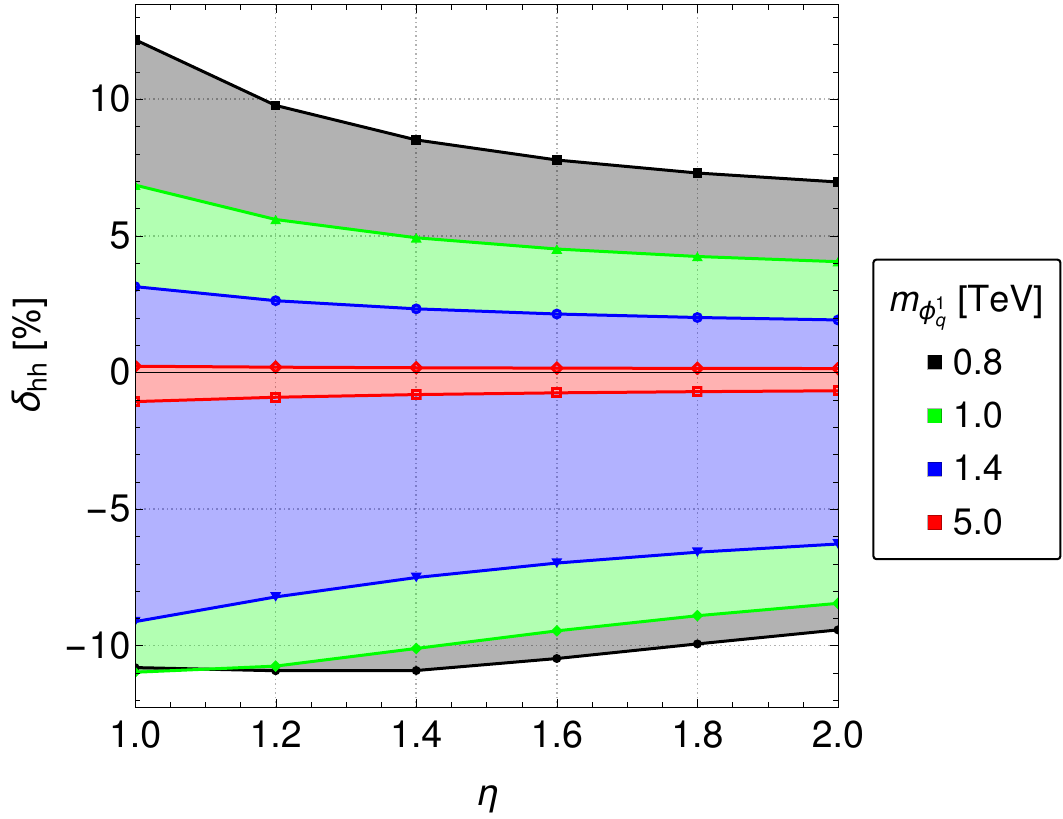}
        \caption[Double Higgs production with two scalars for
        different ratios between the two masses and unconstrained
        couplings.]{ $\delta_{hh}$ as a function of the variable
          $\eta$ defined as $m_{\phi_q^2} = \eta \ m_{\phi_q^1}$. The plot is for $n=2$ and $\eta$ was varied between 1 and 2.  
        This relaxes the condition of equal masses fixing
        $m_{\phi_q^1}$ while increasing
        $m_{\phi_q^2}$. The different colours
        represent the different values for the fixed mass,
        $m_{\phi_q^1}$.
        }
        \label{fig:pairProd:model3:differentMasses}
    \end{figure}

    We very briefly look at the implications of relaxing the
    condition of equal masses. For this, we calculated the full range
    of $\delta_{hh}$ for ratios between the two masses of
	\begin{equation}
m_{\phi_q^2} = \eta \ m_{\phi_q^1} \ ,
	\end{equation}	    
    while scanning over all values for the couplings. Using
    \texttt{HPAIR}, the results for $\delta_{hh}$ are displayed in Fig.~\ref{fig:pairProd:model3:differentMasses}.
    The plot is for $n=2$ and $\eta$ was varied between 1 and 2 and we find similar conclusions to the ones discussed in the previous section. There, we found that, when increasing the two masses equally above 1 TeV, the range of values for $\delta_{hh}$ would always shrink. Naturally, when increasing only one mass, we expect the same to happen, although the effect is milder as can be seen in the figure.
We have checked that for $n=2$,
\begin{equation}
	\delta_{hh}^{\text{max/min}}(m_{\phi_q^1}, m_{\phi_q^2}) 
	\approx 
		\big[
		  \delta_{hh}^{\text{max/min}}(m_{\phi_q^1},m_{\phi_q^1}) 
		+ \delta_{hh}^{\text{max/min}}(m_{\phi_q^2},m_{\phi_q^2}) 
		\big] / 2 
	\ .
\end{equation}
This is a consequence of the more general observation that 
\begin{equation}
	\delta_{hh}(m_{\phi_q^1}, m_{\phi_q^2}, \lambda_{h\phi_q}) 
	\approx 
		\big[ 
 			  \delta_{hh}(m_{\phi_q^1},m_{\phi_q^1},\lambda_{h\phi_q}) 
			+ \delta_{hh}(m_{\phi_q^2},m_{\phi_q^2},\lambda_{h\phi_q}) 
		\big] / 2
	 \ .
\end{equation}
We can conclude that relaxing the equal masses condition will not result in any additional behaviour of note. Reducing one mass has the same effect as reducing both but with the obvious difference that the effect is less significant.
Consequently, we will also not obtain a larger range of values for $\delta_{hh}$ by adding this extra freedom. We can now extrapolate this conclusion for higher values of $n$. This scenario will be discussed in the next section.

\subsection{Models with $n$ Coloured Scalars}
We finalise this chapter by looking in more detail at double Higgs production in the case of an arbitrary number of scalars.
    The parameter space will be comprised of $n$ effective couplings $\lambda_{h\phi_q^k}$
    to the Higgs boson and $n$ scalar masses
    $m_{\phi_q^k}^2$ ($k=1,...,n$), one for each of the coloured scalars, resulting
    in $2n$ input parameters ( $\lambda_k\equiv \lambda_{h\phi_q^k}$ from now on). 
   We again start with the condition of equal masses, $ m_{\phi_q^k}^2
   = m_{\phi_q^l}^2 \equiv \mpq^2$, reducing the input parameters to $n+1$. As discussed in the previous section, this condition should be sufficient in order to fully explore $\delta_{hh}$.
    For single Higgs production, this resulted in a simple dependence of the corrections on only the total sum of the couplings. 
    In the case of Higgs pair production, the amplitude now contains both $\lambda_k$ and $\lambda_k^2$ terms and thus there are now two relevant quantities: the total sum of the couplings and the total sum of the squared couplings.     %
    Naturally, taking these two sums over the couplings as our
    parameters is advantageous, as it allows us to reduce the number
    of input parameters from $n+1$ to only 3, to properly study $\delta_{hh}$ for any model. 
    The cases $n=1,2$ have two and three independent input parameters,
    respectively, and were studied in the previous sections.
    
We will now proceed to write both the cross section  $\sigma_{hh}$
and the relative deviation from
  the SM cross section, $\delta_{hh}$, as a function of the two
effective quantities, $\sum \lambda_k$ and $\sum
\lambda_k^2$. Assuming a common fixed coloured mass
  $\mpq^2$, as we do from now on, we note that because single Higgs
production only depends on $\sum \lambda_k$,
if one is able to write the relative deviations
  $\delta_{h,hh}$ as a function of the same variable, the two results can be combined.
 Even under the simplification of equal masses, we have now
 $\delta_{hh}$ as a function of two parameters, $\delta_{hh} \equiv
 \delta_{hh}\left( \sum \lambda_k, \sum \lambda_k^2
 \right)$. Therefore, the model limits are represented by a
 2-dimensional region in the parameter space of these two
 sums. By taking the approach
 where we consider the sum $\sum \lambda_k$ as the independent
 variable, the limits on this sum are easily obtained. Applying the same
 constraints, $\lambda_{min} \leq \lambda_k \leq \lambda_{max}$, to
 all the couplings of a model with $n$ coloured scalars, the sum of the couplings will be limited by
    \begin{equation}
        n \lambda_{min} \leq \sum \lambda_k \leq n \lambda_{max} \ .
        \label{eq:pairProd:general:modelLimits:sum}
    \end{equation}
As for the limits for $\sum \lambda_k^2$ as a function of $\sum \lambda_k$, we need to find the solution of a conditional extreme problem: the extremes of $\sum \lambda_k^2$ subject to the constraints $\sum \lambda_k = c$ and $\lambda_{min} \leq \lambda_k \leq \lambda_{max}$.
   Within the $n$ dimensional space of the individual couplings, $(\lambda_1, \lambda_2, ... \lambda_n)$, the region of interest is represented by an $n-1$ hyperplane defined by $\sum \lambda_k = c$ but constrained by an $n$-dimensional hypercube resulting from the constrained couplings, $\lambda_{min} \leq \lambda_k \leq \lambda_{max}$.
For a fixed sum ($\sum \lambda_k=c$) the minimum of $\sum \lambda_k^2$ is given when the couplings are all equal
     \begin{equation}
     	 \sum \lambda_k^2
     		\geq \frac{ c^2 }{ n } 
			 = \frac{ \left( \sum \lambda_k \right)^2 }{ n } 
 		\ ,
	    \label{eq:pairProd:general:modelLimits:squared:lower}
     \end{equation}
     which is a just a Cauchy-Schwartz type of inequality.

The determination of the maximum is more elaborated. The solution is given by the edges of the hypercube or, more simply, when all but one coupling are fixed to $\lambda_{min}$ or $\lambda_{max}$. This can be cast in the form,
    \begin{multline}
	    \sum_{k=1}^{n} \lambda_k^2 \leq 
        \sum_{j=0}^{n-1}
        \Bigg\{
            \left[
                j \lambda_{max}^2 + \left( n-1-j \right) \lambda_{min}^2
                + \left(
                      \sum_{k=1}^{n} \lambda_k
                    - \left(
                          j \lambda_{max} + \left( n-1-j \right) \lambda_{min}
                      \right)
                  \right)^2
            \right] \times \\
        \left[ 
        \Theta \left( \sum_{k=1}^{n} \tilde{\lambda}_k - j \right)
       -\Theta \left( \sum_{k=1}^{n} \tilde{\lambda}_k - (j+1) \right)
        \right]
       	\Bigg\} \ ,
        \label{eq:pairProd:general:modelLimits:squared:upperHeaviside}
    \end{multline}
where $\tilde \lambda_k = (\lambda_k - \lambda_{min})/(\lambda_{max} -
\lambda_{min}) $ and $\Theta (x)$ is the Heaviside
function. The derivation of this formula can be found in \cite{Daniel}.

\begin{figure}[!htp]
        \centering
        \begin{subfigure}[c]{.4\linewidth}
        	\centering
            \includegraphics[width=\linewidth]{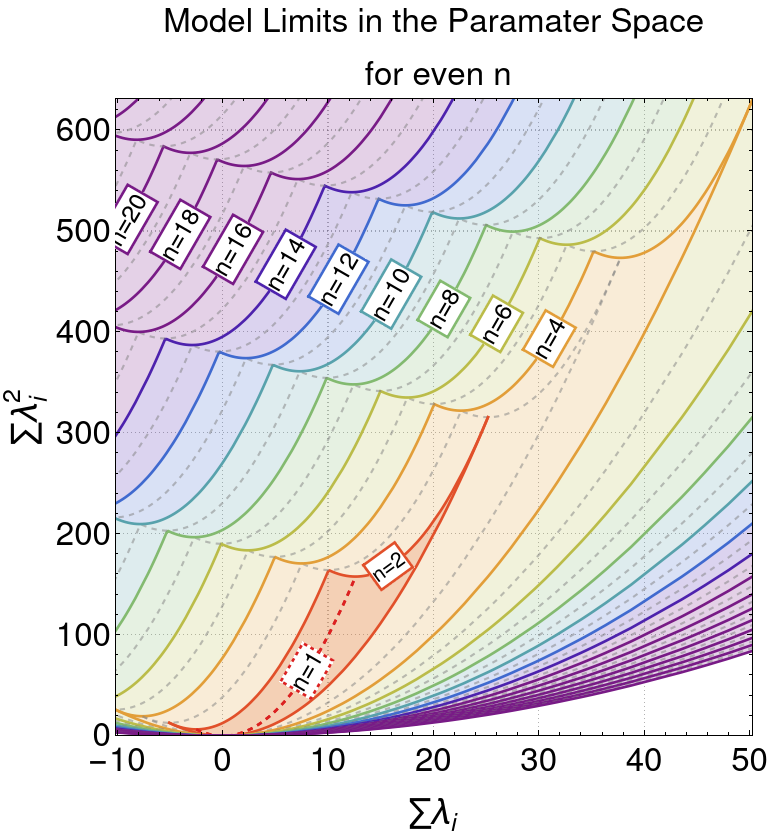}
            \caption{ }%
            \label{subfig:pairProd:general:modelLimits:1}
        \end{subfigure}%
        \hspace{5mm}
        \begin{subfigure}[c]{.4\linewidth}
        	\centering
            \includegraphics[width=\linewidth]{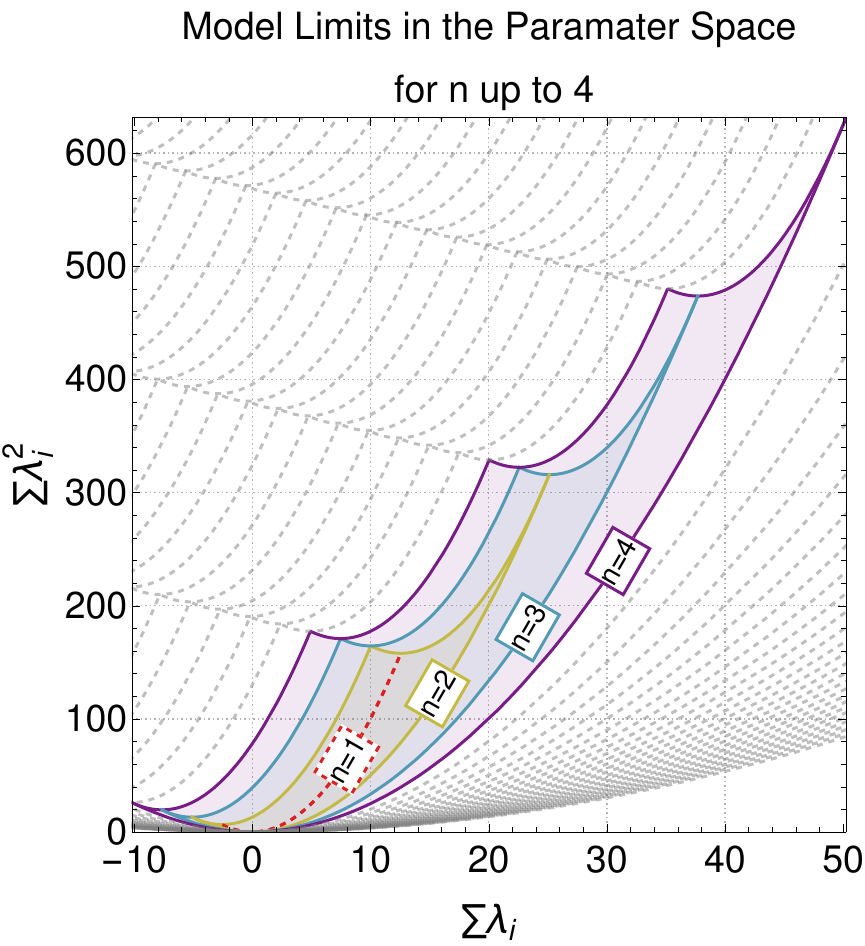}
            \caption{ }%
            \label{subfig:pairProd:general:modelLimits:2}
        \end{subfigure}
        \caption[Models allowed regions in the parameter space of sums for different number of scalars, $n$.]{ 
				Allowed regions for each model
                                defined by the number $n$ of coloured
                                scalars. The left plot depicts the borders of the labelled regions for even values of $n$. The odd values in-between are represented by a dashed grey line. 
				Of note is that, for $n=1$, the limits are not a region but just a single line (represented as a dashed red line). The right plot focuses on the lower values of $n$, representing all up to $n=4$. Higher values are represented by grey dashed lines.       } \label{fig:pairProd:general:modelLimits}
    \end{figure}

	In Fig.~\ref{fig:pairProd:general:modelLimits} we present an example of the region determined by the above conditions. We show the allowed regions for each model defined by the number $n$ of coloured scalars. 
	The left plot depicts the borders of the labelled regions for even values of $n$. The odd values in-between are represented by a dashed grey line. The right plot focuses on the lower values of $n$, representing all up to $n=4$. Higher values are represented by grey dashed lines.

    \begin{figure}[!htp]
        \centering
        \includegraphics[width=0.45\linewidth]{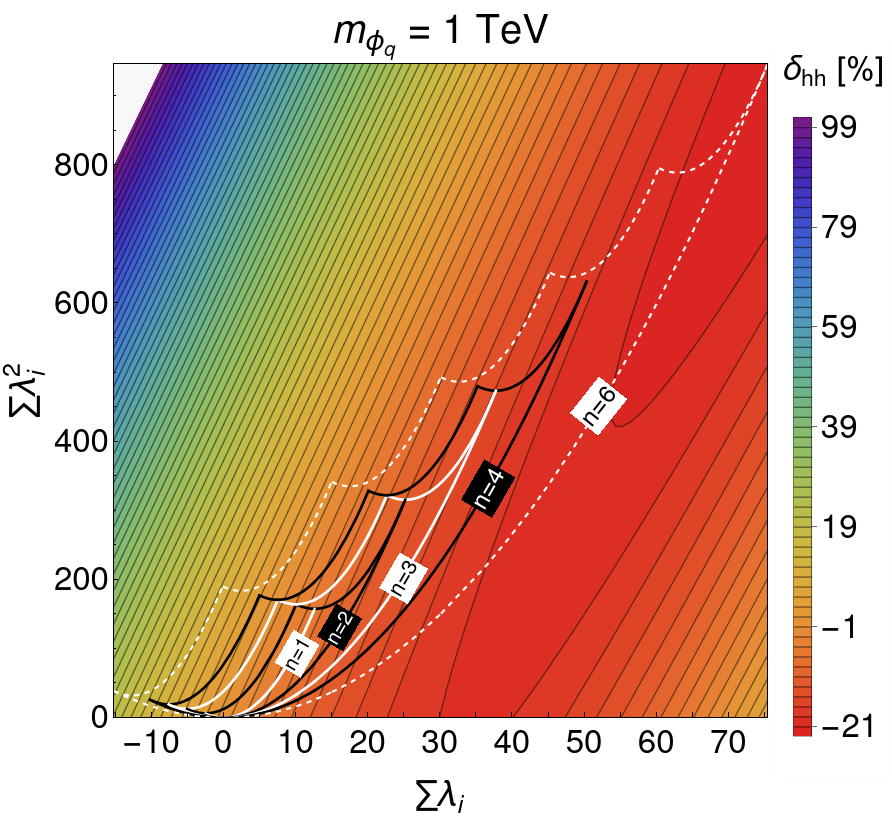}
        \hspace{5mm}
        \includegraphics[width=0.45\linewidth]{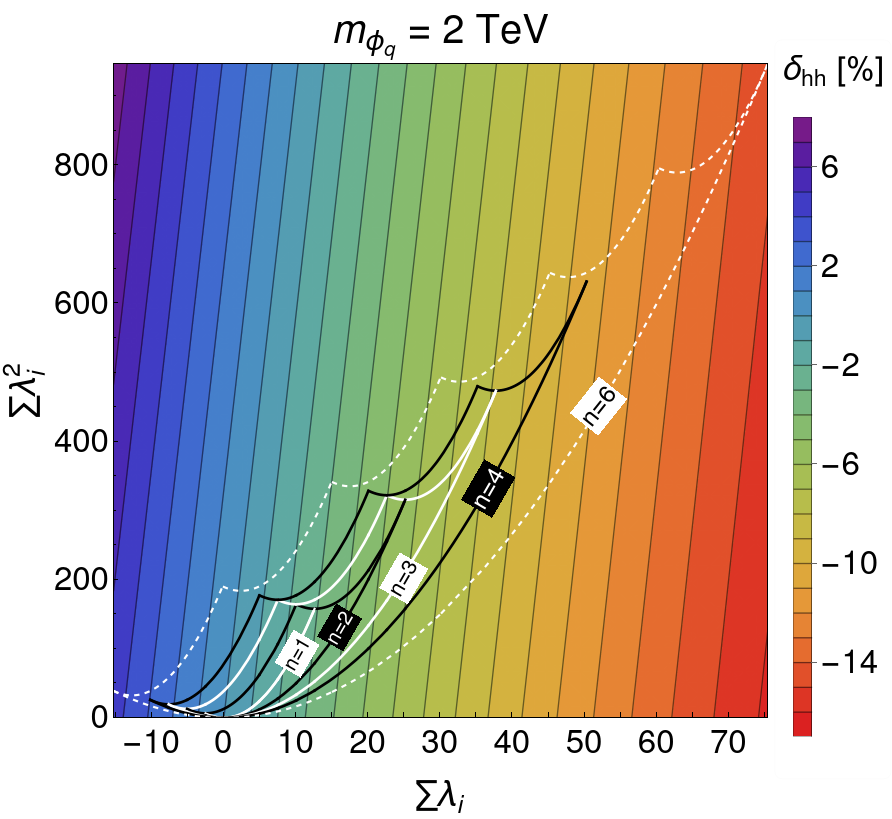}
        \caption[Double Higgs production contributions with Model Regions in the parameter space of sums for equal masses constraint, masses between 0.8 TeV and 2 TeV.]{ $\Sigma \lambda_i^2$ as a function of $\Sigma \lambda_i$ with
        the value of $\delta_{hh}$ in the colour bar. Left (right):
        $m_{\phi_q} =$ 1(2) TeV. Note that the colour scale is not the same in the two figures. The contours represent the allowed values for $\delta_{hh}$  for each value of $n$. 
		The allowed region for $n=5$ has not been represented as it would make the identification of the $n=4$ and $n=6$ regions more difficult. 
		}
        \label{fig:pairProd:general:simulationWithModels:moreMasses}
    \end{figure}

	The next step is to calculate $\delta_{hh}$ as a function of the two sums $\sum \lambda_k$ and $\sum \lambda_k^2$. This can be done by discretising the two variables in $N$ points which would involve a computational time of $\mathcal{O}(N^2)$.
	We will instead present an approach that can recycle the previous results from Fig.~\ref{fig:pairProd:separatedDiag} with $n=2$ and a fixed mass, which can be computed in $\mathcal{O}(N)$ time. We separate the individual contributions to $\delta_{hh}$ into three components as follows:
\beq
\begin{array}{lll}
\delta_{n,\text{ eq}}^{ \ A }(\lambda): 
	& \sim \{
		 F_\triangle^{\phi_q},   |F_\triangle^{\phi_q}|^2, 
		 F_{\square_2}^{\phi_q}, |F_{\square_2}^{\phi_q}|^2,
		 2\mbox{Re} (F_\triangle^{\phi_q}  F_{\square_2}^{\phi_q*}) 
		\}
  	& \sim \{
  		G_{\phi_q}^h, (G_{\phi_q}^h)^2,
  		G_{\phi_q}^{hh}, (G_{\phi_q}^{hh})^2,
  		G_{\phi_q}^h \cdot G_{\phi_q}^{hh} 
  		\} 
  	\\
\delta_{n,\text{ eq}}^{ \ B }(\lambda): 
	& \sim \{
		F_{\square_1}^{\phi_q}, G_{\square_1}^{\phi_q},
        |F_{\square_1}^{\phi_q}|^2, |G_{\square_1}^{\phi_q}|^2
        \}
  	& \sim \{
  		G_{\phi_q}^{h,\, 2}, G_{\phi_q}^{h,\, 2}, 
  		(G_{\phi_q}^{h,\,2})^2, (G_{\phi_q}^{h,\, 2})^2 
  		\} 
  	\\
\delta_{n,\text{ eq}}^{ \ C }(\lambda): 
	& \sim \{ 
		2\mbox{Re} (F_\triangle^{\phi_q} F_{\square_1}^{\phi_q*}) , 
		2\mbox{Re} (F_{\square_1}^{\phi_q} F_{\square_2}^{\phi_q*})
		\} 
	& \sim \{ 
		G_{\phi_q}^h \cdot G_{\phi_q}^{h,\, 2},
		G_{\phi_q}^{h,\, 2} \cdot G_{\phi_q}^{hh} 
		\}
\end{array}  
\eeq
	where the label "eq" indicates the equal coupling condition ($\lambda_{k}=\lambda_{l}\equiv \lambda$) and there is only one independent parameter, $\lambda$, due to this condition. We have already found that all three components can be significant and must be taken into account. The equivalence between these three components for a model with $n$ couplings with an arbitrary model with $n'$ couplings, $\{ \lambda'_1, ..., \lambda'_{n'} \}$, is given by the following formula:
\begin{equation}
        \delta^{n'}_{hh} ( \{ \lambda'_1, ..., \lambda'_{n'} \} ) = 
        	  \delta_{n,\text{ eq}}^{ A } 
        	  	(\lambda)
        	  	\bigg\rvert_{
        	  			\lambda = \frac{1}{n}
        	  				\sum \lambda'_k }
			+ \delta_{n,\text{ eq}}^{ B } 
        	  	(\lambda)
        	  	\bigg\rvert_{
        	  			\lambda = \sqrt{ \frac{1}{n}
        	  				\sum \lambda_k^{'2} } }
            + \delta_{n,\text{ eq}}^{ C }
        	  	(\lambda)
        	  	\bigg\rvert_{
						\lambda = \sqrt[3]{ 
            			\left( \frac{1}{n} \sum \lambda'_k   \right) 
            			\left( \frac{1}{n} \sum \lambda_k^{'2} \right) } 
            				} \ \ \ .
        \label{eq:pairProd:separatedDiag:methodForN}
    \end{equation}
In other words, the $\delta_{hh}^{n'}$ for a model
  with $n'$ couplings can be obtained from the results for a model with
  $n$ equal couplings $\lambda$, by taking the $A$, $B$, and $C$
  contributions at the $\lambda$ values indicated by the vertical bars.
	
In Fig.~\ref{fig:pairProd:general:simulationWithModels:moreMasses} we show $\Sigma \lambda_i^2$ as a function of $\Sigma \lambda_i$ with
        the value of $\delta_{hh}$ in the colour bar. The left plot is for a coloured scalar mass of 1 TeV while the right plot is for 2 TeV. Note that the colour scale is not the same in the two figures. The contours represent the allowed values 
        for $\delta_{hh}$  for each value of $n$.
The comparison of the two
  plots shows that, as expected, the range of variation of $\delta_{hh}$
  decreases with increasing value of $m_{\phi_q}$. Furthermore, the
  dependence of $\delta_{hh}$ on $\sum \lambda_k^2$ decreases with
  increasing coloured mass which is due to the fact that the terms
    proportional to $\sum \lambda_k^2$ are suppressed by a factor of
    $1/m_{\phi_q}^4$.

    \begin{figure}[!htp]
        \centering       
        \includegraphics[width=0.45\linewidth]{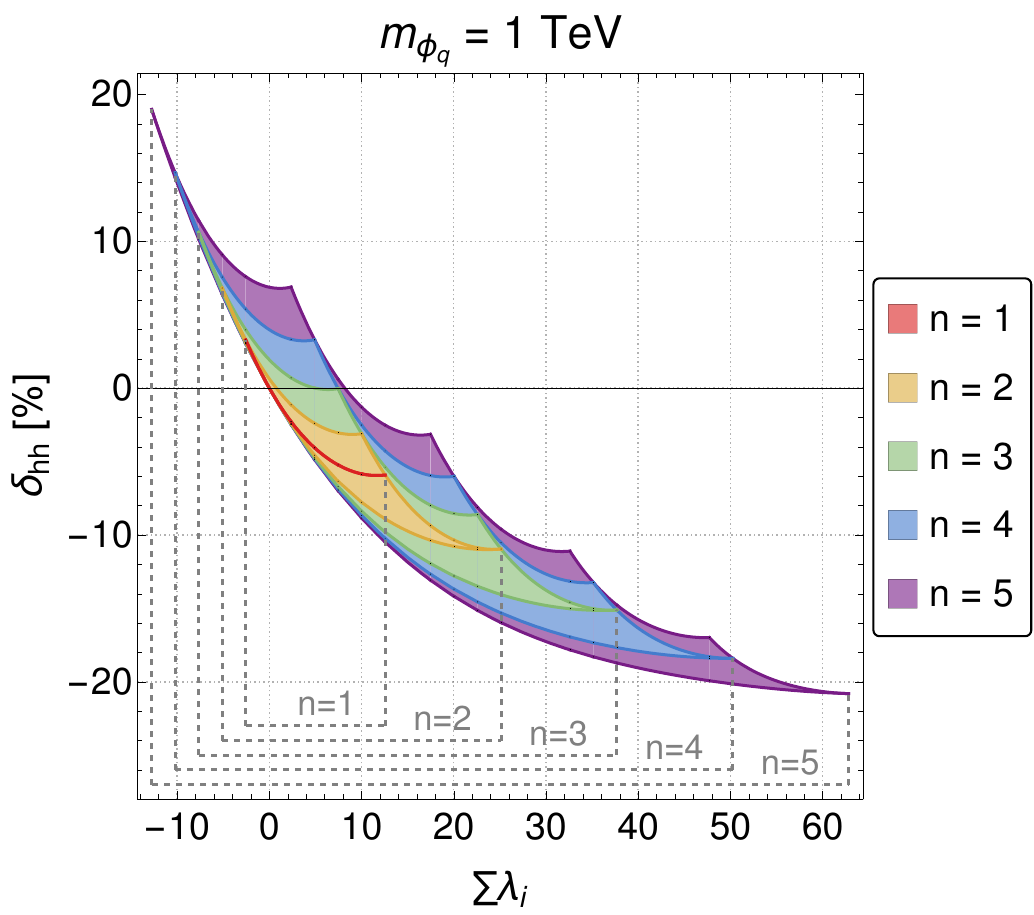}
        \hspace{5mm}
        \includegraphics[width=0.45\linewidth]{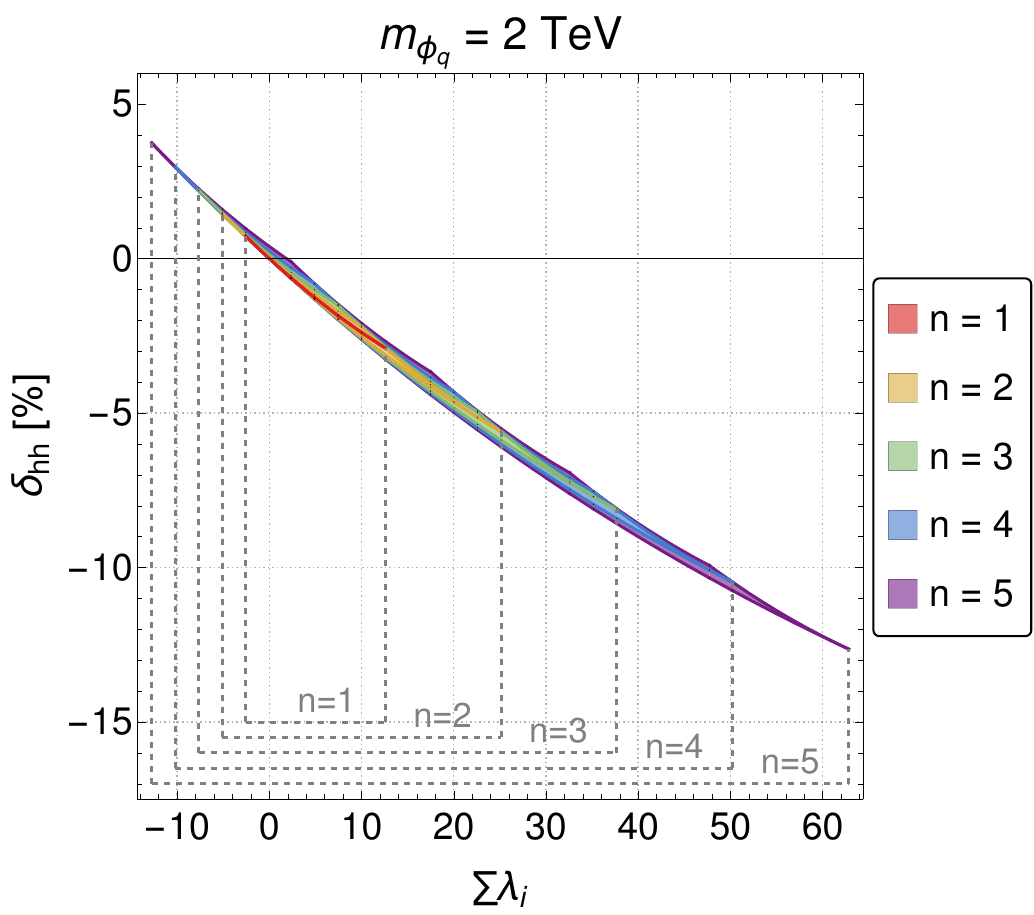}
        \caption[Double Higgs production contributions as a function of only the sum of couplings with equal masses constraint for different number of scalars $N$, masses between 0.8 and 1 TeV. ]
        { 
        $\delta_{hh}$ for a scalar mass of 1 TeV (left) and 2 TeV
        (right) as a function of $\sum \lambda_i$. This
        encompasses the possible range from the freedom in $\sum
        \lambda_i^2$ and was calculated with the results and model
        limits from Figure
        \ref{fig:pairProd:general:simulationWithModels:moreMasses}.
        }
        \label{fig:pairProd:general:results:vssum}
    \end{figure}

    In Fig.~\ref{fig:pairProd:general:results:vssum} we now show
    $\delta_{hh}$ as a function of the sum of couplings, $\sum
    \lambda_k$. The range of variation is related with the freedom in
    $\sum \lambda_k^2$ and was calculated with the results and model
    limits from
    Fig.~\ref{fig:pairProd:general:simulationWithModels:moreMasses}. As the mass grows the term in $\lambda_k$ becomes increasingly important and for a mass of 2 TeV the variation in $\lambda_k^2$ almost vanishes. Therefore, for large masses the dependence 
    for single and double Higgs productions becomes very
    similar. Note, that since the interference is destructive for positive couplings in the case of double Higgs production, the maximum $\delta_{hh}$ occurs
for the smaller (negative) values of the couplings.

    \begin{figure}[!htp]
        \centering       
        \includegraphics[width=0.45\linewidth]{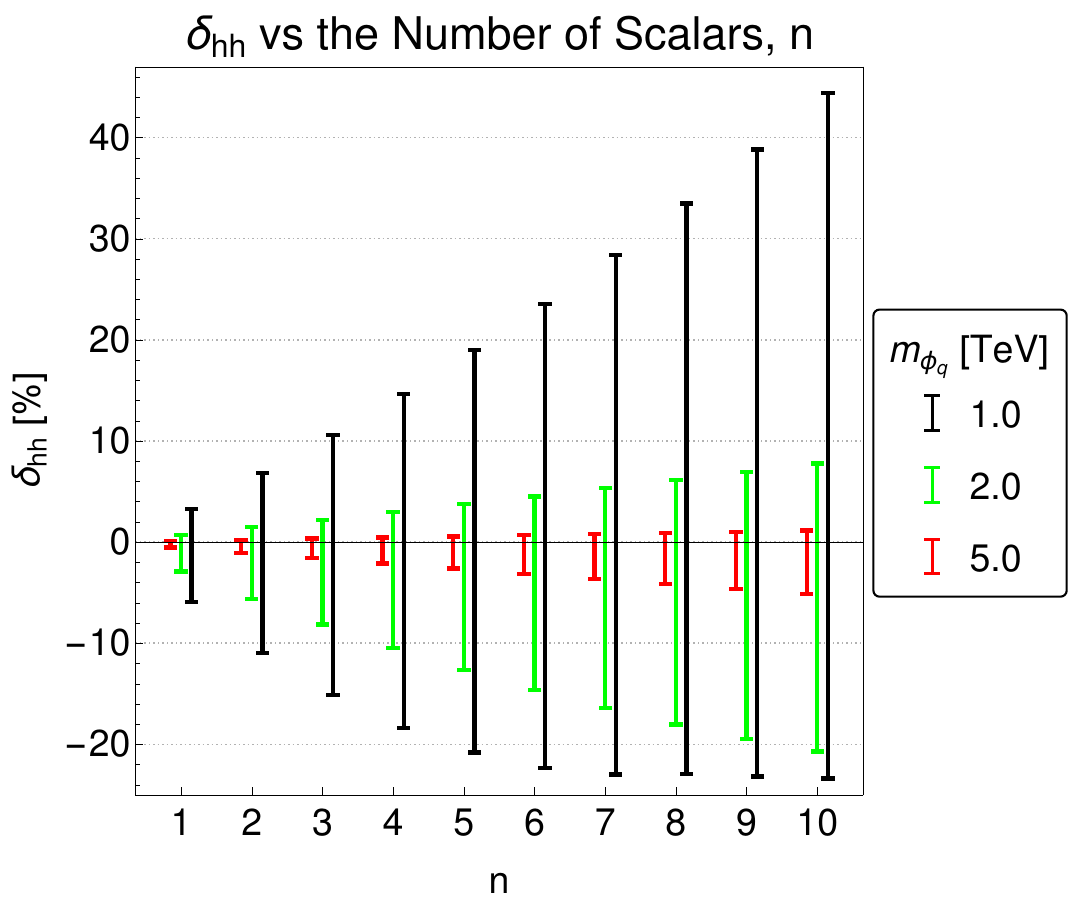}
        \hspace{5mm}
        \includegraphics[width=0.45\linewidth]{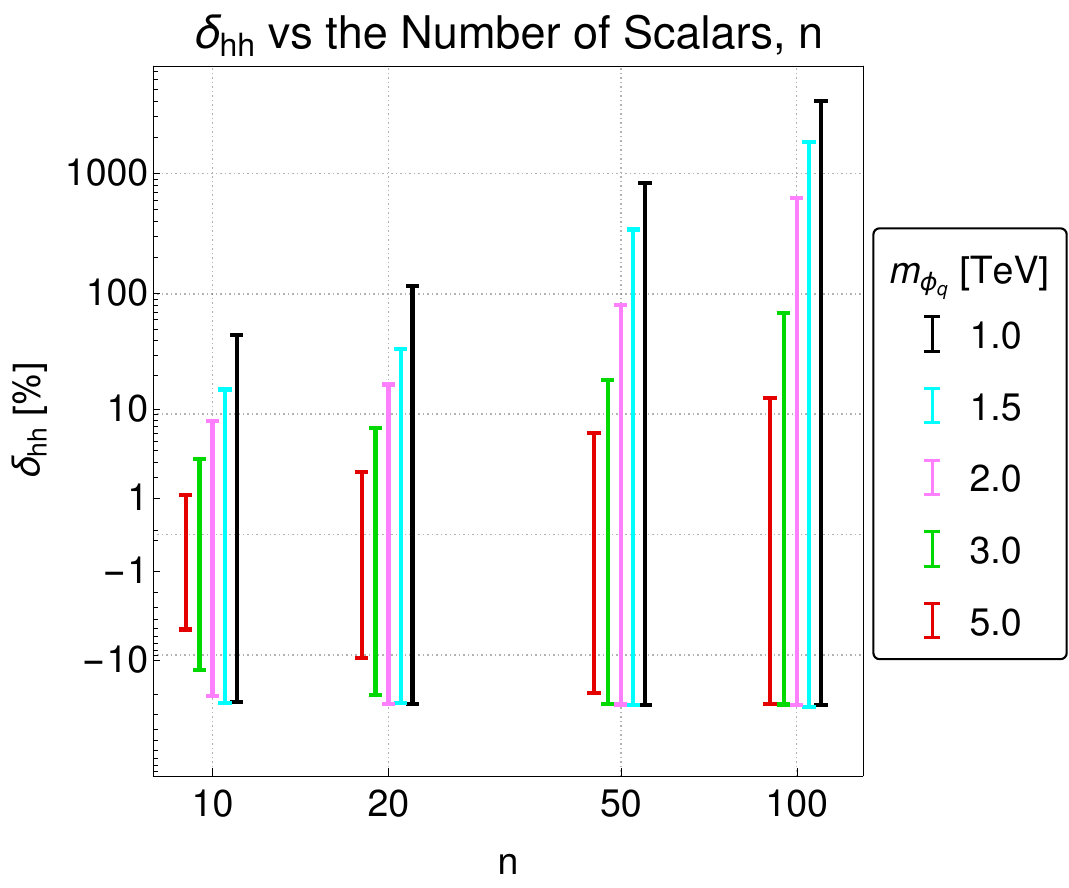}
        \caption[Double Higgs production contributions final results with equal masses constraint in function of the number of scalars, $N$, or scalar masses. ]
        { 
        $\delta_{hh}$ as a function of
        the number of scalars $n$ for three (left) and five (right) coloured scalar
        masses with $n=1,...,10$ (left) and $n=10,20,50,100$ (right).
        }
        \label{fig:pairProd:general:results}
    \end{figure}

Finally, in Fig.~\ref{fig:pairProd:general:results} left (right)  we present  $\delta_{hh}$ as a function of the number of scalars $n$ for three (five) coloured scalar masses. The left plot shows 
the scenarios from $n=1$ to $n=10$ while the right plot shows larger values of $n$. For small $n$ the deviations from the SM are small as we had seen before but they can be extremely large for very large values of $n$, even if the
coloured scalar masses are large. 

 Contrary to single Higgs production, the experimental limits on double Higgs production are very weak and at the moment unlikely to be useful in constraining the parameter space. 
 The lowest observed bound on the limit for double Higgs production,
 as reported by the ATLAS collaboration to be 6.9 times the SM cross section, is equivalent to a $\delta_{hh}$ of $590\%$ \cite{ATLAS:2019qdc} for a c.m.~energy of
 13 TeV. With a mass of 1 TeV, this would apply constraints only above $n\approx 45$.
 	As for possible future improvements we can consider the HL-LHC
  projections \cite{cepeda2019higgs}. For the $hh \rightarrow b \bar{b} b
  \bar{b}$ channel a reported value as low as 1.6 times the SM cross section, equivalent to a $\delta_{hh}$ of $60\%$, can
  be achieved, assuming that the overall uncertainty scales with the
  luminosity as $1/\sqrt{L}$. This would bring the previous threshold value of $n$ down to around 13.
    This means that certain combinations of the masses with the number of scalars will certainly be constrained with future measurements.

 %
    %
   
    %

\section{Single Higgs vs. Double Higgs Production}
\label{sec:4}
In the previous chapters we have discussed in detail the contribution of an arbitrary number of coloured scalars to single Higgs and di-Higgs production processes via gluon fusion at the LHC. We will now discuss the complementarity between the two processes. 
One should note, however, that although we expect a good precision in
the measurement of the single Higgs process this is not the case for di-Higgs production.

The first point to note is that the NP contribution to the single Higgs mode has a constructive interference for positive $\lambda_k$ while for di-Higgs it is negative.       
    The reason for the positive interference term for $\lambda_k>0$, is that both the SM and the NP form factors in single Higgs production are positive. 
    For double Higgs production this is no longer the case. The reason behind this is the destructive interferences between the ($F_\triangle^{Q/\phi_q}$, $F_{\square_2}^{\phi_q}$) and ($F_{\square}^Q$,$F_{\square_1}^{\phi_q}$) form factors. It is already well known that the SM triangle and box form factors interfere destructively as can be read off from their values in the heavy quark limit,
$F_{\triangle}^Q=\frac{2}{3}+\mathcal{O}(m_Q^{-2})$ and
$F_{\square}^Q=-\frac{2}{3}+\mathcal{O}(m_Q^{-2})$.
To understand why this also applies to our coloured scalars we can make use of the Low Energy Theorem as was done in \cite{Agostini:2016vze,Degrassi:2008zj} (for squarks) to deduce the sign of $F_{\square_1}^{\phi_q}$. By this theorem, $F_{\square_1}^{\phi_q}$ is given by the derivative in mass of the term $F_\triangle^{\phi_q} / m_{\phi_q}^2$. 
Since we already know that the triangle form factor for large scalar masses decreases with the mass, the sign of $F_{\square_1}^{\phi_q}$ will be negative.
	Therefore the negative contributions for positive couplings we are observing are due to the interference terms of the NP form factors,
	$F_\triangle^{\phi_q} \cdot F_{\square_1}^{\phi_q}$ and 
	$F_{\square_2}^{\phi_q} \cdot F_{\square_1}^{\phi_q}$,
but also from the interference between SM and NP form factors, 
	$F_\square^Q \cdot F_\triangle^{\phi_q}$,
	$F_\square^Q \cdot F_{\square_2}^{\phi_q}$ and
	$F_\triangle^Q \cdot F_{\square_1}^{\phi_q}$.
The remaining $F \cdot F$ terms involving at least one NP form factor
are positive. As for $G_\square^Q \cdot G_{\square_1}^{\phi_q}$, its
contribution to the amplitude is suppressed by
$(1/m_Q^2)\cdot (1/m_{\phi_q}^6)$, where the latter
  factor stems from the $G_{\square_1}^{\phi_q}$ dependence $\sim
  1/m_{\phi_q}^2$ multiplied by the coupling factor $(g_{\phi_q}^h)^2 \sim
  1/m_{\phi_q}^4$.


        \begin{figure}[h!]
            \centering
            \includegraphics[width=7cm]{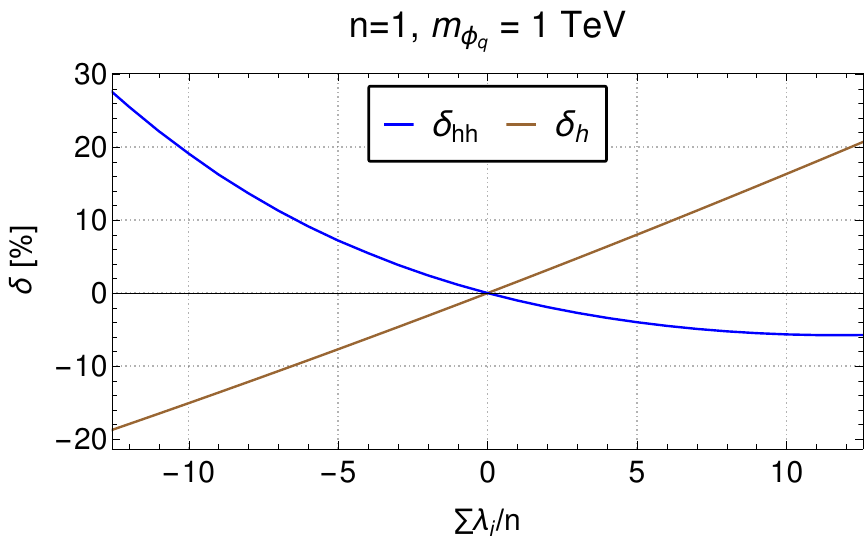}
               \hspace{1cm} 
            \includegraphics[width=7cm]{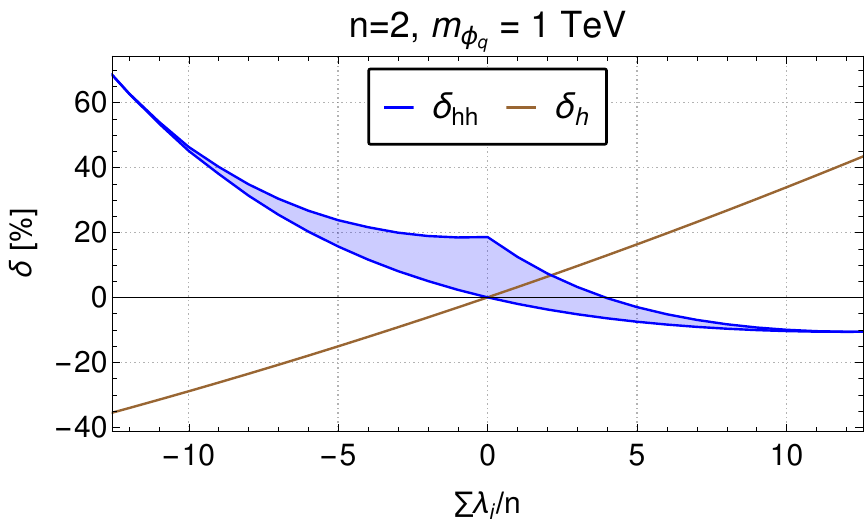}
            \caption{$\delta_{hh}$ (blue) and $\delta_h$ (brown) as a
              function of  the averaged coupling for $n=1$ (left) and
              for $n=2$ (right). For these plots only, the minimum
              coupling used was $-4\pi$ instead of the previous bounded-from-below
   condition.}
            \label{subfig:comp}
        \end{figure}        

In Fig.~\ref{subfig:comp} we present $\delta_{hh}$ (blue) and
$\delta_h$ (brown) as a function of the averaged coloured coupling $\sum 
  \lambda_k/n$ for $n=1$ (left) and $n=2$ (right). The mass of the
  coloured scalars has been chosen 
equal and set to 1 TeV. We note that with the chosen
  input values given above we obtain at $\sqrt{s}=14$~TeV at LO for the SM
  the single Higgs cross section value $\sigma^{h}_{SM}= 15.76$~pb
  calculated with \texttt{HIGLU} including the bottom, charm and top
quark loops, and the double Higgs cross section value
$\sigma^{hh}_{SM}= 16.37$~fb calculated with \texttt{HPAIR} including
the bottom and top quark loops.
The complementarity between the dependence of
  $\delta_h$ and $\delta_{hh}$ w.r.t.~the coupling $\lambda_k$ is
very clear from the figure. We also note that for $n=1$ the $\delta_h$
and $\delta_{hh}$ values are lines while for $n=2$ there is an allowed
region for $\delta_{hh}$  due to the additional dependence on $\sum_k \lambda_k^2$. This leads to the observation that, with a single Higgs measurement very close to the SM value constraining $\sum \lambda_k /n$ to small values, any significant excess of di-Higgs production would provide a strong indication that $n\geq 2$.

        \begin{figure}[h!]
            \centering
            \includegraphics[width=7cm]{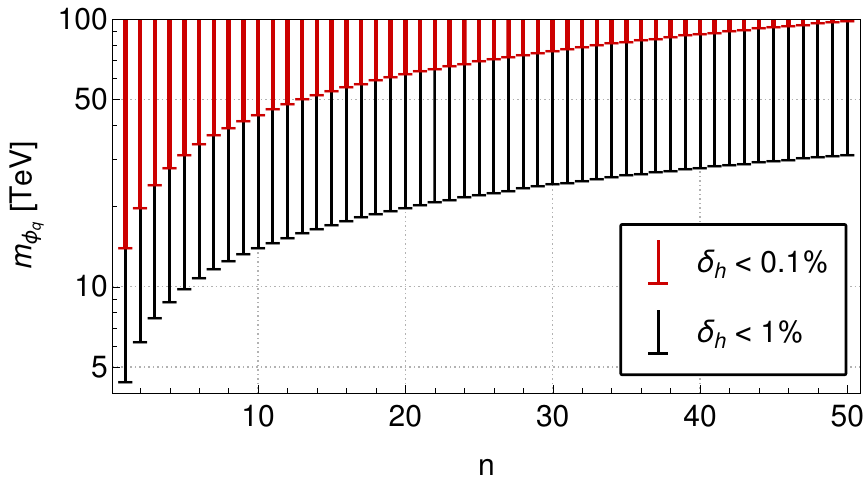}
               \hspace{1cm} 
            \includegraphics[width=7cm]{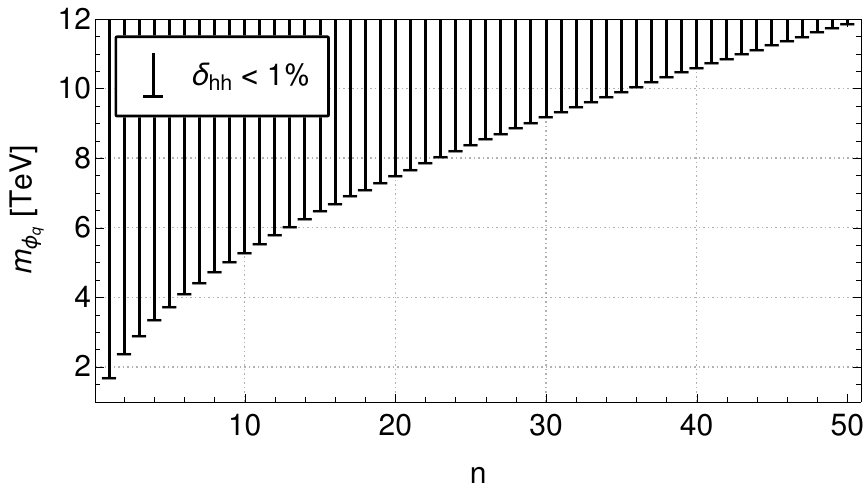}
            \caption{ Regions where $\delta_{h}$ (left)  and $\delta_{hh}$ (right) fall below 1\% as a function of the coloured scalar mass and the number of scalars. For the single Higgs production (left) the threshold of 0.1\% has also been included. For these calculations the previous BFB condition from Eq.~(\ref{eq:bfb}) was used. }
            \label{subfig:comp2}
        \end{figure}        

We finalise this section with a plot (Fig.~\ref{subfig:comp2}) where we show
  the region of the coloured mass versus the number of scalars that
  leads to a maximal deviation of 1\% (black) or 0.1\% (red) in $\delta_h$ (left) and to a maximal deviation of 1\% in $\delta_{hh}$
  (right) of single, respectively, double Higgs production from the
  corresponding SM value,
while varying the couplings within their allowed theoretical bounds. This
gives us a feeling on the region where it will not be
  possible to probe these models even in the long run.
For double Higgs production we indicate the 1\% region only, since the
predictions for the HL-LHC are significantly above this threshold
($\delta_{hh}^{\text{HL-LHC}}=60\%$). For the single Higgs production,
the predictions ($\delta_{h}^{\text{HL-LHC}}=1.6\%$) indicate that a
precision of 1\% could be attainable. Thus we also present the 0.1\%
region in this case as the region that cannot be
  probed by experiment. As single
Higgs production can be constrained more stringently (possibly up to
0.1\%) than Higgs pair 
production this means that larger coloured masses can be probed in
single than in di-Higgs production. Independently of the experimental
precisions, the plots show, that single Higgs production is more
sensitive to coloured scalars than di-Higgs production. For a value of
$n=10$ e.g.~a value of $\delta_h=1$\% probes masses of about 12~TeV, whereas
$\delta_{hh}=1$\% probes masses of 5 TeV only.
	Finally, we have checked that the lower border of the regions, where $\delta = 1\%$ or $0.1 \%$, follows the relationship $n\propto m_{\phi_q}^2$ very closely. The large masses required for these low values of $\delta$ ensure that the terms proportional to $n/m_{\phi_q}^2$ are dominant and hence why this behaviour is observed.

\section{Conclusions}
\label{sec:5}

We have calculated the relative changes $\delta_h$ of SM single and
  $\delta_{hh}$ of SM double Higgs
  production when including new heavy coloured scalars. Our
  calculations are based on the LO cross sections at the LHC using the
  Fortran codes 
  \texttt{HIGLU} and \texttt{HPAIR} where we included our new physics contributions. We have found that for an arbitrary number of scalars and
taking their masses to be equal $\delta_h$ can be written as a
function of only two variables, given by the sum of the couplings of
the coloured particles to the Higgs boson, $\sum_i \lambda_i$, and
their masses $m_i$.  As for the double Higgs case, the $\delta_{hh}$ dependence extends now to three variables, the extra
variable being $\sum_i \lambda_i^2$. We devised a way to find the limits on this new variable in terms of  $\sum_i \lambda_i$, again for equal masses. 
We have discussed the limits on these variables for single Higgs production, where the results already constrain some of the parameter space. For di-Higgs production the bounds
are still very loose and we have to wait until the end of Run3 to hopefully get some bounds on the couplings.

We have shown that if we relax the condition of equal masses what can be said is that the range of allowed values for $\delta_{hh}$ would be smaller than the range obtained by taking all masses equal to the smallest mass of the $n$ coloured scalars. We have also shown that taking equal couplings and performing a scan between their minimum and maximum values is sufficient to obtain the complete range for $\delta_{hh}$.

Another important point to note is the complementarity between single
and di-Higgs processes. Once the value of the coupling is fixed, the
relative deviations from the SM move in different directions. That is, if $\delta_h$
increases $\delta_{hh}$ decreases with the coupling and
vice-versa. The extra freedom of $\delta_{hh}$ also provides another
venue for  determining the number of scalars from observations. An
excess of single or di-Higgs production could indicate the existence of
$n\geq 1$ coloured scalars. But an excess of di-Higgs production paired with a single Higgs measurement close to zero would point to $n>1$.

One final and very important point to note is that in direct searches for DM at the LHC we do not have access to the number of DM fields because we only look for missing energy associated with some SM particle. On the contrary, in our approach the number of fields is a variable that influences the results.

\clearpage

\appendix

\section{HPAIR Extension to Coloured Scalars}
In the following we present our implementation of the contributions
from the coloured scalars to Higgs pair production at leading order in
the code \texttt{HPAIR}. It has been made available at \cite{Gabriel2023BDMHPAIR}.
All changes of the original source code are contained entirely within \lstinline|hpair.f|.
The code is compiled by using \lstinline|make| and then running the executable \lstinline|run| which will assume the input and output files \lstinline|hpair.in| and \lstinline|hpair.out| respectively. For the compilation process the LHAPDF libraries required by \texttt{HPAIR} must be supplied and their installation path indicated with the variable \lstinline|LIBS| in the \lstinline|makefile|.

The BDM input options are contained within the original input file of \texttt{HPAIR}, \lstinline|hpair.in |.
The following lines delimit where these new options are found:
\begin{lstlisting}[language=Fortran, firstnumber=156, escapeinside=||]
!--------------------------------------------------
!BDM OPTIONS:
!=======================|\Suppressnumber|
(...)|\Reactivatenumber{190}|
!---------------------------------BDM OPTIONS END---
\end{lstlisting}
By setting the variable \lstinline|ibdm| to 1, the coloured scalar form factors are added to the SM amplitude. The type of model, characterised by the number of scalars $n$, is selected with the variable \lstinline|ibdmtype|. These are found in the following block:
\begin{lstlisting}[language=Fortran, firstnumber=159]
!IF ibdm=1 THEN THE NEW SCALARS DIAGRAMS WILL BE ADDED TO F1
ibdm     = 1
!SELECT BDM MODEL: 1 - Model n=1, 2 - Model n=2, 3 - Model n >= 2 with equal couplings and equal masses
ibdmtype = 1
\end{lstlisting}
When \lstinline|ibdmtype| is set to either 1 or 2, this corresponds to a model with 1 or 2 scalars, respectively. For each model type the masses in GeV, \lstinline|mphiq|, and couplings to the Higgs, \lstinline|lambHPQ|, must be supplied. These input parameters are set in the following blocks:
\begin{lstlisting}[language=Fortran, firstnumber=164]
!MODEL (n=1) PARAMETERS:
mphiq1   = 2000.D0
lambHPQ1 = 1.0D0

!MODEL (n=2) PARAMETERS:
mphiq1   = 1000.D0
mphiq2   = 1000.D0
lambHPQ1 = 1.000D0
lambHPQ2 = 1.000D0
\end{lstlisting}
On the other hand, setting \lstinline|ibdmtype| to 3, allows for a model with a generic number of scalars $n$. This comes with the limitation that all the scalars must have the same mass, \lstinline|mphiq1|, and coupling, \lstinline|lambHPQ1|. These 3 input parameters can be set in the following block:
\begin{lstlisting}[language=Fortran, firstnumber=174]
!MODEL (n>=2) PARAMETERS:
mphiq1   = 1200.D0
lambHPQ1 = 12.D0
Nscalars = 2
\end{lstlisting}

The final block of parameters relates to the individual form factor selection and it applies to all the three options for \lstinline|ibdmtype|. The relevant lines are:
\begin{lstlisting}[language=Fortran, firstnumber=179]
!DIAGRAMS/FORM FACTORS SELECTION:
!IF FULL=1 THEN FULL FORMULA IS USED INSTEAD
full     = 1
triang   = 1
boxTri   = 1
boxQuad  = 0
\end{lstlisting}
The variable \lstinline|full| is used to select whether all the NP form factors, as presented in Eqs.~(\ref{eq:pairProd:Amplitude:Triangle}) and (\ref{eq:pairProd:Amplitude:Square}), are automatically included in the amplitude (\lstinline|full=1|) or not (\lstinline|full=0|). In the latter case, the following three variables \lstinline|triang|, \lstinline|boxTri| and \lstinline|boxQuad| are used to determine which form factors are to be included in the calculations:
\begin{itemize}
	\item \lstinline|triang=1|: Will include the triangle form factor $g_{\phi_q}^h F_\triangle^{\phi_q}$ (Eq.~(\ref{eq:SingleProd:FormFactorNP})) originating from the triangle diagrams (Fig.~\ref{subfig:pairDiagrams:triangle:BDM1}-\ref{subfig:pairDiagrams:triangle:BDM2})
	\item \lstinline|boxTri=1|: Will include the box form factors
          $(g_{\phi_q}^h)^2 G_{\square_1}^{\phi_q}$ and
          $(g_{\phi_q}^h)^2  F_{\square_1}^{\phi_q}$
          (Eqs.~(\ref{eq:pairProd:Amplitude:Square:G:NP}--\ref{eq:pairProd:Amplitude:Square:F1:NP}))
          originating from the box diagrams with the triple couplings
          between one Higgs and two coloured scalars (Fig.~\ref{subfig:pairDiagrams:box:BDM1}-\ref{subfig:pairDiagrams:box:BDM2})
	\item \lstinline|boxQuad=1|: Will include the box form factor
          $(g_{\phi_q}^{hh}) F_{\square_2}^{\phi_q}$
          (Eq.~(\ref{eq:SingleProd:FormFactorNP}), as $F_{\square_2}^{\phi_q}=F_\triangle ^{\phi_q}$) originating from
          the box diagrams with the quartic couplings between two Higgs and two coloured scalars (Fig.~\ref{subfig:pairDiagrams:box:BDM3}-\ref{subfig:pairDiagrams:box:BDM4})
\end{itemize}

Some care in the formatting must be taken when changing the values of the parameters. The code uses the number of the lines to identify the input parameters and thus they must be preserved. The names of the variables indicated in the input file have no impact. However, the number of characters before the equal sign must always be nine in total:
\begin{lstlisting}[language=Fortran, firstnumber=179]
|$\underbrace{ \text{ibdm\textvisiblespace\textvisiblespace\textvisiblespace\textvisiblespace\textvisiblespace} }_{\text{9 charaters}}$|= 1
\end{lstlisting}
These considerations are important to keep in mind when using a script to automatically change the input values.

The output file suffers no changes from the original \texttt{HPAIR} template, \lstinline|hpair.out|. The NP contributions only affect the LO cross section which can be extracted from the following line:
\begin{lstlisting}[language=Fortran, numbers=none]
SIGMA_BORN = (   2.0478260337291408E-002  +-    2.4261364431116920E-005 ) PB
\end{lstlisting}

%
%
%

\section{Double Higgs Production \label{annex:pairProd:SMforms} }
    For completeness, we repeat here the SM box form factors appearing in Eq.~(\ref{eq:pairProd:Amplitude:Square}), which can
    also be found e.g.~in \cite{Plehn:1996wb}. They are given by:
    \begin{multline}
        F_\square^{Q} = \frac{2 m_Q^2} {s} 
                    \Bigg( 
                    2
                    + ( 4 m_Q^2 ) C_{ab}^{m_Q^2}
                    + \frac{2}{s} ( t - m_h^2 ) ( m_h^2 - 4 m_Q^2 ) C_{ac}^{m_Q^2}\\
                    + \frac{2}{s} ( u - m_h^2 ) ( m_h^2 - 4 m_Q^2 ) C_{bc}^{m_Q^2} 
                    - m_Q^2 ( s + 2 m_h^2 - 8 m_Q^2 ) ( D_{abc}^{m_Q^2} + D_{bac}^{m_Q^2} )\\
                    + ( m_Q^2 ( s + 2 m_h^2 -8 m_Q^2 ) -\frac{1}{s}( m_h^2-4 m_Q^2 ) ( t u - m_h^4 ) ) D_{acb}^{m_Q^2}
                    \Bigg) \ ,
				\label{annex:pairProd:eq:SMforms:Fsquare}
    \end{multline}
    \begin{multline}
        G_\square^{Q} = \frac{2 m_Q^2} {s}
                        \left( \frac{1}{t u - m_h^4 } \right)
                    \Bigg( 
                        \frac{s}{2}( t^2+u^2 + 2 m_h^4 -8 m_Q^2 (t+u) ) C_{ab}^{m_Q^2} \\
                        + ( t^2 + m_h^4 -8 t m_Q^2 ) ( t - m_h^2) C_{ac}^{m_Q^2}
                        + ( u^2 + m_h^4 -8 u m_Q^2 ) ( u - m_h^2) C_{bc}^{m_Q^2}\\
                        - \frac{1}{2} ( t^2+u^2 - 2 m_h^4 ) ( t+u - 8 m_Q^2) C_{cd}^{m_Q^2} \\
                        - ( s t ( t^2/2 + m_h^4/2 -4 t m_Q^2 ) + m_Q^2 ( t+u-8 m_Q^2 ) (t u - m_h^4 )) D_{bac}^{m_Q^2} \\
                        - ( s u ( u^2/2 + m_h^4/2 -4 u m_Q^2 ) + m_Q^2 ( t+u-8 m_Q^2 ) (t u - m_h^4 )) D_{abc}^{m_Q^2} \\
                        -m_Q^2 (t+u-8 m_Q^2) ( t u - m_h^4 ) D_{acb}^{m_Q^2}
                    \Bigg) \ ,
                \label{annex:pairProd:eq:SMforms:Gsquare}
    \end{multline}
    where the Mandelstam variables $s,t,u$ are defined as:
    \begin{equation}
    	s = (p_a + p_b)^2 \ , \quad
    	t = (p_a - p_c)^2 \ , \quad
    	u = (p_b - p_c)^2 \ .
    \end{equation}
	They also involve the following scalar integrals:
	\begin{gather}
		C_{ab}= \int \frac{d^4 q}{i \pi^2} 
					\frac{1}{ 
						\left(
							q^2 - m_X^2
						\right)
						\left(
							(q+p_a)^2 - m_X^2
						\right)
						\left(
							(q+p_a+p_b)^2 - m_X^2
						\right)						
							 } \ , \\
		D_{abc}= \int \frac{d^4 q}{i \pi^2} 
					\frac{1}{ 
						\left(
							q^2 - m_X^2
						\right)
						\left(
							(q+p_a)^2 - m_X^2
						\right)
						\left(
							(q+p_a+p_b)^2 - m_X^2
						\right)	
						\left(
							(q+p_a+p_b+p_c)^2 - m_X^2
						\right)						
							 } \ ,
			\label{annex:pairProd:eq:Coeficients}
	\end{gather}
	where $X$ stands for the quark or coloured scalar as appropriate. The exact formula for $C_{ab}$ has been determined and is given by:
	\begin{equation}
		C_{ab} = -\frac{2}{s} f(\tau) \ ,
	\end{equation}
	where $f_{\tau}$ is given in Eq.~(\ref{eq:ftau}) and $\tau = \frac{4 m_X^2}{s}$.

\vspace*{1cm}
\subsubsection*{Acknowledgements}

PG is supported by the Portuguese Foundation for
Science and Technology (FCT) with a PhD Grant No. 2022.11377.BD.
PG, DN and RS are partially supported by FCT under Contracts no. UIDB/00618/2020,
UIDP/00618/2020, PTDC/FIS-PAR/31000/2017 and
CERN/FIS-PAR/0014 /2019. The work of MM is supported by the DFG Collaborative Research Center TRR257 “Particle Physics Phenomenology after the Higgs Discovery”.


\vspace*{1cm}
\bibliographystyle{JHEP}
\bibliography{model3.bib}

\providecommand{\href}[2]{#2}\begingroup\raggedright\begin{thebibliography}{10}

\bibitem{Spira_1995}
M.~Spira, A.~Djouadi, D.~Graudenz and R.~Zerwas, \emph{{Higgs} boson production
  at the {LHC}},
  \href{http://dx.doi.org/10.1016/0550-3213(95)00379-7}{\emph{Nuclear Physics
  B} \textbf{ 453} (oct, 1995) 17--82}.

\bibitem{Harlander_2005}
R.~V. Harlander and P.~Kant, \emph{{Higgs} production and decay: analytic
  results at next-to-leading order {QCD}},
  \href{http://dx.doi.org/10.1088/1126-6708/2005/12/015}{\emph{Journal of High
  Energy Physics} \textbf{ 2005} (dec, 2005) 015--015}.

\bibitem{Anastasiou_2007}
C.~Anastasiou, S.~Beerli, S.~Bucherer, A.~Daleo and Z.~Kunszt, \emph{Two-loop
  amplitudes and master integrals for the production of a {Higgs} boson via a
  massive quark and a scalar-quark loop},
  \href{http://dx.doi.org/10.1088/1126-6708/2007/01/082}{\emph{Journal of High
  Energy Physics} \textbf{ 2007} (jan, 2007) 082--082}.

\bibitem{Aglietti_2007}
U.~Aglietti, R.~Bonciani, G.~Degrassi and A.~Vicini, \emph{Analytic results for
  virtual {QCD} corrections to {Higgs} production and decay},
  \href{http://dx.doi.org/10.1088/1126-6708/2007/01/021}{\emph{Journal of High
  Energy Physics} \textbf{ 2007} (jan, 2007) 021--021}.

\bibitem{DJOUADI1991440}
A.~Djouadi, M.~Spira and P.~Zerwas, \emph{Production of {Higgs} bosons in
  proton colliders. {QCD} corrections},
  \href{http://dx.doi.org/https://doi.org/10.1016/0370-2693(91)90375-Z}{\emph{Physics
  Letters B} \textbf{ 264} (1991) 440--446}.

\bibitem{DAWSON1991283}
S.~Dawson, \emph{Radiative corrections to {Higgs} boson production},
  \href{http://dx.doi.org/https://doi.org/10.1016/0550-3213(91)90061-2}{\emph{Nuclear
  Physics B} \textbf{ 359} (1991) 283--300}.

\bibitem{PhysRevLett_70_1372}
D.~Graudenz, M.~Spira and P.~M. Zerwas, \emph{{QCD} corrections to
  {Higgs}-boson production at proton-proton colliders},
  \href{http://dx.doi.org/10.1103/PhysRevLett.70.1372}{\emph{Phys. Rev. Lett.}
  \textbf{ 70} (Mar, 1993) 1372--1375}.

\bibitem{Anastasiou_2009}
C.~Anastasiou, S.~Bucherer and Z.~Kunszt, \emph{{HPro}: A {NLO} monte-carlo for
  {Higgs} production via gluon fusion with finite heavy quark masses},
  \href{http://dx.doi.org/10.1088/1126-6708/2009/10/068}{\emph{Journal of High
  Energy Physics} \textbf{ 2009} (oct, 2009) 068--068}.

\bibitem{Catani_2001}
S.~Catani, D.~de~Florian and M.~Grazzini, \emph{{Higgs} production in hadron
  collisions: soft and virtual {QCD} corrections at {NNLO}},
  \href{http://dx.doi.org/10.1088/1126-6708/2001/05/025}{\emph{Journal of High
  Energy Physics} \textbf{ 2001} (may, 2001) 025--025}.

\bibitem{PhysRevD_64_013015}
R.~V. Harlander and W.~B. Kilgore, \emph{Soft and virtual corrections to
  $p\vec{p}{H}+{X}$ at next-to-next-to-leading order},
  \href{http://dx.doi.org/10.1103/PhysRevD.64.013015}{\emph{Phys. Rev. D}
  \textbf{ 64} (Jun, 2001) 013015}.

\bibitem{PhysRevLett_88_201801}
R.~V. Harlander and W.~B. Kilgore, \emph{Next-to-next-to-leading order {Higgs}
  production at hadron colliders},
  \href{http://dx.doi.org/10.1103/PhysRevLett.88.201801}{\emph{Phys. Rev.
  Lett.} \textbf{ 88} (May, 2002) 201801}.

\bibitem{Anastasiou_2002}
C.~Anastasiou and K.~Melnikov, \emph{{Higgs} boson production at hadron
  colliders in {NNLO} {QCD}},
  \href{http://dx.doi.org/10.1016/s0550-3213(02)00837-4}{\emph{Nuclear Physics
  B} \textbf{ 646} (dec, 2002) 220--256}.

\bibitem{RAVINDRAN2003325}
V.~Ravindran, J.~Smith and W.~{van Neerven}, \emph{{NNLO} corrections to the
  total cross section for {Higgs} boson production in hadron-hadron
  collisions},
  \href{http://dx.doi.org/https://doi.org/10.1016/S0550-3213(03)00457-7}{\emph{Nuclear
  Physics B} \textbf{ 665} (2003) 325--366}.

\bibitem{MARZANI2008127}
S.~Marzani, R.~D. Ball, V.~{Del Duca}, S.~Forte and A.~Vicini, \emph{{Higgs}
  production via gluon-gluon fusion with finite top mass beyond next-to-leading
  order},
  \href{http://dx.doi.org/https://doi.org/10.1016/j.nuclphysb.2008.03.016}{\emph{Nuclear
  Physics B} \textbf{ 800} (2008) 127--145}.

\bibitem{Harlander_2010}
R.~V. Harlander, H.~Mantler, S.~Marzani and K.~J. Ozeren, \emph{Higgs
  production in gluon fusion at next-to-next-to-leading order {QCD} for finite
  top mass}, \href{http://dx.doi.org/10.1140/epjc/s10052-010-1258-x}{\emph{The
  European Physical Journal C} \textbf{ 66} (feb, 2010) 359--372}.

\bibitem{Pak_2010}
A.~Pak, M.~Rogal and M.~Steinhauser, \emph{Finite top quark mass effects in
  {NNLO} higgs boson production at {LHC}},
  \href{http://dx.doi.org/10.1007/jhep02(2010)025}{\emph{Journal of High Energy
  Physics} \textbf{ 2010} (feb, 2010) }.

\bibitem{Gehrmann_2012}
T.~Gehrmann, M.~Jaquier, E.~W.~N. Glover and A.~Koukoutsakis, \emph{Two-loop
  {QCD} corrections to the helicity amplitudes for
  {H}{\hspace{0.167em}}$\rightarrow${\hspace{0.167em}}3 partons},
  \href{http://dx.doi.org/10.1007/jhep02(2012)056}{\emph{Journal of High Energy
  Physics} \textbf{ 2012} (feb, 2012) }.

\bibitem{Anastasiou_2013_Soft}
C.~Anastasiou, C.~Duhr, F.~Dulat and B.~Mistlberger, \emph{Soft triple-real
  radiation for {Higgs} production at {N3LO}},
  \href{http://dx.doi.org/10.1007/jhep07(2013)003}{\emph{Journal of High Energy
  Physics} \textbf{ 2013} (jul, 2013) }.

\bibitem{Anastasiou_2013_Real_virtual}
C.~Anastasiou, C.~Duhr, F.~Dulat, F.~Herzog and B.~Mistlberger,
  \emph{Real-virtual contributions to the inclusive {Higgs} cross-section at
  {N3LO}}, \href{http://dx.doi.org/10.1007/jhep12(2013)088}{\emph{Journal of
  High Energy Physics} \textbf{ 2013} (dec, 2013) }.

\bibitem{PhysRevD_89_073008}
W.~B. Kilgore, \emph{One-loop single-real-emission contributions to
  $pp\ensuremath{\rightarrow}{H}+x$ at next-to-next-to-next-to-leading order},
  \href{http://dx.doi.org/10.1103/PhysRevD.89.073008}{\emph{Phys. Rev. D}
  \textbf{ 89} (Apr, 2014) 073008}.

\bibitem{PhysRevD_90_053006}
Y.~Li, A.~von Manteuffel, R.~M. Schabinger and H.~X. Zhu,
  \emph{${N}^{3}\mathrm{LO}$ {Higgs} boson and drell-yan production at
  threshold: The one-loop two-emission contribution},
  \href{http://dx.doi.org/10.1103/PhysRevD.90.053006}{\emph{Phys. Rev. D}
  \textbf{ 90} (Sep, 2014) 053006}.

\bibitem{anastasiou2014higgs}
C.~Anastasiou, C.~Duhr, F.~Dulat, E.~Furlan, T.~Gehrmann, F.~Herzog et~al.,
  \emph{{Higgs} boson gluon-fusion production beyond threshold in {N3LO QCD}},
  \href{http://dx.doi.org/10.1007/JHEP03(2015)091}{\emph{J. High Energ. Phys.}
  \textbf{ 2015} (March, 2015) 91}.

\bibitem{PhysRevLett_114_212001}
C.~Anastasiou, C.~Duhr, F.~Dulat, F.~Herzog and B.~Mistlberger, \emph{{Higgs}
  boson gluon-fusion production in {QCD} at three loops},
  \href{http://dx.doi.org/10.1103/PhysRevLett.114.212001}{\emph{Phys. Rev.
  Lett.} \textbf{ 114} (May, 2015) 212001}.

\bibitem{anastasiou2016high}
C.~Anastasiou, C.~Duhr, F.~Dulat, E.~Furlan, T.~Gehrmann, F.~Herzog et~al.,
  \emph{High precision determination of the gluon fusion {Higgs} boson
  cross-section at the {LHC}},
  \href{http://dx.doi.org/10.1007/JHEP05(2016)058}{\emph{J. High Energ. Phys.}
  \textbf{ 2016} (May, 2016) 58}.

\bibitem{Mistlberger_2018}
B.~Mistlberger, \emph{Higgs boson production at hadron colliders at {N3LO} in
  {QCD}}, \href{http://dx.doi.org/10.1007/jhep05(2018)028}{\emph{Journal of
  High Energy Physics} \textbf{ 2018} (may, 2018) }.

\bibitem{Duhr_2022}
C.~Duhr, B.~Mistlberger and G.~Vita, \emph{Soft integrals and soft anomalous
  dimensions at n3lo and beyond},
  \href{http://dx.doi.org/10.1007/jhep09(2022)155}{\emph{Journal of High Energy
  Physics} \textbf{ 2022} (sep, 2022) }.

\bibitem{Baglio_2022}
J.~Baglio, C.~Duhr, B.~Mistlberger and R.~Szafron, \emph{Inclusive production
  cross sections at {N3LO}},
  \href{http://dx.doi.org/10.1007/jhep12(2022)066}{\emph{Journal of High Energy
  Physics} \textbf{ 2022} (dec, 2022) }.

\bibitem{Frellesvig_2020}
H.~Frellesvig, M.~Hidding, L.~Maestri, F.~Moriello and G.~Salvatori, \emph{The
  complete set of two-loop master integrals for higgs + jet production in
  {QCD}}, \href{http://dx.doi.org/10.1007/jhep06(2020)093}{\emph{Journal of
  High Energy Physics} \textbf{ 2020} (jun, 2020) }.

\bibitem{Prausa_2021}
M.~Prausa and J.~Usovitsch, \emph{The analytic leading color contribution to
  the higgs-gluon form factor in {QCD} at {NNLO}},
  \href{http://dx.doi.org/10.1007/jhep03(2021)127}{\emph{Journal of High Energy
  Physics} \textbf{ 2021} (mar, 2021) }.

\bibitem{DEFLORIAN2012117}
D.~{de Florian} and M.~Grazzini, \emph{{Higgs} production at the {LHC}: Updated
  cross sections at $\sqrt{s}=8$ {TeV}},
  \href{http://dx.doi.org/https://doi.org/10.1016/j.physletb.2012.10.019}{\emph{Physics
  Letters B} \textbf{ 718} (2012) 117--120}.

\bibitem{Bonvini_2014}
M.~Bonvini and S.~Marzani, \emph{Resummed {Higgs} cross section at {N3LL}},
  \href{http://dx.doi.org/10.1007/jhep09(2014)007}{\emph{Journal of High Energy
  Physics} \textbf{ 2014} (sep, 2014) }.

\bibitem{PhysRevD_93_014022}
T.~Schmidt and M.~Spira, \emph{{Higgs} boson production via gluon fusion:
  Soft-gluon resummation including mass effects},
  \href{http://dx.doi.org/10.1103/PhysRevD.93.014022}{\emph{Phys. Rev. D}
  \textbf{ 93} (Jan, 2016) 014022}.

\bibitem{Bonvini_2016}
M.~Bonvini, S.~Marzani, C.~Muselli and L.~Rottoli, \emph{On the {Higgs} cross
  section at {N3LO+N3LL} and its uncertainty},
  \href{http://dx.doi.org/10.1007/jhep08(2016)105}{\emph{Journal of High Energy
  Physics} \textbf{ 2016} (aug, 2016) }.

\bibitem{KRAMER1998523}
M.~Krämer, E.~Laenen and M.~Spira, \emph{Soft gluon radiation in {Higgs} boson
  production at the {LHC}},
  \href{http://dx.doi.org/https://doi.org/10.1016/S0550-3213(97)00679-2}{\emph{Nuclear
  Physics B} \textbf{ 511} (1998) 523--549}.

\bibitem{Stefano_Catani_2003}
S.~Catani, D.~de~Florian, M.~Grazzini and P.~Nason, \emph{Soft-gluon
  resummation for {Higgs} boson production at hadron colliders},
  \href{http://dx.doi.org/10.1088/1126-6708/2003/07/028}{\emph{Journal of High
  Energy Physics} \textbf{ 2003} (jul, 2003) 028}.

\bibitem{MOCH200548}
S.~Moch and A.~Vogt, \emph{Higher-order soft corrections to lepton pair and
  {Higgs} boson production},
  \href{http://dx.doi.org/https://doi.org/10.1016/j.physletb.2005.09.061}{\emph{Physics
  Letters B} \textbf{ 631} (2005) 48--57}.

\bibitem{RAVINDRAN200658}
V.~Ravindran, \emph{On sudakov and soft resummations in {QCD}},
  \href{http://dx.doi.org/https://doi.org/10.1016/j.nuclphysb.2006.04.008}{\emph{Nuclear
  Physics B} \textbf{ 746} (2006) 58--76}.

\bibitem{RAVINDRAN2006173}
V.~Ravindran, \emph{Higher-order threshold effects to inclusive processes in
  {QCD}},
  \href{http://dx.doi.org/https://doi.org/10.1016/j.nuclphysb.2006.06.025}{\emph{Nuclear
  Physics B} \textbf{ 752} (2006) 173--196}.

\bibitem{PhysRevD_73_077501}
A.~Idilbi, X.~Ji, J.-P. Ma and F.~Yuan, \emph{Threshold resummation for {Higgs}
  production in effective field theory},
  \href{http://dx.doi.org/10.1103/PhysRevD.73.077501}{\emph{Phys. Rev. D}
  \textbf{ 73} (Apr, 2006) 077501}.

\bibitem{ahrens2009renormalization}
V.~Ahrens, T.~Becher, M.~Neubert and L.~L. Yang, \emph{Renormalization-group
  improved prediction for {Higgs} production at hadron colliders},
  \href{http://dx.doi.org/10.1140/epjc/s10052-009-1030-2}{\emph{The European
  Physical Journal C} \textbf{ 62} (2009) 333--353}.

\bibitem{DEFLORIAN2009291}
D.~{de Florian} and M.~Grazzini, \emph{{Higgs} production through gluon fusion:
  Updated cross sections at the tevatron and the {LHC}},
  \href{http://dx.doi.org/https://doi.org/10.1016/j.physletb.2009.03.033}{\emph{Physics
  Letters B} \textbf{ 674} (2009) 291--294}.

\bibitem{de_Florian_2014}
D.~de~Florian, J.~Mazzitelli, S.~Moch and A.~Vogt, \emph{Approximate {N3LO}
  {Higgs}-boson production cross section using physical-kernel constraints},
  \href{http://dx.doi.org/10.1007/jhep10(2014)176}{\emph{Journal of High Energy
  Physics} \textbf{ 2014} (oct, 2014) }.

\bibitem{PhysRevD_91_051301}
M.~Bonvini and L.~Rottoli, \emph{Three loop soft function for {N3LL'} gluon
  fusion {Higgs} boson production in soft-collinear effective theory},
  \href{http://dx.doi.org/10.1103/PhysRevD.91.051301}{\emph{Phys. Rev. D}
  \textbf{ 91} (Mar, 2015) 051301}.

\bibitem{CATANI201475}
S.~Catani, L.~Cieri, D.~{de Florian}, G.~Ferrera and M.~Grazzini,
  \emph{Threshold resummation at {N3LL} accuracy and soft-virtual cross
  sections at {N3LO}},
  \href{http://dx.doi.org/https://doi.org/10.1016/j.nuclphysb.2014.09.012}{\emph{Nuclear
  Physics B} \textbf{ 888} (2014) 75--91}.

\bibitem{Bonvini_2018}
M.~Bonvini and S.~Marzani, \emph{Double resummation for higgs production},
  \href{http://dx.doi.org/10.1103/physrevlett.120.202003}{\emph{Physical Review
  Letters} \textbf{ 120} (may, 2018) }.

\bibitem{PhysRevLett_78_594}
K.~G. Chetyrkin, B.~A. Kniehl and M.~Steinhauser, \emph{Virtual top-quark
  effects on the
  $\mathit{H}\ensuremath{\rightarrow}\mathit{b}\overline{\mathit{b}}$ decay at
  next-to-leading order in {QCD}},
  \href{http://dx.doi.org/10.1103/PhysRevLett.78.594}{\emph{Phys. Rev. Lett.}
  \textbf{ 78} (Jan, 1997) 594--597}.

\bibitem{PhysRevLett_73_2528}
A.~Djouadi and P.~Gambino, \emph{Leading electroweak correction to {Higgs}
  boson production at proton colliders},
  \href{http://dx.doi.org/10.1103/PhysRevLett.73.2528}{\emph{Phys. Rev. Lett.}
  \textbf{ 73} (Nov, 1994) 2528--2531}.

\bibitem{CHETYRKIN199719}
K.~Chetyrkin, B.~Kniehl and M.~Steinhauser, \emph{Three-loop ${O}(\alpha_s^2
  {G}_{F} {M}_t^2)$ corrections to hadronic {Higgs} decays},
  \href{http://dx.doi.org/https://doi.org/10.1016/S0550-3213(97)00051-5}{\emph{Nuclear
  Physics B} \textbf{ 490} (1997) 19--39}.

\bibitem{AGLIETTI2004432}
U.~Aglietti, R.~Bonciani, G.~Degrassi and A.~Vicini, \emph{Two-loop light
  fermion contribution to {Higgs} production and decays},
  \href{http://dx.doi.org/https://doi.org/10.1016/j.physletb.2004.06.063}{\emph{Physics
  Letters B} \textbf{ 595} (2004) 432--441}.

\bibitem{aglietti2006twoloop}
U.~Aglietti, R.~Bonciani, G.~Degrassi and A.~Vicini, \emph{{Two-loop
  electroweak corrections to Higgs production in proton-proton collisions}},
  in \emph{{TeV4LHC Workshop: 2nd Meeting}}, 10, 2006.
\newblock \href{https://arxiv.org/abs/hep-ph/0610033}{{\texttt
  hep-ph/0610033}}.

\bibitem{DEGRASSI2004255}
G.~Degrassi and F.~Maltoni, \emph{Two-loop electroweak corrections to {Higgs}
  production at hadron colliders},
  \href{http://dx.doi.org/https://doi.org/10.1016/j.physletb.2004.09.008}{\emph{Physics
  Letters B} \textbf{ 600} (2004) 255--260}.

\bibitem{ACTIS200812}
S.~Actis, G.~Passarino, C.~Sturm and S.~Uccirati, \emph{{NLO} electroweak
  corrections to {Higgs} boson production at hadron colliders},
  \href{http://dx.doi.org/https://doi.org/10.1016/j.physletb.2008.10.018}{\emph{Physics
  Letters B} \textbf{ 670} (2008) 12--17}.

\bibitem{ACTIS2009182}
S.~Actis, G.~Passarino, C.~Sturm and S.~Uccirati, \emph{{NNLO} computational
  techniques: The cases ${H}\rightarrow\gamma\gamma$ and ${H}\rightarrow g g$},
  \href{http://dx.doi.org/https://doi.org/10.1016/j.nuclphysb.2008.11.024}{\emph{Nuclear
  Physics B} \textbf{ 811} (2009) 182--273}.

\bibitem{Bonetti_2017}
M.~Bonetti, K.~Melnikov and L.~Tancredi, \emph{Two-loop electroweak corrections
  to higgs{\textendash}gluon couplings to higher orders in the dimensional
  regularization parameter},
  \href{http://dx.doi.org/10.1016/j.nuclphysb.2017.01.020}{\emph{Nuclear
  Physics B} \textbf{ 916} (mar, 2017) 709--726}.

\bibitem{Dawson_1998}
S.~Dawson, S.~Dittmaier and M.~Spira, \emph{Neutral {Higgs}-boson pair
  production at hadron colliders: {QCD} corrections},
  \href{http://dx.doi.org/10.1103/physrevd.58.115012}{\emph{Physical Review D}
  \textbf{ 58} (nov, 1998) }.

\bibitem{Borowka_2016_Higgs}
S.~Borowka, N.~Greiner, G.~Heinrich, S.~Jones, M.~Kerner, J.~Schlenk et~al.,
  \emph{{Higgs} boson pair production in gluon fusion at next-to-leading order
  with full top-quark mass dependence},
  \href{http://dx.doi.org/10.1103/physrevlett.117.012001}{\emph{Physical Review
  Letters} \textbf{ 117} (jun, 2016) }.

\bibitem{Baglio_2019}
J.~Baglio, F.~Campanario, S.~Glaus, M.~Mühlleitner, M.~Spira and J.~Streicher,
  \emph{Gluon fusion into {Higgs} pairs at {NLO} {QCD} and the top mass
  scheme}, \href{http://dx.doi.org/10.1140/epjc/s10052-019-6973-3}{\emph{The
  European Physical Journal C} \textbf{ 79} (may, 2019) }.

\bibitem{Baglio_2022wkx}
J.~Baglio, F.~Campanario, S.~Glaus, M.~M. Mühlleitner, J.~Ronca and M.~Spira,
  \emph{Full {NLO QCD} corrections to higgs-pair production in the standard
  model and beyond}, \href{http://dx.doi.org/10.22323/1.380.0393}{\emph{PoS}
  \textbf{ PANIC2021} (2022) 393}.

\bibitem{Borowka_2016_Full}
S.~Borowka, N.~Greiner, G.~Heinrich, S.~Jones, M.~Kerner, J.~Schlenk et~al.,
  \emph{Full top quark mass dependence in {Higgs} boson pair production at
  {NLO}}, \href{http://dx.doi.org/10.1007/jhep10(2016)107}{\emph{Journal of
  High Energy Physics} \textbf{ 2016} (oct, 2016) }.

\bibitem{Maltoni_2014}
F.~Maltoni, E.~Vryonidou and M.~Zaro, \emph{Top-quark mass effects in double
  and triple {Higgs} production in gluon-gluon fusion at {NLO}},
  \href{http://dx.doi.org/10.1007/jhep11(2014)079}{\emph{Journal of High Energy
  Physics} \textbf{ 2014} (nov, 2014) }.

\bibitem{Bonciani_2018}
R.~Bonciani, G.~Degrassi, P.~P. Giardino and R.~Gröber, \emph{Analytical
  method for next-to-leading-order {QCD} corrections to double-higgs
  production},
  \href{http://dx.doi.org/10.1103/physrevlett.121.162003}{\emph{Physical Review
  Letters} \textbf{ 121} (oct, 2018) }.

\bibitem{de_Florian_2013_Higgs}
D.~de~Florian and J.~Mazzitelli, \emph{{Higgs} boson pair production at
  next-to-next-to-leading order in {QCD}},
  \href{http://dx.doi.org/10.1103/physrevlett.111.201801}{\emph{Physical Review
  Letters} \textbf{ 111} (nov, 2013) }.

\bibitem{de_Florian_2013_Two_Loop}
D.~de~Florian and J.~Mazzitelli, \emph{Two-loop virtual corrections to {Higgs}
  pair production},
  \href{http://dx.doi.org/10.1016/j.physletb.2013.06.046}{\emph{Physics Letters
  B} \textbf{ 724} (jul, 2013) 306--309}.

\bibitem{Grigo_2014}
J.~Grigo, K.~Melnikov and M.~Steinhauser, \emph{Virtual corrections to {Higgs}
  boson pair production in the large top quark mass limit},
  \href{http://dx.doi.org/10.1016/j.nuclphysb.2014.09.003}{\emph{Nuclear
  Physics B} \textbf{ 888} (nov, 2014) 17--29}.

\bibitem{Grazzini_2018}
M.~Grazzini, G.~Heinrich, S.~Jones, S.~Kallweit, M.~Kerner, J.~M. Lindert
  et~al., \emph{{Higgs} boson pair production at {NNLO} with top quark mass
  effects}, \href{http://dx.doi.org/10.1007/jhep05(2018)059}{\emph{Journal of
  High Energy Physics} \textbf{ 2018} (may, 2018) }.

\bibitem{Chen_2020_Higgs}
L.-B. Chen, H.~T. Li, H.-S. Shao and J.~Wang, \emph{{Higgs} boson pair
  production via gluon fusion at {N3LO} in {QCD}},
  \href{http://dx.doi.org/10.1016/j.physletb.2020.135292}{\emph{Physics Letters
  B} \textbf{ 803} (apr, 2020) 135292}.

\bibitem{Chen_2020_The_Gluon}
L.-B. Chen, H.~T. Li, H.-S. Shao and J.~Wang, \emph{The gluon-fusion production
  of {Higgs} boson pair: {N3LO} {QCD} corrections and top-quark mass effects},
  \href{http://dx.doi.org/10.1007/jhep03(2020)072}{\emph{Journal of High Energy
  Physics} \textbf{ 2020} (mar, 2020) }.

\bibitem{Baglio_2021}
J.~Baglio, F.~Campanario, S.~Glaus, M.~Mühlleitner, J.~Ronca and M.~Spira,
  \emph{$gg \rightarrow {HH}$ : Combined uncertainties},
  \href{http://dx.doi.org/10.1103/physrevd.103.056002}{\emph{Physical Review D}
  \textbf{ 103} (mar, 2021) }.

\end{thebibliography}\endgroup


\providecommand{\href}[2]{#2}\begingroup\raggedright\begin{thebibliography}{10}

\bibitem{LHCb:2021trn}
{\scshape LHCb} collaboration, R.~Aaij et~al., \emph{{Test of lepton
  universality in beauty-quark decays}},
  \href{http://dx.doi.org/10.1038/s41567-021-01478-8}{\emph{Nature Phys.}
  \textbf{ 18} (2022) 277--282},
  [\href{https://arxiv.org/abs/2103.11769}{{\texttt 2103.11769}}].

\bibitem{LHCb:2019hip}
{\scshape LHCb} collaboration, R.~Aaij et~al., \emph{{Search for
  lepton-universality violation in $B^+\to K^+\ell^+\ell^-$ decays}},
  \href{http://dx.doi.org/10.1103/PhysRevLett.122.191801}{\emph{Phys. Rev.
  Lett.} \textbf{ 122} (2019) 191801},
  [\href{https://arxiv.org/abs/1903.09252}{{\texttt 1903.09252}}].

\bibitem{Muong-2:2023cdq}
{\scshape Muon g-2} collaboration, D.~P. Aguillard et~al., \emph{{Measurement
  of the Positive Muon Anomalous Magnetic Moment to 0.20 ppm}},
  \href{https://arxiv.org/abs/2308.06230}{{\texttt 2308.06230}}.

\bibitem{ParticleDataGroup:2018ovx}
{\scshape Particle Data Group} collaboration, M.~Tanabashi et~al.,
  \emph{{Review of Particle Physics}},
  \href{http://dx.doi.org/10.1103/PhysRevD.98.030001}{\emph{Phys. Rev. D}
  \textbf{ 98} (2018) 030001}.

\bibitem{Gorringe:2015cma}
T.~P. Gorringe and D.~W. Hertzog, \emph{{Precision Muon Physics}},
  \href{http://dx.doi.org/10.1016/j.ppnp.2015.06.001}{\emph{Prog. Part. Nucl.
  Phys.} \textbf{ 84} (2015) 73--123},
  [\href{https://arxiv.org/abs/1506.01465}{{\texttt 1506.01465}}].

\bibitem{Aoyama:2020ynm}
T.~Aoyama et~al., \emph{{The anomalous magnetic moment of the muon in the
  Standard Model}},
  \href{http://dx.doi.org/10.1016/j.physrep.2020.07.006}{\emph{Phys. Rept.}
  \textbf{ 887} (2020) 1--166},
  [\href{https://arxiv.org/abs/2006.04822}{{\texttt 2006.04822}}].

\bibitem{Muong-2:2006rrc}
{\scshape Muon g-2} collaboration, G.~W. Bennett et~al., \emph{{Final Report of
  the Muon E821 Anomalous Magnetic Moment Measurement at BNL}},
  \href{http://dx.doi.org/10.1103/PhysRevD.73.072003}{\emph{Phys. Rev. D}
  \textbf{ 73} (2006) 072003},
  [\href{https://arxiv.org/abs/hep-ex/0602035}{{\texttt hep-ex/0602035}}].

\bibitem{LHCb:2022zom}
{\scshape LHCb} collaboration, \emph{{Measurement of lepton universality
  parameters in $B^+\to K^+\ell^+\ell^-$ and $B^0\to K^{*0}\ell^+\ell^-$
  decays}},  \href{https://arxiv.org/abs/2212.09153}{{\texttt 2212.09153}}.

\bibitem{LHCb:2022qnv}
{\scshape LHCb} collaboration, \emph{{Test of lepton universality in $b
  \rightarrow s \ell^+ \ell^-$ decays}},
  \href{https://arxiv.org/abs/2212.09152}{{\texttt 2212.09152}}.

\bibitem{Huang:2020ris}
D.~Huang, A.~P. Morais and R.~Santos, \emph{{Anomalies in $B$-meson decays and
  the muon $g-2$ from dark loops}},
  \href{http://dx.doi.org/10.1103/PhysRevD.102.075009}{\emph{Phys. Rev. D}
  \textbf{ 102} (2020) 075009},
  [\href{https://arxiv.org/abs/2007.05082}{{\texttt 2007.05082}}].

\bibitem{Capucha:2022kwo}
R.~Capucha, D.~Huang, T.~Lopes and R.~Santos, \emph{{Impact of electroweak
  group representation in models for B and g-2 anomalies from dark loops}},
  \href{http://dx.doi.org/10.1103/PhysRevD.106.095032}{\emph{Phys. Rev. D}
  \textbf{ 106} (2022) 095032},
  [\href{https://arxiv.org/abs/2207.11556}{{\texttt 2207.11556}}].

\bibitem{Cerdeno:2019vpd}
D.~G. Cerde\~no, A.~Cheek, P.~Mart\'\i{}n-Ramiro and J.~M. Moreno, \emph{{B
  anomalies and dark matter: a complex connection}},
  \href{http://dx.doi.org/10.1140/epjc/s10052-019-6979-x}{\emph{Eur. Phys. J.
  C} \textbf{ 79} (2019) 517},
  [\href{https://arxiv.org/abs/1902.01789}{{\texttt 1902.01789}}].

\bibitem{ATLAS:2017hoo}
{\scshape ATLAS} collaboration, M.~Aaboud et~al., \emph{{Search for dark matter
  produced in association with bottom or top quarks in $\sqrt{s}=13$ TeV pp
  collisions with the ATLAS detector}},
  \href{http://dx.doi.org/10.1140/epjc/s10052-017-5486-1}{\emph{Eur. Phys. J.
  C} \textbf{ 78} (2018) 18}, [\href{https://arxiv.org/abs/1710.11412}{{\texttt
  1710.11412}}].

\bibitem{Georgi:1977gs}
H.~M. Georgi, S.~L. Glashow, M.~E. Machacek and D.~V. Nanopoulos, \emph{{Higgs
  Bosons from Two Gluon Annihilation in Proton Proton Collisions}},
  \href{http://dx.doi.org/10.1103/PhysRevLett.40.692}{\emph{Phys. Rev. Lett.}
  \textbf{ 40} (1978) 692}.

\bibitem{Spira:2016ztx}
M.~Spira, \emph{{Higgs Boson Production and Decay at Hadron Colliders}},
  \href{http://dx.doi.org/10.1016/j.ppnp.2017.04.001}{\emph{Prog. Part. Nucl.
  Phys.} \textbf{ 95} (2017) 98--159},
  [\href{https://arxiv.org/abs/1612.07651}{{\texttt 1612.07651}}].

\bibitem{Spira:1995mt}
M.~Spira, \emph{{HIGLU: A program for the calculation of the total Higgs
  production cross-section at hadron colliders via gluon fusion including QCD
  corrections}},  \href{https://arxiv.org/abs/hep-ph/9510347}{{\texttt
  hep-ph/9510347}}.

\bibitem{DJOUADI1991440}
A.~Djouadi, M.~Spira and P.~Zerwas, \emph{Production of {Higgs} bosons in
  proton colliders. {QCD} corrections},
  \href{http://dx.doi.org/https://doi.org/10.1016/0370-2693(91)90375-Z}{\emph{Physics
  Letters B} \textbf{ 264} (1991) 440--446}.

\bibitem{DAWSON1991283}
S.~Dawson, \emph{Radiative corrections to {Higgs} boson production},
  \href{http://dx.doi.org/https://doi.org/10.1016/0550-3213(91)90061-2}{\emph{Nuclear
  Physics B} \textbf{ 359} (1991) 283--300}.

\bibitem{Graudenz:1992pv}
D.~Graudenz, M.~Spira and P.~M. Zerwas, \emph{{QCD corrections to Higgs boson
  production at proton proton colliders}},
  \href{http://dx.doi.org/10.1103/PhysRevLett.70.1372}{\emph{Phys. Rev. Lett.}
  \textbf{ 70} (1993) 1372--1375}.

\bibitem{Spira:1995rr}
M.~Spira, A.~Djouadi, D.~Graudenz and P.~M. Zerwas, \emph{{Higgs boson
  production at the LHC}},
  \href{http://dx.doi.org/10.1016/0550-3213(95)00379-7}{\emph{Nucl. Phys. B}
  \textbf{ 453} (1995) 17--82},
  [\href{https://arxiv.org/abs/hep-ph/9504378}{{\texttt hep-ph/9504378}}].

\bibitem{Harlander:2005rq}
R.~Harlander and P.~Kant, \emph{{Higgs production and decay: Analytic results
  at next-to-leading order QCD}},
  \href{http://dx.doi.org/10.1088/1126-6708/2005/12/015}{\emph{JHEP} \textbf{
  12} (2005) 015}, [\href{https://arxiv.org/abs/hep-ph/0509189}{{\texttt
  hep-ph/0509189}}].

\bibitem{Anastasiou:2006hc}
C.~Anastasiou, S.~Beerli, S.~Bucherer, A.~Daleo and Z.~Kunszt, \emph{{Two-loop
  amplitudes and master integrals for the production of a Higgs boson via a
  massive quark and a scalar-quark loop}},
  \href{http://dx.doi.org/10.1088/1126-6708/2007/01/082}{\emph{JHEP} \textbf{
  01} (2007) 082}, [\href{https://arxiv.org/abs/hep-ph/0611236}{{\texttt
  hep-ph/0611236}}].

\bibitem{Aglietti:2006tp}
U.~Aglietti, R.~Bonciani, G.~Degrassi and A.~Vicini, \emph{{Analytic Results
  for Virtual QCD Corrections to Higgs Production and Decay}},
  \href{http://dx.doi.org/10.1088/1126-6708/2007/01/021}{\emph{JHEP} \textbf{
  01} (2007) 021}, [\href{https://arxiv.org/abs/hep-ph/0611266}{{\texttt
  hep-ph/0611266}}].

\bibitem{Anastasiou:2009kn}
C.~Anastasiou, S.~Bucherer and Z.~Kunszt, \emph{{HPro: A NLO Monte-Carlo for
  Higgs production via gluon fusion with finite heavy quark masses}},
  \href{http://dx.doi.org/10.1088/1126-6708/2009/10/068}{\emph{JHEP} \textbf{
  10} (2009) 068}, [\href{https://arxiv.org/abs/0907.2362}{{\texttt
  0907.2362}}].

\bibitem{Catani:2001ic}
S.~Catani, D.~de~Florian and M.~Grazzini, \emph{{Higgs production in hadron
  collisions: Soft and virtual QCD corrections at NNLO}},
  \href{http://dx.doi.org/10.1088/1126-6708/2001/05/025}{\emph{JHEP} \textbf{
  05} (2001) 025}, [\href{https://arxiv.org/abs/hep-ph/0102227}{{\texttt
  hep-ph/0102227}}].

\bibitem{Harlander:2001is}
R.~V. Harlander and W.~B. Kilgore, \emph{{Soft and virtual corrections to
  proton proton ---\ensuremath{>} H + x at NNLO}},
  \href{http://dx.doi.org/10.1103/PhysRevD.64.013015}{\emph{Phys. Rev. D}
  \textbf{ 64} (2001) 013015},
  [\href{https://arxiv.org/abs/hep-ph/0102241}{{\texttt hep-ph/0102241}}].

\bibitem{Harlander:2002wh}
R.~V. Harlander and W.~B. Kilgore, \emph{{Next-to-next-to-leading order Higgs
  production at hadron colliders}},
  \href{http://dx.doi.org/10.1103/PhysRevLett.88.201801}{\emph{Phys. Rev.
  Lett.} \textbf{ 88} (2002) 201801},
  [\href{https://arxiv.org/abs/hep-ph/0201206}{{\texttt hep-ph/0201206}}].

\bibitem{Anastasiou:2002yz}
C.~Anastasiou and K.~Melnikov, \emph{{Higgs boson production at hadron
  colliders in NNLO QCD}},
  \href{http://dx.doi.org/10.1016/S0550-3213(02)00837-4}{\emph{Nucl. Phys. B}
  \textbf{ 646} (2002) 220--256},
  [\href{https://arxiv.org/abs/hep-ph/0207004}{{\texttt hep-ph/0207004}}].

\bibitem{RAVINDRAN2003325}
V.~Ravindran, J.~Smith and W.~{van Neerven}, \emph{{NNLO} corrections to the
  total cross section for {Higgs} boson production in hadron-hadron
  collisions},
  \href{http://dx.doi.org/https://doi.org/10.1016/S0550-3213(03)00457-7}{\emph{Nuclear
  Physics B} \textbf{ 665} (2003) 325--366}.

\bibitem{MARZANI2008127}
S.~Marzani, R.~D. Ball, V.~{Del Duca}, S.~Forte and A.~Vicini, \emph{{Higgs}
  production via gluon-gluon fusion with finite top mass beyond next-to-leading
  order},
  \href{http://dx.doi.org/https://doi.org/10.1016/j.nuclphysb.2008.03.016}{\emph{Nuclear
  Physics B} \textbf{ 800} (2008) 127--145}.

\bibitem{Gehrmann:2011aa}
T.~Gehrmann, M.~Jaquier, E.~W.~N. Glover and A.~Koukoutsakis, \emph{{Two-Loop
  QCD Corrections to the Helicity Amplitudes for $H \to$ 3 partons}},
  \href{http://dx.doi.org/10.1007/JHEP02(2012)056}{\emph{JHEP} \textbf{ 02}
  (2012) 056}, [\href{https://arxiv.org/abs/1112.3554}{{\texttt 1112.3554}}].

\bibitem{Anastasiou:2013srw}
C.~Anastasiou, C.~Duhr, F.~Dulat and B.~Mistlberger, \emph{{Soft triple-real
  radiation for Higgs production at N3LO}},
  \href{http://dx.doi.org/10.1007/JHEP07(2013)003}{\emph{JHEP} \textbf{ 07}
  (2013) 003}, [\href{https://arxiv.org/abs/1302.4379}{{\texttt 1302.4379}}].

\bibitem{Anastasiou:2013mca}
C.~Anastasiou, C.~Duhr, F.~Dulat, F.~Herzog and B.~Mistlberger,
  \emph{{Real-virtual contributions to the inclusive Higgs cross-section at
  $N^3LO$}}, \href{http://dx.doi.org/10.1007/JHEP12(2013)088}{\emph{JHEP}
  \textbf{ 12} (2013) 088}, [\href{https://arxiv.org/abs/1311.1425}{{\texttt
  1311.1425}}].

\bibitem{Kilgore:2013gba}
W.~B. Kilgore, \emph{{One-loop single-real-emission contributions to $pp\to H +
  X$ at next-to-next-to-next-to-leading order}},
  \href{http://dx.doi.org/10.1103/PhysRevD.89.073008}{\emph{Phys. Rev. D}
  \textbf{ 89} (2014) 073008}, [\href{https://arxiv.org/abs/1312.1296}{{\texttt
  1312.1296}}].

\bibitem{Li:2014bfa}
Y.~Li, A.~von Manteuffel, R.~M. Schabinger and H.~X. Zhu, \emph{{N$^3$LO Higgs
  boson and Drell-Yan production at threshold: The one-loop two-emission
  contribution}},
  \href{http://dx.doi.org/10.1103/PhysRevD.90.053006}{\emph{Phys. Rev. D}
  \textbf{ 90} (2014) 053006}, [\href{https://arxiv.org/abs/1404.5839}{{\texttt
  1404.5839}}].

\bibitem{Anastasiou:2014lda}
C.~Anastasiou, C.~Duhr, F.~Dulat, E.~Furlan, T.~Gehrmann, F.~Herzog et~al.,
  \emph{{Higgs Boson GluonFfusion Production Beyond Threshold in N$^{3}LO$
  QCD}}, \href{http://dx.doi.org/10.1007/JHEP03(2015)091}{\emph{JHEP} \textbf{
  03} (2015) 091}, [\href{https://arxiv.org/abs/1411.3584}{{\texttt
  1411.3584}}].

\bibitem{Anastasiou:2015vya}
C.~Anastasiou, C.~Duhr, F.~Dulat, F.~Herzog and B.~Mistlberger, \emph{{Higgs
  Boson Gluon-Fusion Production in QCD at Three Loops}},
  \href{http://dx.doi.org/10.1103/PhysRevLett.114.212001}{\emph{Phys. Rev.
  Lett.} \textbf{ 114} (2015) 212001},
  [\href{https://arxiv.org/abs/1503.06056}{{\texttt 1503.06056}}].

\bibitem{Anastasiou:2016cez}
C.~Anastasiou, C.~Duhr, F.~Dulat, E.~Furlan, T.~Gehrmann, F.~Herzog et~al.,
  \emph{{High precision determination of the gluon fusion Higgs boson
  cross-section at the LHC}},
  \href{http://dx.doi.org/10.1007/JHEP05(2016)058}{\emph{JHEP} \textbf{ 05}
  (2016) 058}, [\href{https://arxiv.org/abs/1602.00695}{{\texttt 1602.00695}}].

\bibitem{Mistlberger:2018etf}
B.~Mistlberger, \emph{{Higgs boson production at hadron colliders at N$^{3}$LO
  in QCD}}, \href{http://dx.doi.org/10.1007/JHEP05(2018)028}{\emph{JHEP}
  \textbf{ 05} (2018) 028}, [\href{https://arxiv.org/abs/1802.00833}{{\texttt
  1802.00833}}].

\bibitem{Duhr:2022cob}
C.~Duhr, B.~Mistlberger and G.~Vita, \emph{{Soft integrals and soft anomalous
  dimensions at N$^{3}$LO and beyond}},
  \href{http://dx.doi.org/10.1007/JHEP09(2022)155}{\emph{JHEP} \textbf{ 09}
  (2022) 155}, [\href{https://arxiv.org/abs/2205.04493}{{\texttt 2205.04493}}].

\bibitem{Baglio:2022wzu}
J.~Baglio, C.~Duhr, B.~Mistlberger and R.~Szafron, \emph{{Inclusive production
  cross sections at N$^{3}$LO}},
  \href{http://dx.doi.org/10.1007/JHEP12(2022)066}{\emph{JHEP} \textbf{ 12}
  (2022) 066}, [\href{https://arxiv.org/abs/2209.06138}{{\texttt 2209.06138}}].

\bibitem{Harlander:2009bw}
R.~V. Harlander and K.~J. Ozeren, \emph{{Top mass effects in Higgs production
  at next-to-next-to-leading order QCD: Virtual corrections}},
  \href{http://dx.doi.org/10.1016/j.physletb.2009.08.012}{\emph{Phys. Lett. B}
  \textbf{ 679} (2009) 467--472},
  [\href{https://arxiv.org/abs/0907.2997}{{\texttt 0907.2997}}].

\bibitem{Harlander:2009mq}
R.~V. Harlander and K.~J. Ozeren, \emph{{Finite top mass effects for hadronic
  Higgs production at next-to-next-to-leading order}},
  \href{http://dx.doi.org/10.1088/1126-6708/2009/11/088}{\emph{JHEP} \textbf{
  11} (2009) 088}, [\href{https://arxiv.org/abs/0909.3420}{{\texttt
  0909.3420}}].

\bibitem{Harlander:2009my}
R.~V. Harlander, H.~Mantler, S.~Marzani and K.~J. Ozeren, \emph{{Higgs
  production in gluon fusion at next-to-next-to-leading order QCD for finite
  top mass}},
  \href{http://dx.doi.org/10.1140/epjc/s10052-010-1258-x}{\emph{Eur. Phys. J.
  C} \textbf{ 66} (2010) 359--372},
  [\href{https://arxiv.org/abs/0912.2104}{{\texttt 0912.2104}}].

\bibitem{Pak:2009dg}
A.~Pak, M.~Rogal and M.~Steinhauser, \emph{{Finite top quark mass effects in
  NNLO Higgs boson production at LHC}},
  \href{http://dx.doi.org/10.1007/JHEP02(2010)025}{\emph{JHEP} \textbf{ 02}
  (2010) 025}, [\href{https://arxiv.org/abs/0911.4662}{{\texttt 0911.4662}}].

\bibitem{Djouadi:1994ge}
A.~Djouadi and P.~Gambino, \emph{{Leading electroweak correction to Higgs boson
  production at proton colliders}},
  \href{http://dx.doi.org/10.1103/PhysRevLett.73.2528}{\emph{Phys. Rev. Lett.}
  \textbf{ 73} (1994) 2528--2531},
  [\href{https://arxiv.org/abs/hep-ph/9406432}{{\texttt hep-ph/9406432}}].

\bibitem{Chetyrkin:1996ke}
K.~G. Chetyrkin, B.~A. Kniehl and M.~Steinhauser, \emph{{Three loop O
  (alpha-s**2 G(F) M(t)**2) corrections to hadronic Higgs decays}},
  \href{http://dx.doi.org/10.1016/S0550-3213(97)00051-5}{\emph{Nucl. Phys. B}
  \textbf{ 490} (1997) 19--39},
  [\href{https://arxiv.org/abs/hep-ph/9701277}{{\texttt hep-ph/9701277}}].

\bibitem{Chetyrkin:1996wr}
K.~G. Chetyrkin, B.~A. Kniehl and M.~Steinhauser, \emph{{Virtual top quark
  effects on the H ---\ensuremath{>} b anti-b decay at next-to-leading order in
  QCD}}, \href{http://dx.doi.org/10.1103/PhysRevLett.78.594}{\emph{Phys. Rev.
  Lett.} \textbf{ 78} (1997) 594--597},
  [\href{https://arxiv.org/abs/hep-ph/9610456}{{\texttt hep-ph/9610456}}].

\bibitem{Degrassi:2004mx}
G.~Degrassi and F.~Maltoni, \emph{{Two-loop electroweak corrections to Higgs
  production at hadron colliders}},
  \href{http://dx.doi.org/10.1016/j.physletb.2004.09.008}{\emph{Phys. Lett. B}
  \textbf{ 600} (2004) 255--260},
  [\href{https://arxiv.org/abs/hep-ph/0407249}{{\texttt hep-ph/0407249}}].

\bibitem{Aglietti:2006yd}
U.~Aglietti, R.~Bonciani, G.~Degrassi and A.~Vicini, \emph{{Two-loop
  electroweak corrections to Higgs production in proton-proton collisions}},
  in \emph{{TeV4LHC Workshop: 2nd Meeting}}, 10, 2006.
\newblock \href{https://arxiv.org/abs/hep-ph/0610033}{{\texttt
  hep-ph/0610033}}.

\bibitem{Actis:2008ug}
S.~Actis, G.~Passarino, C.~Sturm and S.~Uccirati, \emph{{NLO Electroweak
  Corrections to Higgs Boson Production at Hadron Colliders}},
  \href{http://dx.doi.org/10.1016/j.physletb.2008.10.018}{\emph{Phys. Lett. B}
  \textbf{ 670} (2008) 12--17},
  [\href{https://arxiv.org/abs/0809.1301}{{\texttt 0809.1301}}].

\bibitem{Actis:2008ts}
S.~Actis, G.~Passarino, C.~Sturm and S.~Uccirati, \emph{{NNLO Computational
  Techniques: The Cases H ---\ensuremath{>} gamma gamma and H ---\ensuremath{>}
  g g}}, \href{http://dx.doi.org/10.1016/j.nuclphysb.2008.11.024}{\emph{Nucl.
  Phys. B} \textbf{ 811} (2009) 182--273},
  [\href{https://arxiv.org/abs/0809.3667}{{\texttt 0809.3667}}].

\bibitem{Anastasiou:2008tj}
C.~Anastasiou, R.~Boughezal and F.~Petriello, \emph{{Mixed QCD-electroweak
  corrections to Higgs boson production in gluon fusion}},
  \href{http://dx.doi.org/10.1088/1126-6708/2009/04/003}{\emph{JHEP} \textbf{
  04} (2009) 003}, [\href{https://arxiv.org/abs/0811.3458}{{\texttt
  0811.3458}}].

\bibitem{Butterworth:2015oua}
J.~Butterworth et~al., \emph{{PDF4LHC recommendations for LHC Run II}},
  \href{http://dx.doi.org/10.1088/0954-3899/43/2/023001}{\emph{J. Phys. G}
  \textbf{ 43} (2016) 023001},
  [\href{https://arxiv.org/abs/1510.03865}{{\texttt 1510.03865}}].

\bibitem{NNPDF:2021njg}
{\scshape NNPDF} collaboration, R.~D. Ball et~al., \emph{{The path to proton
  structure at 1\% accuracy}},
  \href{http://dx.doi.org/10.1140/epjc/s10052-022-10328-7}{\emph{Eur. Phys. J.
  C} \textbf{ 82} (2022) 428},
  [\href{https://arxiv.org/abs/2109.02653}{{\texttt 2109.02653}}].

\bibitem{Crivellin:2021ejk}
A.~Crivellin and L.~Schnell, \emph{{Complete Lagrangian and set of Feynman
  rules for scalar leptoquarks}},
  \href{http://dx.doi.org/10.1016/j.cpc.2021.108188}{\emph{Comput. Phys.
  Commun.} \textbf{ 271} (2022) 108188},
  [\href{https://arxiv.org/abs/2105.04844}{{\texttt 2105.04844}}].

\bibitem{Ferreira:2004yd}
P.~M. Ferreira, R.~Santos and A.~Barroso, \emph{{Stability of the tree-level
  vacuum in two Higgs doublet models against charge or CP spontaneous
  violation}},
  \href{http://dx.doi.org/10.1016/j.physletb.2004.10.022}{\emph{Phys. Lett. B}
  \textbf{ 603} (2004) 219--229},
  [\href{https://arxiv.org/abs/hep-ph/0406231}{{\texttt hep-ph/0406231}}].

\bibitem{ATLAS:2019nkf}
{\scshape ATLAS} collaboration, G.~Aad et~al., \emph{{Combined measurements of
  Higgs boson production and decay using up to $80$ fb$^{-1}$ of proton-proton
  collision data at $\sqrt{s}=$ 13 TeV collected with the ATLAS experiment}},
  \href{http://dx.doi.org/10.1103/PhysRevD.101.012002}{\emph{Phys. Rev. D}
  \textbf{ 101} (2020) 012002},
  [\href{https://arxiv.org/abs/1909.02845}{{\texttt 1909.02845}}].

\bibitem{CMS:2018uag}
{\scshape CMS} collaboration, A.~M. Sirunyan et~al., \emph{{Combined
  measurements of Higgs boson couplings in proton\textendash{}proton collisions
  at $\sqrt{s}=13\,\text {Te}\text {V} $}},
  \href{http://dx.doi.org/10.1140/epjc/s10052-019-6909-y}{\emph{Eur. Phys. J.
  C} \textbf{ 79} (2019) 421},
  [\href{https://arxiv.org/abs/1809.10733}{{\texttt 1809.10733}}].

\bibitem{cepeda2019higgs}
M.~Cepeda, S.~Gori, P.~Ilten, M.~Kado, F.~Riva, R.~A. Khalek et~al.,
  \emph{Higgs physics at the {HL-LHC} and {HE-LHC}},
  \href{https://arxiv.org/abs/1902.00134}{{\texttt 1902.00134}}.

\bibitem{Glover:1987nx}
E.~W.~N. Glover and J.~J. van~der Bij, \emph{{HIGGS BOSON PAIR PRODUCTION VIA
  GLUON FUSION}},
  \href{http://dx.doi.org/10.1016/0550-3213(88)90083-1}{\emph{Nucl. Phys. B}
  \textbf{ 309} (1988) 282--294}.

\bibitem{Plehn:1996wb}
T.~Plehn, M.~Spira and P.~M. Zerwas, \emph{{Pair production of neutral Higgs
  particles in gluon-gluon collisions}},
  \href{http://dx.doi.org/10.1016/0550-3213(96)00418-X}{\emph{Nucl. Phys. B}
  \textbf{ 479} (1996) 46--64},
  [\href{https://arxiv.org/abs/hep-ph/9603205}{{\texttt hep-ph/9603205}}].

\bibitem{HAHN2001418}
T.~Hahn, \emph{Generating feynman diagrams and amplitudes with {FeynArts} 3},
  \href{http://dx.doi.org/https://doi.org/10.1016/S0010-4655(01)00290-9}{\emph{Computer
  Physics Communications} \textbf{ 140} (2001) 418--431}.

\bibitem{feynarts_manual_link}
{Thomas Hahn}, ``{{FeynArts} 3.11 {User’s} {Guide}}.''
  \url{https://feynarts.de/FA3Guide.pdf}, (accessed 2023-07-26).

\bibitem{Kribs:2012kz}
G.~D. Kribs and A.~Martin, \emph{{Enhanced di-Higgs Production through Light
  Colored Scalars}},
  \href{http://dx.doi.org/10.1103/PhysRevD.86.095023}{\emph{Phys. Rev. D}
  \textbf{ 86} (2012) 095023}, [\href{https://arxiv.org/abs/1207.4496}{{\texttt
  1207.4496}}].

\bibitem{Enkhbat:2013oba}
T.~Enkhbat, \emph{{Scalar leptoquarks and Higgs pair production at the LHC}},
  \href{http://dx.doi.org/10.1007/JHEP01(2014)158}{\emph{JHEP} \textbf{ 01}
  (2014) 158}, [\href{https://arxiv.org/abs/1311.4445}{{\texttt 1311.4445}}].

\bibitem{Dawson:1998py}
S.~Dawson, S.~Dittmaier and M.~Spira, \emph{{Neutral Higgs boson pair
  production at hadron colliders: QCD corrections}},
  \href{http://dx.doi.org/10.1103/PhysRevD.58.115012}{\emph{Phys. Rev. D}
  \textbf{ 58} (1998) 115012},
  [\href{https://arxiv.org/abs/hep-ph/9805244}{{\texttt hep-ph/9805244}}].

\bibitem{Borowka:2016ehy}
S.~Borowka, N.~Greiner, G.~Heinrich, S.~P. Jones, M.~Kerner, J.~Schlenk et~al.,
  \emph{{Higgs Boson Pair Production in Gluon Fusion at Next-to-Leading Order
  with Full Top-Quark Mass Dependence}},
  \href{http://dx.doi.org/10.1103/PhysRevLett.117.079901,
  10.1103/PhysRevLett.117.012001}{\emph{Phys. Rev. Lett.} \textbf{ 117} (2016)
  012001}, [\href{https://arxiv.org/abs/1604.06447}{{\texttt 1604.06447}}].

\bibitem{Borowka:2016ypz}
S.~Borowka, N.~Greiner, G.~Heinrich, S.~P. Jones, M.~Kerner, J.~Schlenk et~al.,
  \emph{{Full top quark mass dependence in Higgs boson pair production at
  NLO}}, \href{http://dx.doi.org/10.1007/JHEP10(2016)107}{\emph{JHEP} \textbf{
  10} (2016) 107}, [\href{https://arxiv.org/abs/1608.04798}{{\texttt
  1608.04798}}].

\bibitem{Baglio:2018lrj}
J.~Baglio, F.~Campanario, S.~Glaus, M.~MÃŒhlleitner, M.~Spira and
  J.~Streicher, \emph{{Gluon fusion into Higgs pairs at NLO QCD and the top
  mass scheme}},
  \href{http://dx.doi.org/10.1140/epjc/s10052-019-6973-3}{\emph{Eur. Phys. J.}
  \textbf{ C79} (2019) 459}, [\href{https://arxiv.org/abs/1811.05692}{{\texttt
  1811.05692}}].

\bibitem{Baglio:2020ini}
J.~Baglio, F.~Campanario, S.~Glaus, M.~M\"uhlleitner, J.~Ronca, M.~Spira
  et~al., \emph{{Higgs-Pair Production via Gluon Fusion at Hadron Colliders:
  NLO QCD Corrections}},
  \href{http://dx.doi.org/10.1007/JHEP04(2020)181}{\emph{JHEP} \textbf{ 04}
  (2020) 181}, [\href{https://arxiv.org/abs/2003.03227}{{\texttt 2003.03227}}].

\bibitem{Baglio:2020wgt}
J.~Baglio, F.~Campanario, S.~Glaus, M.~M\"uhlleitner, J.~Ronca and M.~Spira,
  \emph{{$gg\to HH$ : Combined uncertainties}},
  \href{http://dx.doi.org/10.1103/PhysRevD.103.056002}{\emph{Phys. Rev. D}
  \textbf{ 103} (2021) 056002},
  [\href{https://arxiv.org/abs/2008.11626}{{\texttt 2008.11626}}].

\bibitem{deFlorian:2013jea}
D.~de~Florian and J.~Mazzitelli, \emph{{Higgs Boson Pair Production at
  Next-to-Next-to-Leading Order in QCD}},
  \href{http://dx.doi.org/10.1103/PhysRevLett.111.201801}{\emph{Phys. Rev.
  Lett.} \textbf{ 111} (2013) 201801},
  [\href{https://arxiv.org/abs/1309.6594}{{\texttt 1309.6594}}].

\bibitem{Grigo:2014jma}
J.~Grigo, K.~Melnikov and M.~Steinhauser, \emph{{Virtual corrections to Higgs
  boson pair production in the large top quark mass limit}},
  \href{http://dx.doi.org/10.1016/j.nuclphysb.2014.09.003}{\emph{Nucl. Phys. B}
  \textbf{ 888} (2014) 17--29},
  [\href{https://arxiv.org/abs/1408.2422}{{\texttt 1408.2422}}].

\bibitem{Shao:2013bz}
D.~Y. Shao, C.~S. Li, H.~T. Li and J.~Wang, \emph{{Threshold resummation
  effects in Higgs boson pair production at the LHC}},
  \href{http://dx.doi.org/10.1007/JHEP07(2013)169}{\emph{JHEP} \textbf{ 07}
  (2013) 169}, [\href{https://arxiv.org/abs/1301.1245}{{\texttt 1301.1245}}].

\bibitem{deFlorian:2015moa}
D.~de~Florian and J.~Mazzitelli, \emph{{Higgs pair production at
  next-to-next-to-leading logarithmic accuracy at the LHC}},
  \href{http://dx.doi.org/10.1007/JHEP09(2015)053}{\emph{JHEP} \textbf{ 09}
  (2015) 053}, [\href{https://arxiv.org/abs/1505.07122}{{\texttt 1505.07122}}].

\bibitem{Banerjee:2018lfq}
P.~Banerjee, S.~Borowka, P.~K. Dhani, T.~Gehrmann and V.~Ravindran,
  \emph{{Two-loop massless QCD corrections to the $g + g \to H + H$ four-point
  amplitude}}, \href{http://dx.doi.org/10.1007/JHEP11(2018)130}{\emph{JHEP}
  \textbf{ 11} (2018) 130}, [\href{https://arxiv.org/abs/1809.05388}{{\texttt
  1809.05388}}].

\bibitem{Chen:2019lzz}
L.-B. Chen, H.~T. Li, H.-S. Shao and J.~Wang, \emph{{Higgs boson pair
  production via gluon fusion at N$^3$LO in QCD}},
  \href{http://dx.doi.org/10.1016/j.physletb.2020.135292}{\emph{Phys. Lett. B}
  \textbf{ 803} (2020) 135292},
  [\href{https://arxiv.org/abs/1909.06808}{{\texttt 1909.06808}}].

\bibitem{Chen:2019fhs}
L.-B. Chen, H.~T. Li, H.-S. Shao and J.~Wang, \emph{{The gluon-fusion
  production of Higgs boson pair: N$^3$LO QCD corrections and top-quark mass
  effects}}, \href{http://dx.doi.org/10.1007/JHEP03(2020)072}{\emph{JHEP}
  \textbf{ 03} (2020) 072}, [\href{https://arxiv.org/abs/1912.13001}{{\texttt
  1912.13001}}].

\bibitem{Ajjath:2022kpv}
A.~H. Ajjath and H.-S. Shao, \emph{{N$^{3}$LO+N$^{3}$LL QCD improved Higgs pair
  cross sections}},
  \href{http://dx.doi.org/10.1007/JHEP02(2023)067}{\emph{JHEP} \textbf{ 02}
  (2023) 067}, [\href{https://arxiv.org/abs/2209.03914}{{\texttt 2209.03914}}].

\bibitem{DiMicco:2019ngk}
J.~Alison et~al., \emph{{Higgs boson potential at colliders: Status and
  perspectives}},
  \href{http://dx.doi.org/10.1016/j.revip.2020.100045}{\emph{Rev. Phys.}
  \textbf{ 5} (2020) 100045}, [\href{https://arxiv.org/abs/1910.00012}{{\texttt
  1910.00012}}].

\bibitem{Muhlleitner:2022ijf}
M.~M\"uhlleitner, J.~Schlenk and M.~Spira, \emph{{Top-Yukawa-induced
  corrections to Higgs pair production}},
  \href{http://dx.doi.org/10.1007/JHEP10(2022)185}{\emph{JHEP} \textbf{ 10}
  (2022) 185}, [\href{https://arxiv.org/abs/2207.02524}{{\texttt 2207.02524}}].

\bibitem{Davies:2022ram}
J.~Davies, G.~Mishima, K.~Sch\"onwald, M.~Steinhauser and H.~Zhang,
  \emph{{Higgs boson contribution to the leading two-loop Yukawa corrections to
  gg $rightarrow$ HH}},
  \href{http://dx.doi.org/10.1007/JHEP08(2022)259}{\emph{JHEP} \textbf{ 08}
  (2022) 259}, [\href{https://arxiv.org/abs/2207.02587}{{\texttt 2207.02587}}].

\bibitem{Davies:2023obx}
J.~Davies, K.~Sch\"onwald and M.~Steinhauser, \emph{{Towards $gg\to HH$ at
  next-to-next-to-leading order: light-fermionic three-loop corrections}},
  \href{https://arxiv.org/abs/2307.04796}{{\texttt 2307.04796}}.

\bibitem{Davies:2023npk}
J.~Davies, K.~Sch\"onwald, M.~Steinhauser and H.~Zhang, \emph{{Next-to-leading
  order electroweak corrections to $gg \to HH$ and $gg \to gH$ in the
  large-$m_t$ limit}},  \href{https://arxiv.org/abs/2308.01355}{{\texttt
  2308.01355}}.

\bibitem{Daniel}
D.~Neacsu, \emph{{Higgs Production at the LHC with Colored Scalars from B-meson
  Decays Models}},  Master's thesis, IST, University of Lisbon, 9, 2023.

\bibitem{ATLAS:2019qdc}
{\scshape ATLAS} collaboration, G.~Aad et~al., \emph{{Combination of searches
  for Higgs boson pairs in $pp$ collisions at $\sqrt{s} = $13 TeV with the
  ATLAS detector}},
  \href{http://dx.doi.org/10.1016/j.physletb.2019.135103}{\emph{Phys. Lett. B}
  \textbf{ 800} (2020) 135103},
  [\href{https://arxiv.org/abs/1906.02025}{{\texttt 1906.02025}}].

\bibitem{Agostini:2016vze}
A.~Agostini, G.~Degrassi, R.~Gr\"ober and P.~Slavich, \emph{{NLO-QCD
  corrections to Higgs pair production in the MSSM}},
  \href{http://dx.doi.org/10.1007/JHEP04(2016)106}{\emph{JHEP} \textbf{ 04}
  (2016) 106}, [\href{https://arxiv.org/abs/1601.03671}{{\texttt 1601.03671}}].

\bibitem{Degrassi:2008zj}
G.~Degrassi and P.~Slavich, \emph{{On the NLO QCD corrections to Higgs
  production and decay in the MSSM}},
  \href{http://dx.doi.org/10.1016/j.nuclphysb.2008.07.022}{\emph{Nucl. Phys. B}
  \textbf{ 805} (2008) 267--286},
  [\href{https://arxiv.org/abs/0806.1495}{{\texttt 0806.1495}}].

\bibitem{Gabriel2023BDMHPAIR}
P.~Gabriel, M.~Mühlleitner, D.~Neacsu and R.~Santos, ``{BDM-HPAIR}.''
  \url{https://gitlab.com/bdm-models/higgs-production/bdm-hpair}, 2023.

\end{thebibliography}\endgroup

\end{document}